%% file: main.tex
\documentclass[nopreprintline,3p]{elsarticle}

\usepackage[caption=false]{subfig}

\usepackage{amssymb}
\usepackage{amsmath}
\usepackage{tabularx}
\usepackage{booktabs}

\usepackage{bm}
\usepackage{textcomp, gensymb}
\usepackage{lmodern}
\usepackage{multicol}
\usepackage{tikz}
\usetikzlibrary{spy}
\usetikzlibrary{calc} 
\usepackage{siunitx}
\usepackage{xcolor}

\def\draftmode{false}

\newcommand{\imgfolder}{figures_jpeg}

\newcommand{\wss}{\text{WSS}}
\newcommand{\tawss}{\text{TAWSS}}
\newcommand{\osi}{\text{OSI}}
\newcommand{\rrt}{\text{RRT}}

\begin{document}

\begin{frontmatter}



\title{In-silico Analysis of Hemodynamic Indicators in Idealized Stented Coronary Arteries for Varying Stent Indentation}

\author[label1]{A. M.~Ranno\corref{cor1}}
\author[label2]{K. Manjunatha}
\author[label3]{A. Glitz}
\author[label3]{N. Schaaps}
\author[label2]{S. Reese}
\author[label3]{F. Vogt}
\author[label1]{M. Behr}

\cortext[cor1]{Corresponding author. Email: ranno@cats.rwth-aachen.de}

\affiliation[label1]{organization={Chair for Computational Analysis of Technical Systems (CATS), RWTH Aachen University, Aachen, Germany}}
\affiliation[label2]{organization={Institute of Applied Mechanics (IFAM), RWTH Aachen University, Aachen, Germany}}
\affiliation[label3]{organization={Department of Cardiology, Vascular Medicine and Intensive Care (CARD), RWTH Aachen University, Aachen, Germany}}

\begin{abstract}
\input{sections/0-abstract}
\end{abstract}


\end{frontmatter}

\section{Introduction}
\label{1Intro}
\input{sections/1-introduction}
\section{Model and method}
\label{2ModelMethod}
In sections \ref{2aMethod} and \ref{2aMethodIndentation}, we present the workflow to obtain the computational domain of an idealized coronary artery lumen with \textit{Xience V} stent at different indentation levels. The computational model for blood flow and the derivation of hemodynamic indicators are described in section \ref{2bMethod}.
\input{sections/2a-stentedArtery}
\input{sections/2b-method}
\section{Numerical results}
\label{3Results}
\input{sections/3-results}
\section{Discussion}
\label{4Discussion}
\input{sections/4-discussion}
\section{Conclusions}
\label{5Conclusion}
\input{sections/5-conclusion}

\section*{Acknowledgements}
The authors especially thank Jana Sasse for her work on blood model investigations, Svenja Nerzak for her numerous contributions to the geometry, mesh modeling and model testing, Henrik Volgmann for his work during his project thesis and Michael Aichmüller for fruitful discussions. We especially thank Ulrich Fuchs for his guidance in mesh generation with Pointwise.\\

The authors gratefully acknowledge the computing time granted through JARA-HPC on the supercomputer JURECA at Forschungszentrum J\"ulich and on the supercomputer CLAIX at RWTH Aachen University. The authors gratefully acknowledge the computing time provided to them on the high-performance computer Lichtenberg at the NHR Centers NHR4CES at TU Darmstadt. This is funded by the Federal Ministry of Education and Research, and the state governments participating on the basis of the resolutions of the GWK for national high performance computing at universities (www.nhr-verein.de/unsere-partner).
\section*{Funding}
This work has been funded through the financial support of German Research Foundation (DFG) for the project “Drug-eluting coronary stents in stenosed arteries: medical investigation and computational modelling” (project number 395712048:BE 3689/15, RE 1057/44, VO1624/5) and for the subproject “In-stent restenosis in coronary arteries - in silico investigations based on patient-specific data and meta modeling” of SPP2311. This work was also partly supported by the IRTG-2379 Modern Inverse Problems which is funded by the DFG – 333849990/GRK2379.
\bibliographystyle{elsarticle-num}
\bibliography{main}


\end{document}

%% file: sections/0-abstract.tex
In this work, we investigate the effects of stent indentation on hemodynamic indicators in stented coronary arteries. Our aim is to assess in-silico risk factors for in-stent restenosis (ISR) and thrombosis after stent implantation. The proposed model is applied to an idealized artery with \textit{Xience V} stent for four indentation percentages and three mesh refinements. We analyze the patterns of hemodynamic indicators arising from different stent indentations and propose an empirical frequency analysis of time-averaged WSS (TAWSS), oscillatory shear index (OSI), and relative residence time (RRT). We observe that higher indentations display higher frequency of critically low TAWSS and non-physiological OSI and RRT. Furthermore, an appropriate mesh refinement is needed for accurate representation of hemodynamics in the stent vicinity. The results provide physics-based evidence for the correlation between high indentation and ISR.

%% file: sections/1-introduction.tex
Cardiovascular diseases are the most common cause of death worldwide. More specifically, 30\% of global mortality is due to coronary artery disease \cite{townsend2016cardiovascular, roth2020global}. Its treatment largely consists in a minimally invasive surgical procedure: a coronary stent implantation. During the stent implantation, the artery wall will be injured. This triggers an inflammatory reaction which implies a number of side-effects: endothelial damage, consequent neo-intima hyperplasia (NIH), in-stent restenosis (ISR) and long term lesions with 40\% occurrence rate, accumulation of platelets and possible thrombus formation, which is fatal in 45\% of the cases \cite{farooq2011restenosis,holmes2010stent}. Drug-eluting stents (DES) are coated with a layer of polymer and infused with drug, designed to mitigate such side-effects \cite{khan2012drug}. However, even after the introduction of DES, the percentage of stent implantation side-effects such as ISR and thrombosis is still significant \cite{douglas2012drug, buccheri2016understanding, MCQUEEN2021do,takayama2011stent}.\\

Research on stent implantation ranges from balloon pressure, to drug elution, vessel injury, optimal implantation and long term effects such as ISR \cite{zunino2009numerical, antoniadis2015biomechanical, wang2015three}. Complex coupled models include modeling the physical and biochemical species to ISR, \cite{yang2017ale, zun2017comparison, manjunatha2022multiphysics}, a structural framework for the artery wall \cite{romarowski2019novel, silva2020modeling}, and fluid-structure interaction (FSI) between the blood and the wall \cite{forti2017monolithic}. Numerous studies have shown the importance of altered hemodynamics in stented arteries \cite{ladisa2003three, zunino2009numerical, hachem2023reinforcement}, the influence of stent designs and vessel irregularities \cite{de2013computational} and flow patterns before and after stent implantation \cite{chiastra2013computational}. Furthermore, flow patterns and recirculation areas, in particular, greatly influence drug release in DES close to the stent struts \cite{calo2008multiphysics, kolachalama2009luminal, vergara2008multiscale}.\\

Blood is approximated as a homogeneous Newtonian fluid in large arteries with physiological flow \cite{gijsen1999influence, leuprecht2001computer}. Various studies have been conducted to assess shear thinning effects and blood damage in capillaries or pathological conditions \cite{behr2006models, behbahani2009review,hassler2019variational,marsden2014recent, Sasse}. Wall shear-stress (WSS) is often used as indicator of the well-being of the cardiovascular system \cite{hsiao2012hemodynamic, morbiducci2020wall, park2016vivo, zingaro2021hemodynamics, kronborg2023triple}. Blood recirculation occurs near areas of low WSS making them particularly dangerous: they are more prone to platelet aggregation, and therefore thrombus formation \cite{rayz2010flow}, endothelial dysfunction and induced inflammation resulting in NIH \cite{nakazawa2008delayed, cecchi2011role, koskinas2012role, brindise2017hemodynamics,jenei2016wall,wentzel2001relationship, stone2003effect}. In particular, prior studies have shown a correlation link between the WSS and ISR \cite{he2020mechanistic}. Hemodynamic indicators derived from WSS are time averaged WSS (TAWSS), oscillatory shear index (OSI), and relative residence time (RRT) \cite{soulis2011relative, john2017influence}.\\

One of the biggest challenges to model hemodynamics in stented arteries is to obtain the geometry of the artery lumen after stent implantation. Stent geometries can be very complex and difficult to reproduce \cite{auricchio2013patient, zunino2016integrated} and the arterial wall is inhomogeneous due to, e.g., calcified or necrotic areas. The artery model can be patient-specific \cite{kim2010patient}, obtained from Magnetic Resonance Imaging or CT scans \cite{takizawa2012patient,colombo2020computing,conti2016carotid} or an idealized cylindrical surface. Stents can also be implanted into the artery wall at different levels of depth, which depends on the pressure applied during the implantation. With "stent indentation" we refer to the impression of struts into the vessel wall after stent implantation \cite{Cornelissen2023Development}. To achieve a satisfactory minimum stent area, e.g., with calcified areas, the stent has to be expanded with a high pressure which results in a higher indentation. Qualitative injury scores, such as the Schwartz score \cite{schwartz1992restenosis} are used for estimation of ISR risk post-mortem. Low injury scores have been correlated with little NIH and healthy neo-intima growth. Different levels of indentation are associated with an injury score from 0 to 3, depending on the expected side-effects. A 0 injury score is associated with 0 to 10\% indentation, for 10\% to 25\% indentation injury score is 1, 25\% to 50\% indentation results in injury score 2 and 3 is reserved for higher indentation. If the stent is only 10\% indented into the artery wall, a successful stent implantation and few side effects such as ISR are expected. High indentation percentage means that a lot of pressure was applied during stent implantation and higher inflammation and vessel injury are expected. Yet the effect of indentation on hemodynamics remains uninvestigated.\\

In this work, we propose a 3D model of stented coronary artery for varying stent indentations and an in-silico analysis of hemodynamic indicators. We obtain a 3D CAD model of the \textit{Xience V} stent \cite{ding2009xience} scanned under electronic microscope and then computationally insert it into a cylindrical idealized artery. Blood flow is modeled with Navier Stokes equations, solved by means of stabilized Finite Element Method (FEM) and Backward Differentiation Formula (BDF) for time discretization. We investigate how indentation can influence blood flow indicators. In particular, critical areas are detected analyzing WSS, TAWSS, OSI, RRT and their empirical frequency. In the next section, we describe the in-silico model for hemodynamics in stented arteries. In particular, sections \ref{2aMethod} and \ref{2aMethodIndentation} introduce the geometry and computational domain of an idealized coronary artery lumen with a \textit{Xience V} for varying indentations and the choice of mesh refinements. In section \ref{2bMethod}, we derive the blood flow model and the hemodynamic indicators. We show the numerical results obtained for four indentation levels and three mesh refinements in section \ref{3Results}. In particular qualitative results of WSS and TAWSS are shown in sections \ref{3aWSSandTAWSS} and \ref{3bTAWSSDev} and critical threshold, OSI and RRT are highlighted in sections \ref{3cCriticalTAWSS} and \ref{3dOSIandRRT}. We discuss the numerical results and draw conclusions in sections \ref{4Discussion} and \ref{5Conclusion}, respectively.

%% file: sections/2a-stentedArtery.tex
\subsection{Geometry of expanded \textit{Xience V} stent}
\label{2aMethod}

\begin{figure}
	\centering
	\subfloat[Stent mounted on catheter and delivery balloon before implantation.]{%
		\resizebox*{5.3cm}{!}{
            \includegraphics[draft=\draftmode]{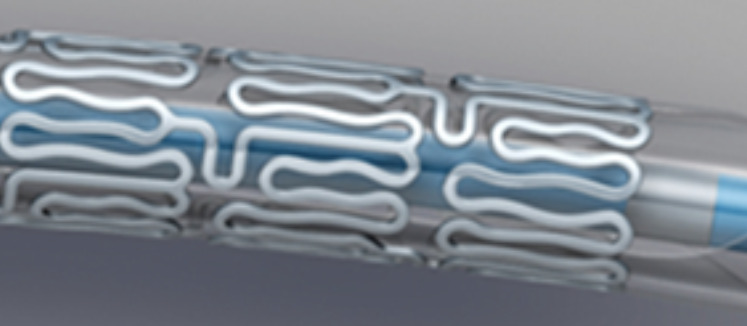}}\label{fig:stentCath}}
    \hspace{3pt}
	\subfloat[Crimped struts under scanning electron microscope.]{%
		\resizebox*{4.85cm}{!}{\includegraphics[draft=\draftmode]{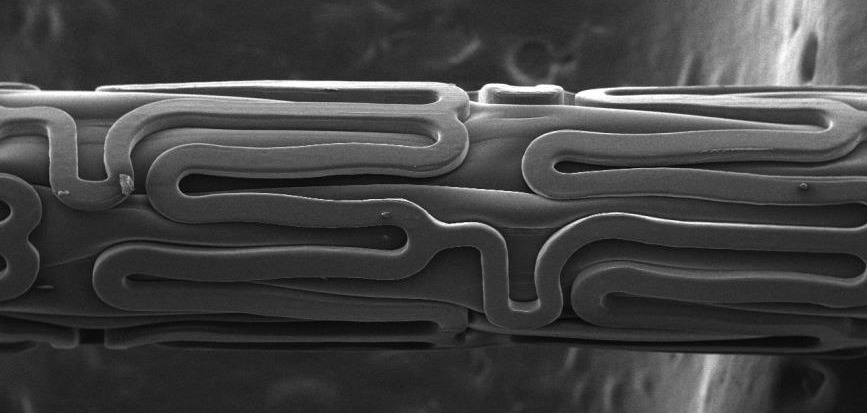}}\label{fig:crimpStent}}
    \hspace{3pt}
	\subfloat[Expanded struts under scanning electron microscope.]{%
		\resizebox*{3.5cm}{!}{\includegraphics[draft=\draftmode]{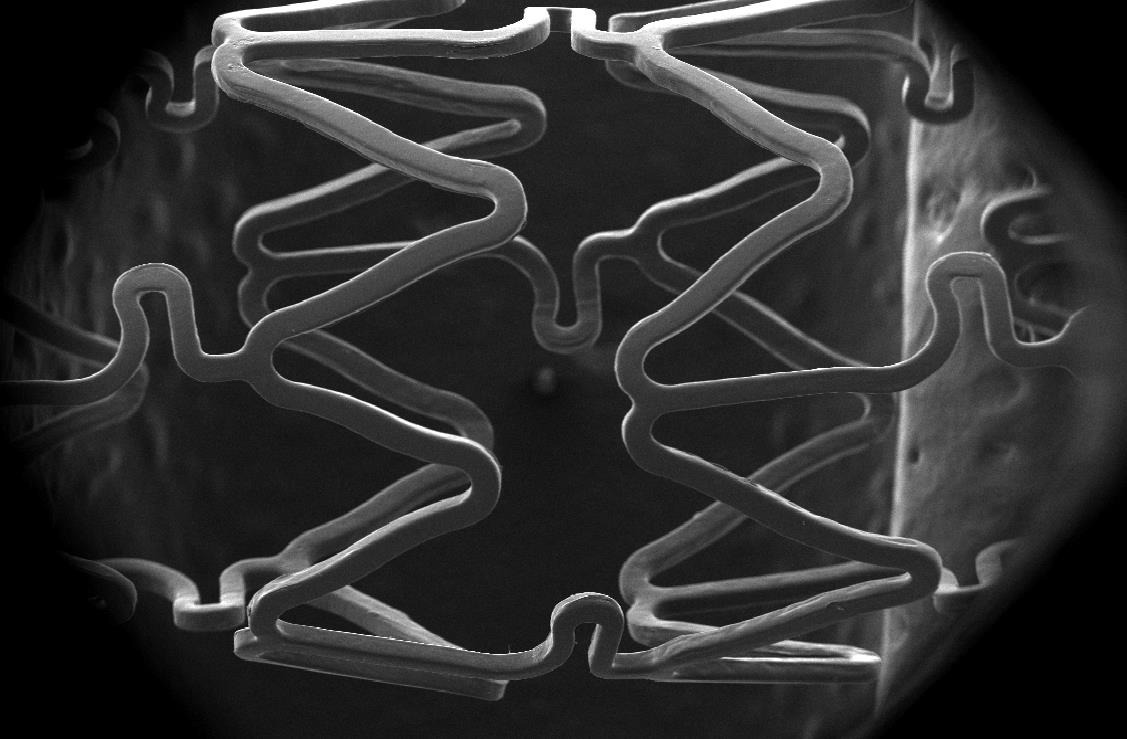}}\label{fig:expStent}}\\
	\subfloat[Discretized stent in the nominal configuration.]{%
		\resizebox*{7cm}{!}{\includegraphics[draft=\draftmode]{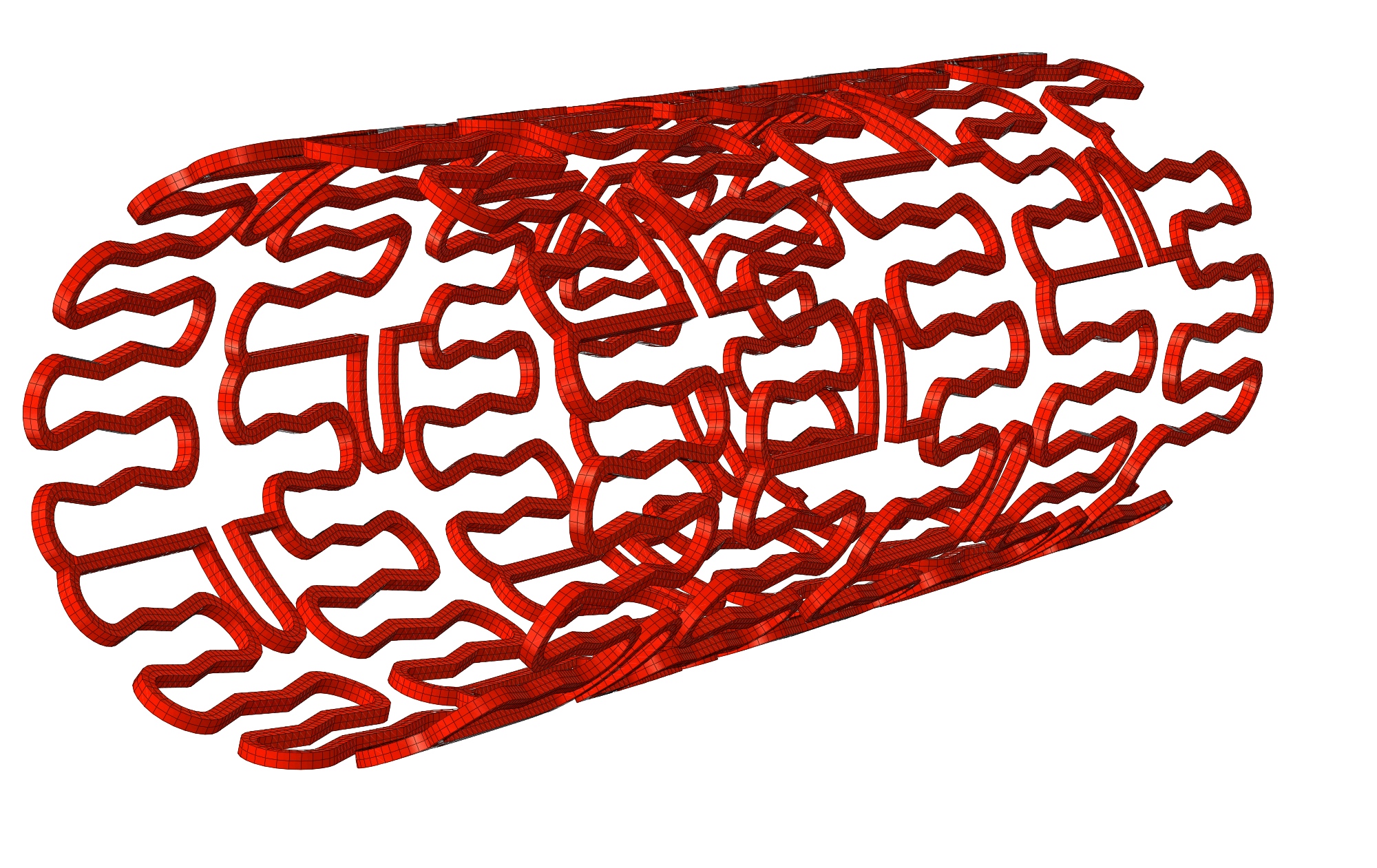}}\label{fig:xiencev_nominal}}
    \hspace{3pt}
	\subfloat[Discrete rigid cylinder utilized for crimping.]{%
		\resizebox*{7cm}{!}{\includegraphics[draft=\draftmode]{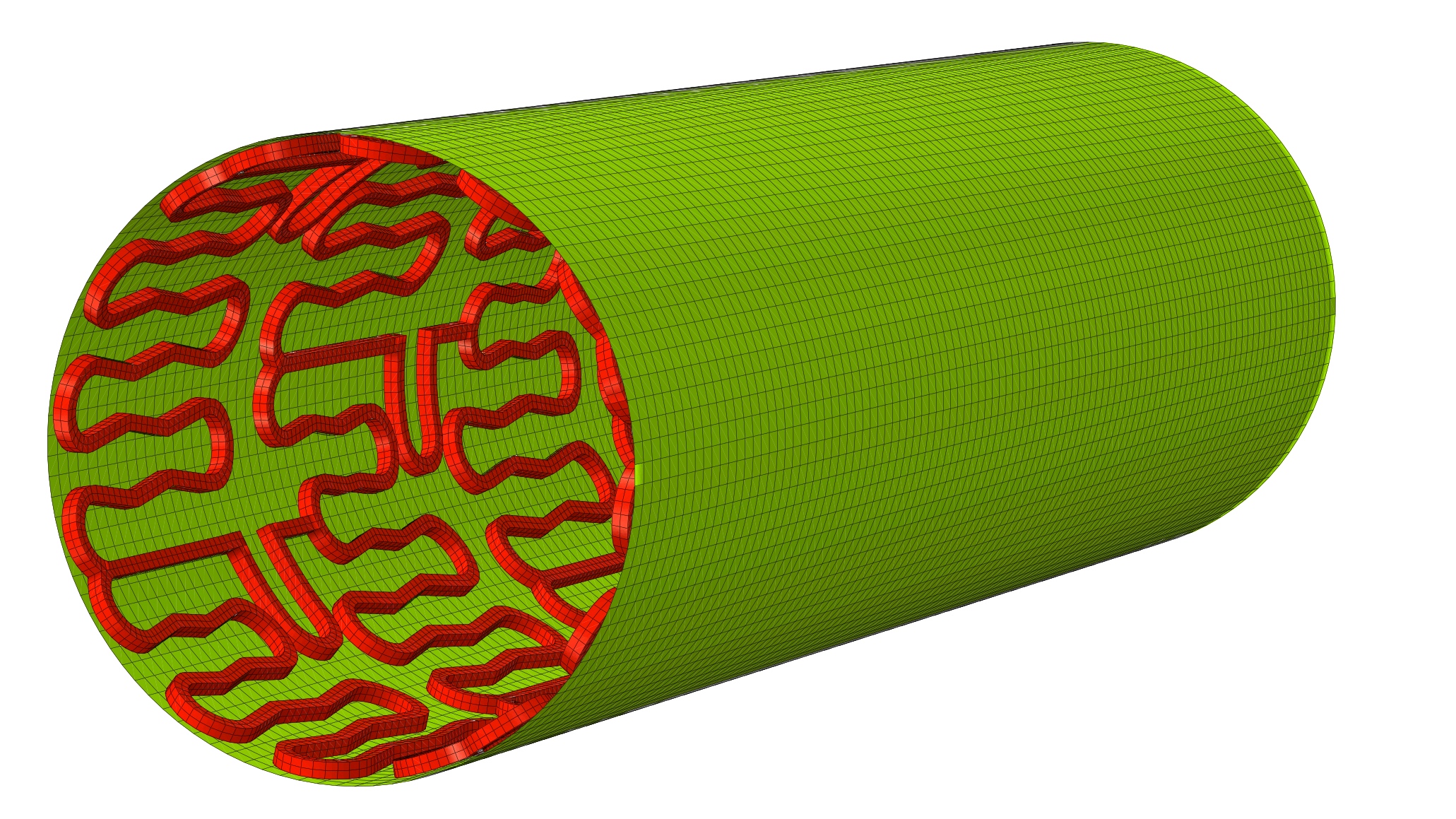}}\label{fig:crimpcylinder}}\\
    \subfloat[Plastic strain distribution in the crimped state.]{%
			\resizebox*{13cm}{!}{\includegraphics[draft=\draftmode]{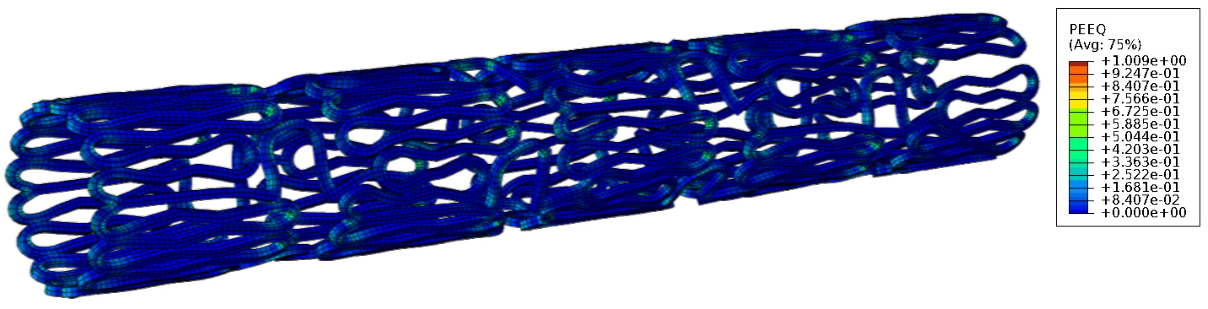}}\label{fig:stent_crimp_peeq}}\\
    \subfloat[After crimping and expansion in ABAQUS.]{%
		\resizebox*{3.5cm}{!}{\includegraphics[draft=\draftmode]{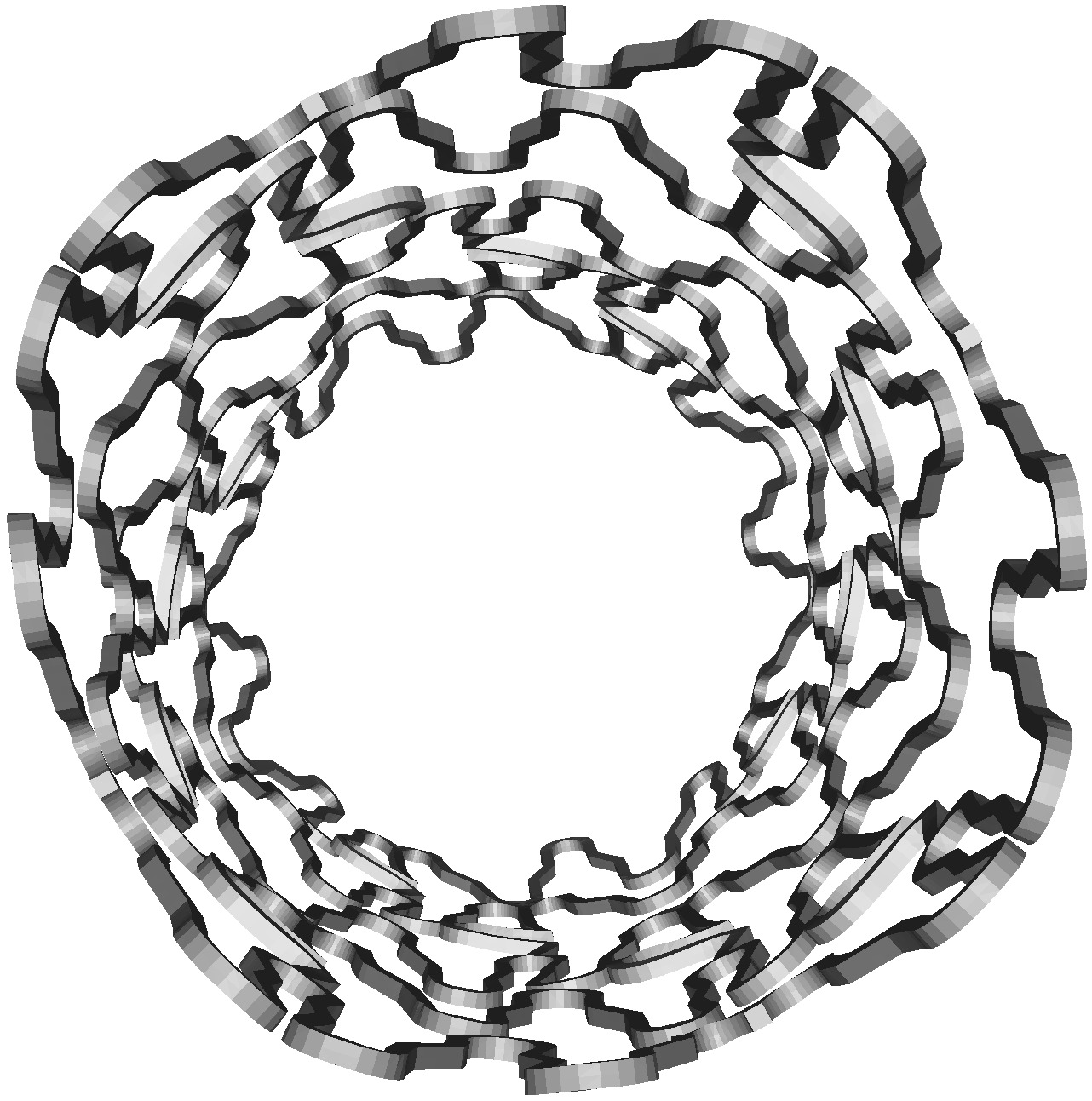}}\label{fig:stentVABAQUS}}
    \hspace{3pt}
    \subfloat[Side view of \ref{fig:stentVABAQUS}.]{%
		\resizebox*{9.8cm}{!}{\includegraphics[draft=\draftmode]{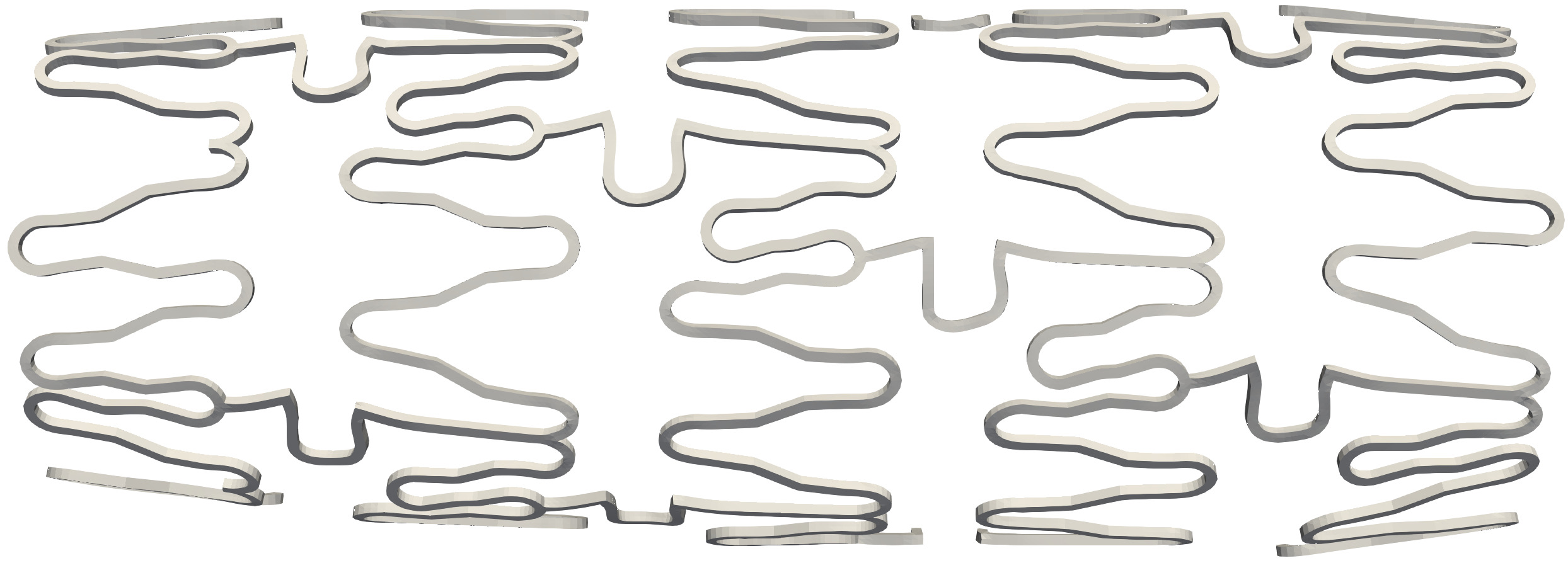}}\label{fig:stentVABAQUSSide}}\\
	\subfloat[Stent after adaptation in Pointwise.]{%
		\resizebox*{3.5cm}{!}{\includegraphics[draft=\draftmode]{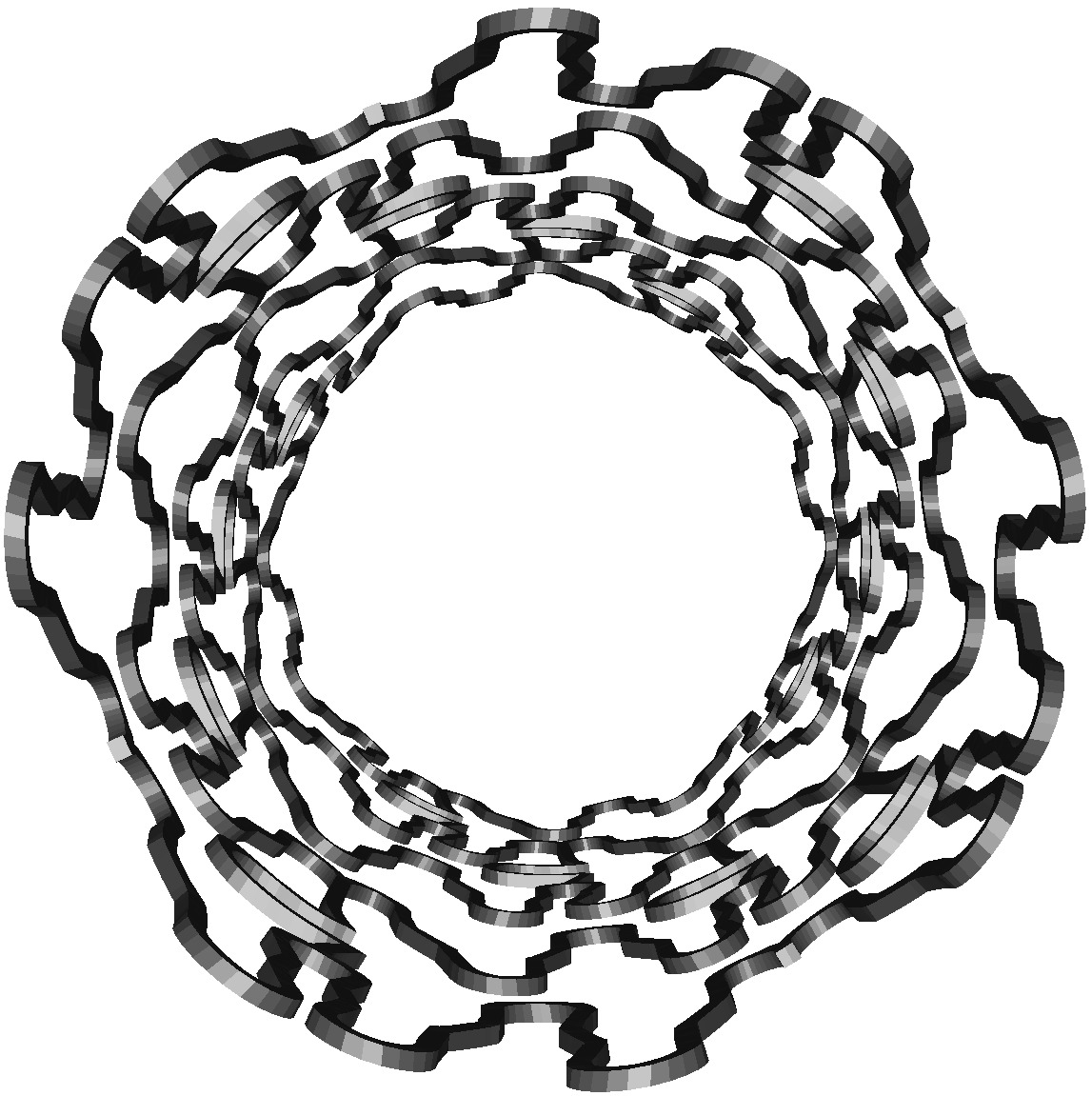}}\label{fig:XienceVGeom}}
	\hspace{3pt}
	\subfloat[Side view of \ref{fig:XienceVGeom}.]{%
		\resizebox*{9.8cm}{!}{\includegraphics[draft=\draftmode]{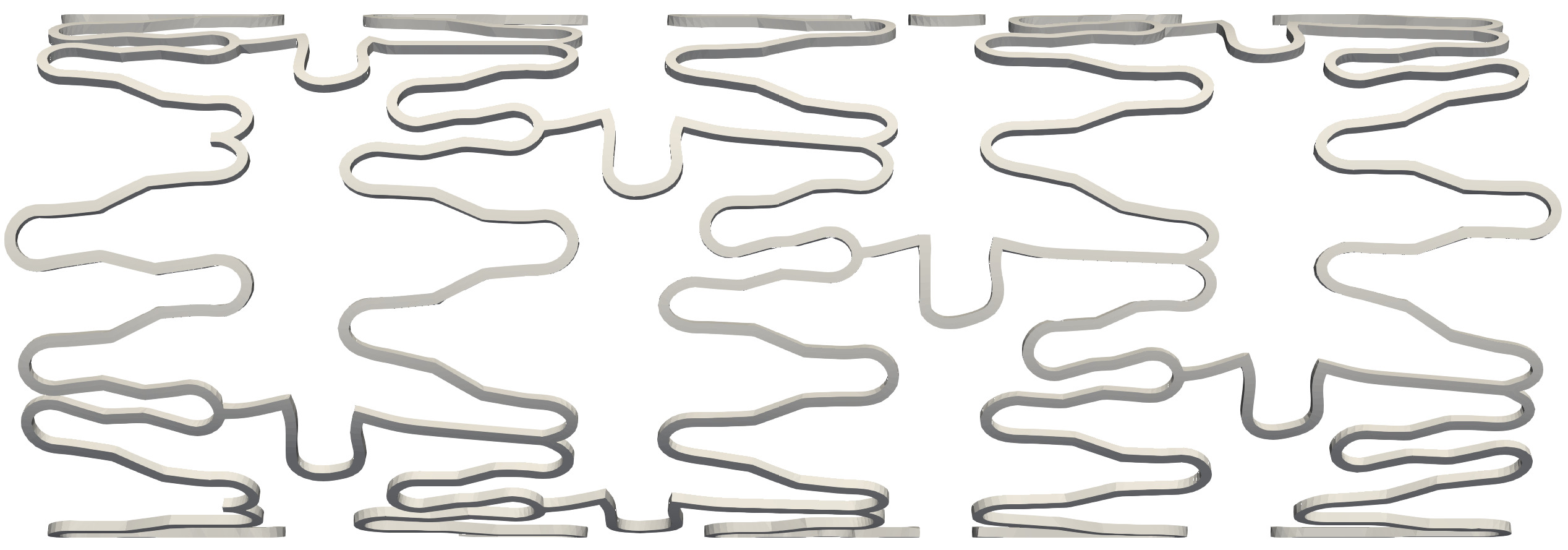}}\label{fig:XienceVGeomSide}}
    \caption{\textit{Xience V} stent.}\label{xienceVABAQUSexpansion}
\end{figure}

The complex geometry of a \textit{Xience V} stent is shown in Fig. \ref{xienceVABAQUSexpansion}. In particular, Fig. \ref{fig:stentCath} shows the \textit{Xience V} stent mounted on a delivery balloon. Figures \ref{fig:crimpStent} and \ref{fig:expStent} show the \textit{Xience V} stent under scanning electron microscope, in crimped and expanded state, respectively. The strut thickness is around 80 $\mu$m and the stent length is 10.3 mm. A CAD model of the nominal configuration (factory geometry before crimping) is obtained by CARD and consequently virtually expanded to a diameter of 3.5 mm in ABAQUS by IFAM.\\

The \textit{Xience V} stent geometry is discretized with 22,216 trilinear hexahedral elements with embedded incompatible modes (C3D8I) as shown in Figure \ref{fig:xiencev_nominal}. Young's modulus of $E = 222$ [G Pa] and Poisson's ratio of $\nu = 0.29$ are used to model the elastic behavior of the cobalt-chromium (Co-Cr) alloy that constitutes the stent struts. The model is also endowed with an elastoplastic constitutive law with isotropic hardening via the prescription of the yield stress-plastic strain tuples extracted for the Co-Cr alloy from \cite{Poncin2004}. A discrete rigid cylinder (see Fig. \ref{fig:crimpcylinder}) is used in ABAQUS to crimp the stent to the configuration shown in Fig. \ref{fig:stent_crimp_peeq} employing the penalty formulation for the contact between the cylinder and the stent, as well as stent self-contact. Plastic hinging is observed along the curvatures present within each cell of the  strut configuration. The crimped stent is then allowed to spring back in an intermediate step. Finally, the sprung-back stent is expanded by the prescription of pressure loading on the luminal surface to obtain the configuration in Fig. \ref{fig:stentVABAQUS}. The non-uniform expanded profile of the stent is attributed to the non-uniform plastic hinging induced by the crimping process. The ends of the stents are devoid of the intermediate connecting arches, thereby resulting in less plastification. This then results in the ends behaving more stiffly than the interior under pressure loading, as can be observed from Fig. \ref{fig:stentVABAQUSSide}.\\

To adapt the stent geometry to an idealized cylindrical artery, we apply the following workflow using the commercial meshing software Pointwise: first, we extract the outer-most surface of the stent, then we project it onto a cylinder, and finally, we extrude the projected surface inwards to retrieve the original stent thickness of 80 $\mu$m. The final geometry is shown in Figures \ref{fig:XienceVGeom} and \ref{fig:XienceVGeomSide}.

\subsection{Stent indentation in lumen of idealized artery: geometry and mesh}
\label{2aMethodIndentation}

For the artery geometry, we consider a 20 mm long segment, approximated as an ideal cylinder with a 3.6 mm diameter. The diameter is chosen based on the average values of the right coronary artery \cite{dodge1992lumen}, and the distal length is chosen such that the artery segment completely contains the stent. This way, we can analyze the effects of stent implantation in the stent proximity, with a focus on the downstream portion.\\

\begin{figure}
	\centering
	\subfloat[10\%.]{
		\resizebox*{3.29cm}{!}{\includegraphics[draft=\draftmode,trim={75cm 46cm 36.3cm 8.4cm},clip=true]{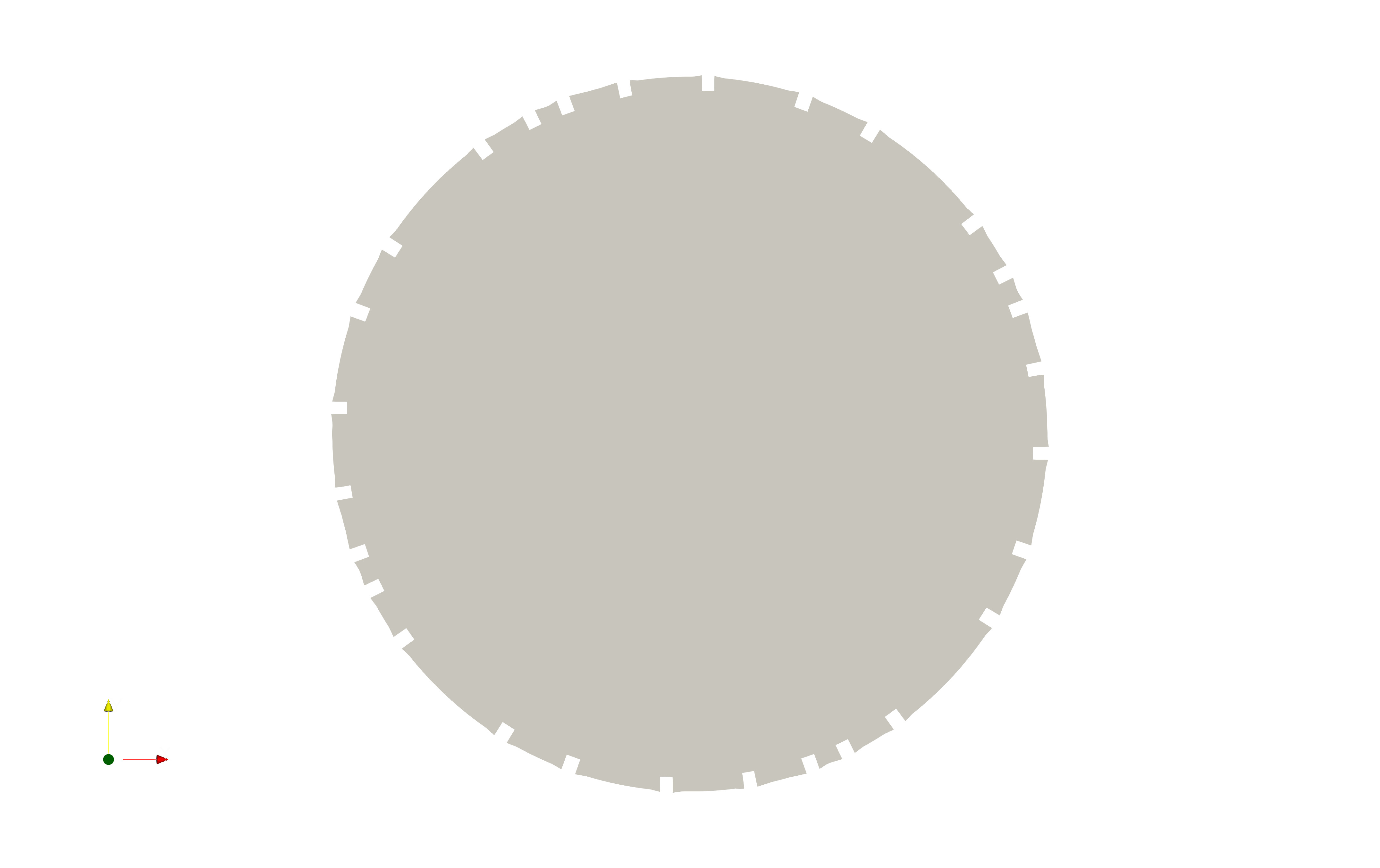}}\label{indentation10}}\hspace{3pt}
	\subfloat[25\%.]{
		\resizebox*{3.29cm}{!}{\includegraphics[draft=\draftmode,trim={75cm 46cm 36.3cm 8.4cm},clip=true]{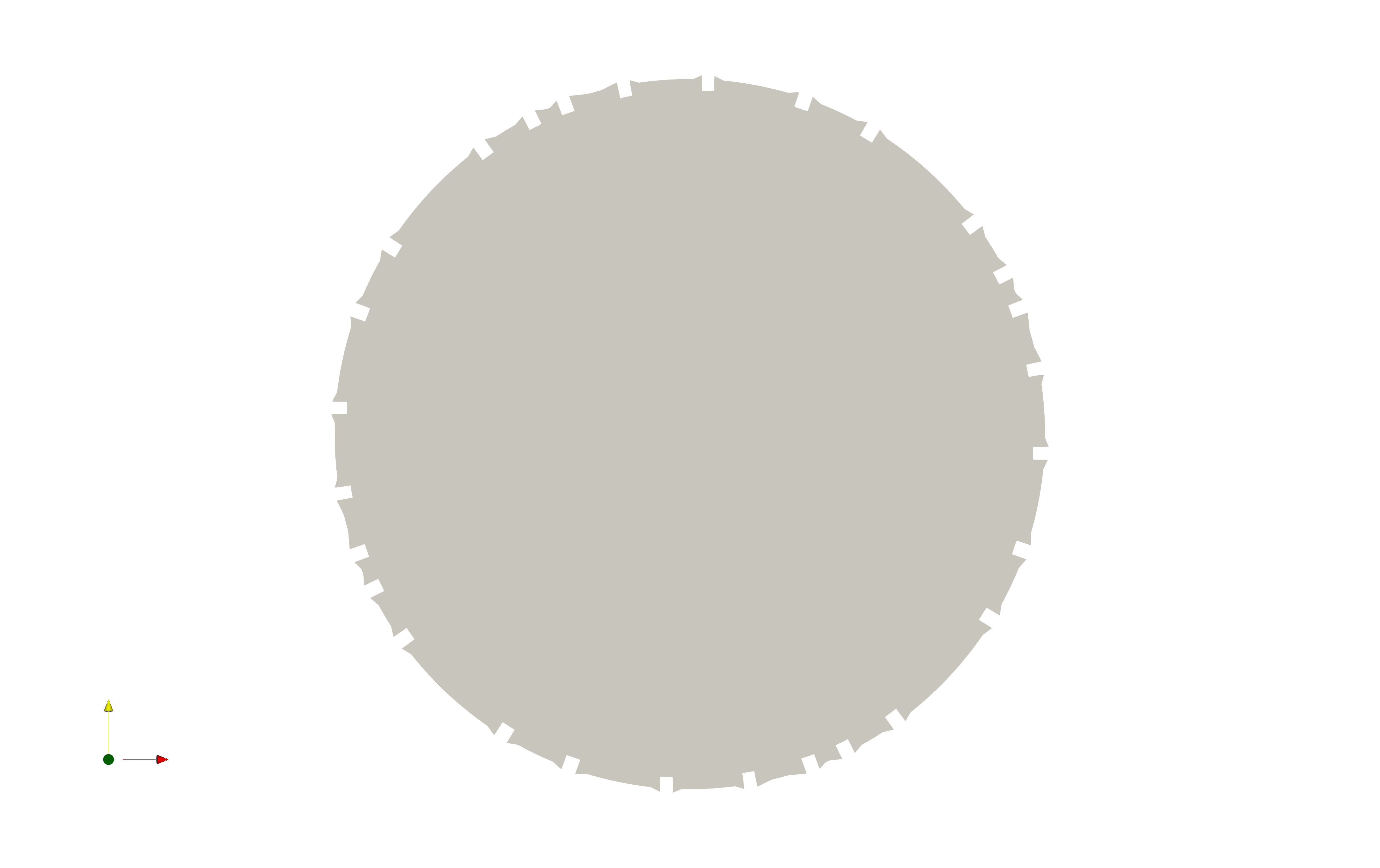}}\label{indentation25}}\hspace{3pt}
	\subfloat[50\%.]{
		\resizebox*{3.29cm}{!}{\includegraphics[draft=\draftmode,trim={75cm 46cm 36.3cm 8.4cm},clip=true]{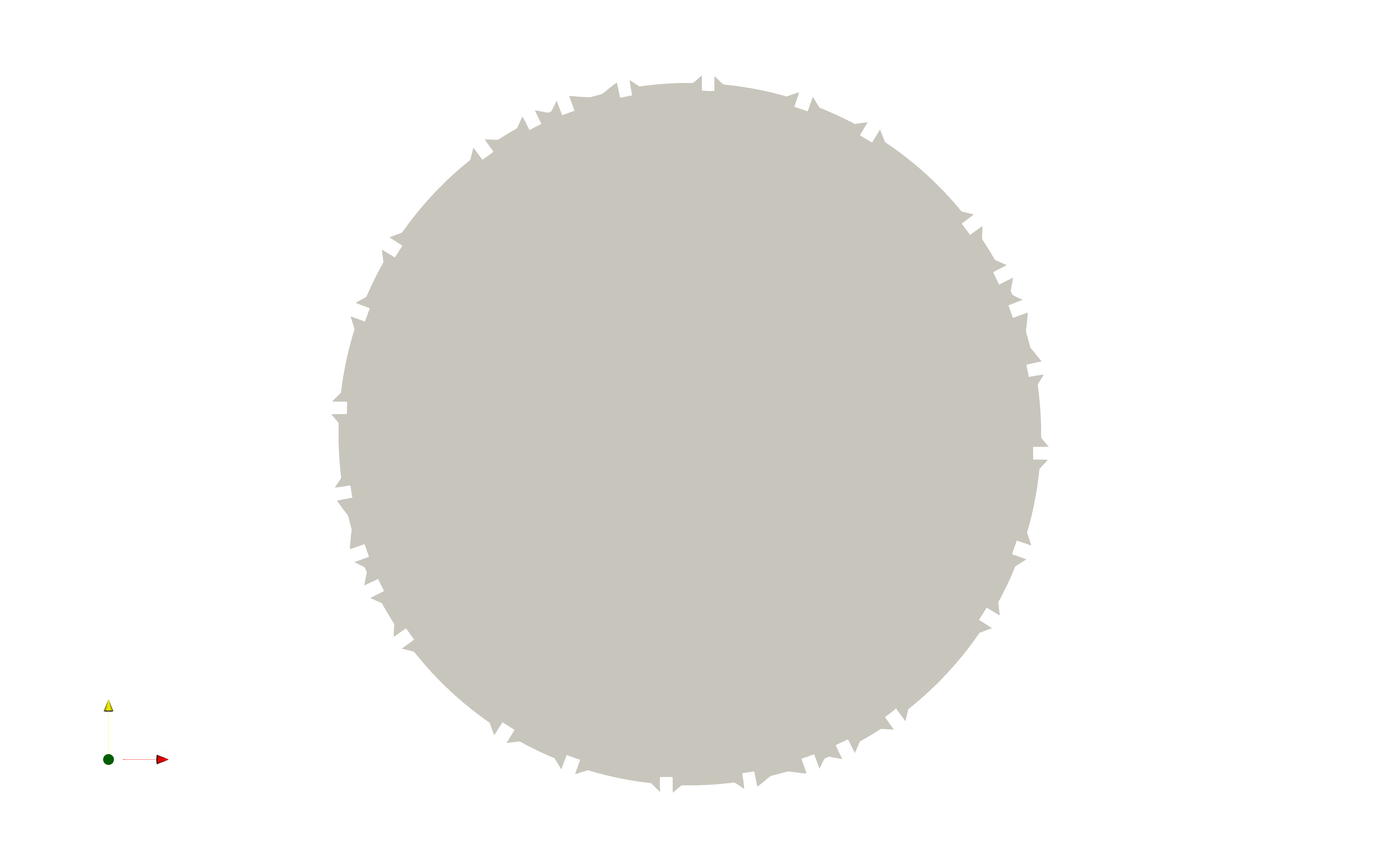}}\label{indentation50}}\hspace{3pt}
	\subfloat[75\%.]{
		\resizebox*{3.29cm}{!}{\includegraphics[draft=\draftmode,trim={75cm 46cm 36.3cm 8.4cm},clip=true]{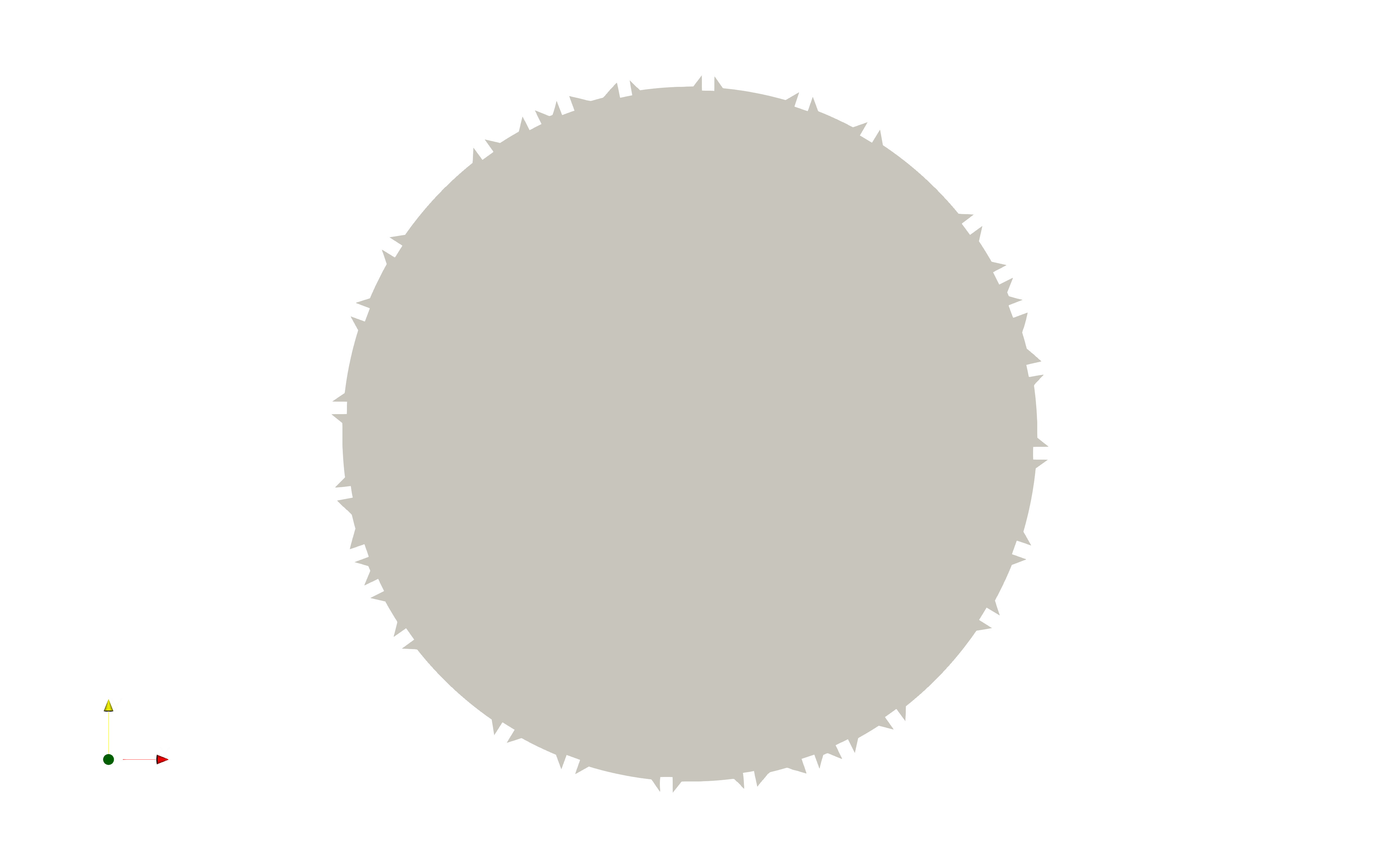}}\label{indentation75}}\\
    \subfloat[10\%.]{
		\resizebox*{14cm}{!}{\includegraphics[draft=\draftmode,trim={0cm 2cm 0cm 0cm},clip=true]{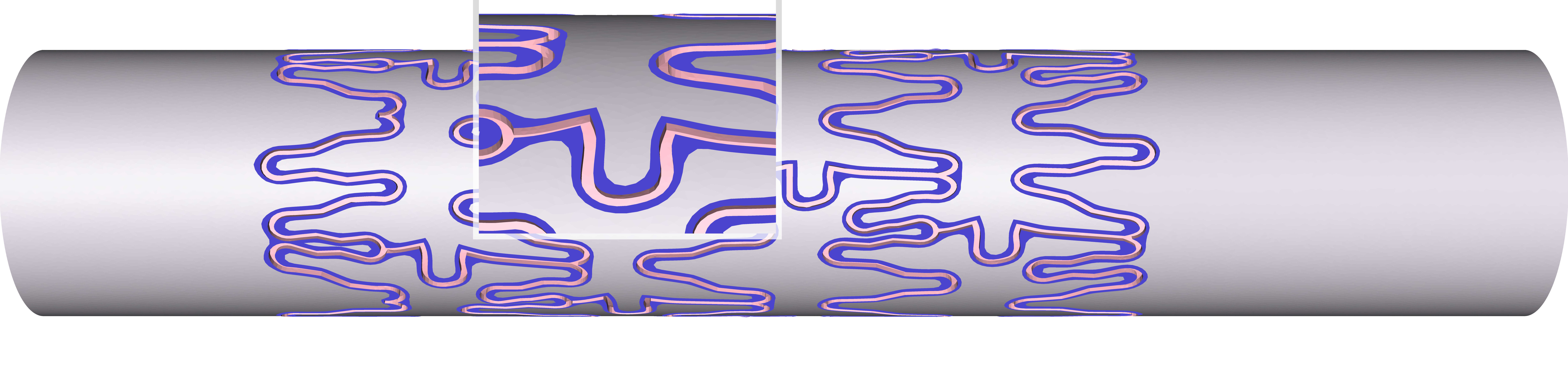}}\label{fig:p90Geom}}\\
    \centering
	\subfloat[75\%.]{%
		\resizebox*{14cm}{!}{\includegraphics[draft=\draftmode,trim={0cm 2cm 0cm 0cm},clip=true]{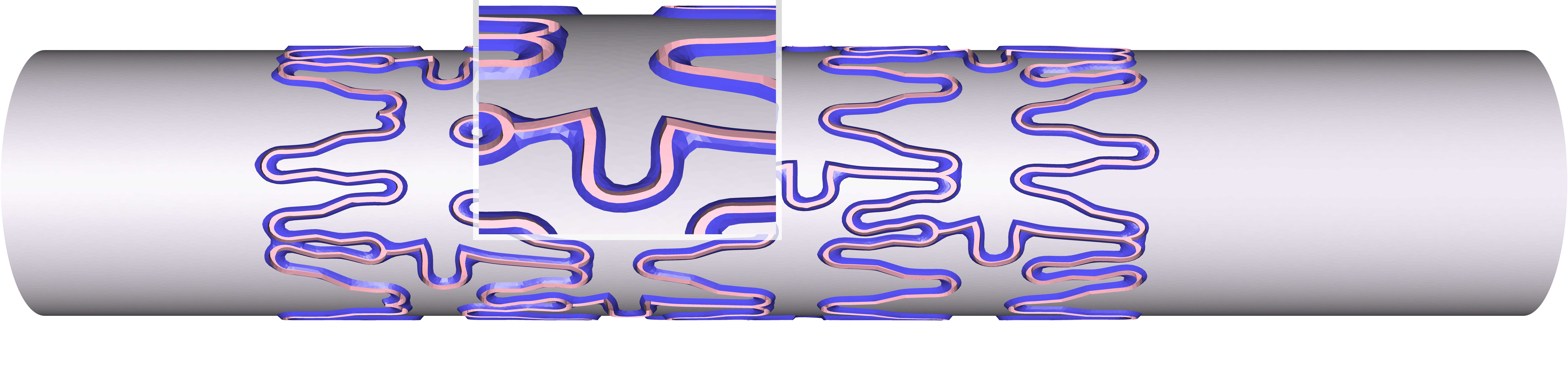}}\label{fig:p25Geom}}\\
    \subfloat[10\%.]{
		\resizebox*{2.4cm}{!}
        {\begin{tikzpicture}
        \node[anchor=south west]{\includegraphics[draft=\draftmode]{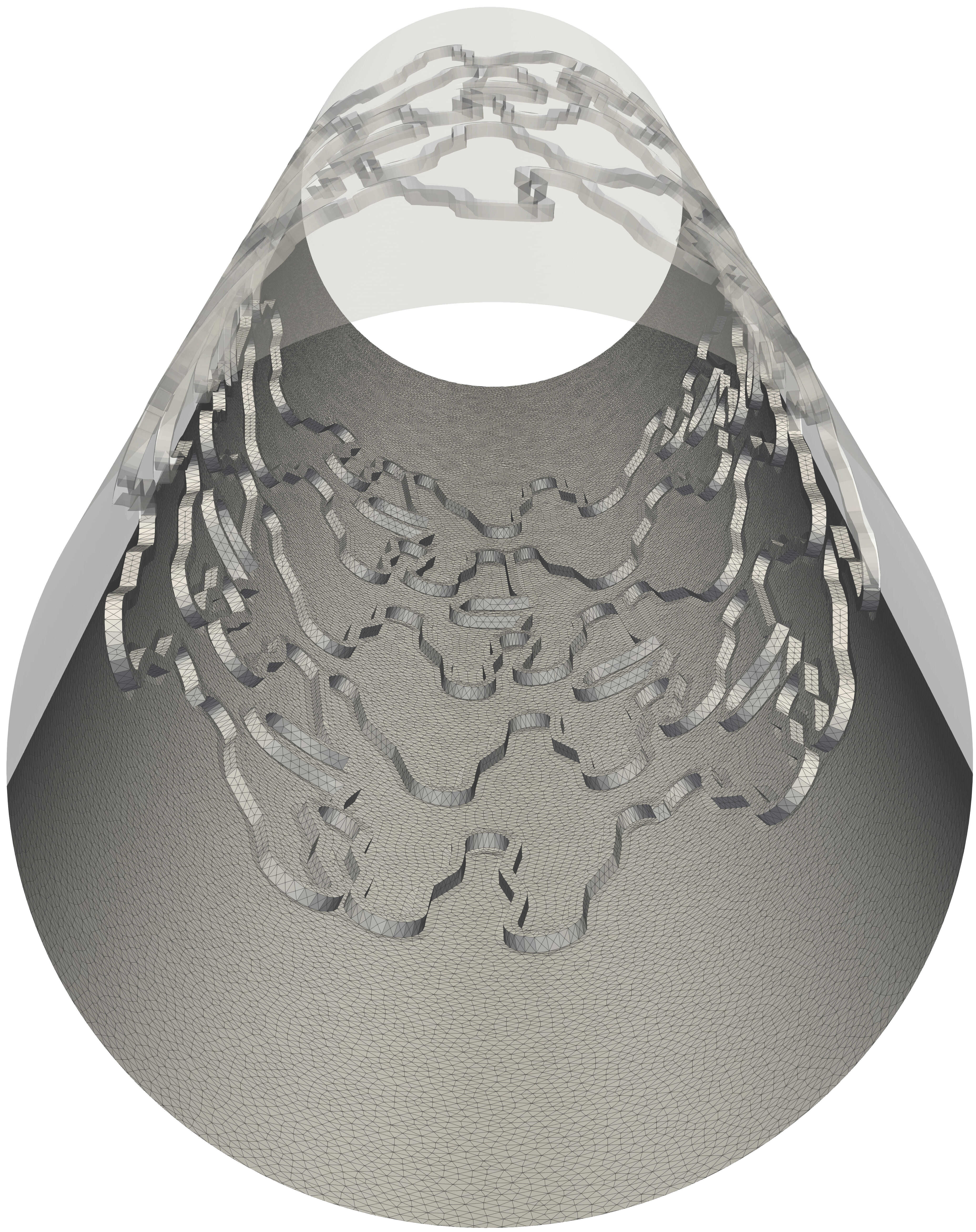}};
        \draw[red, dashed, dash pattern = on 3cm off 3cm, line width = 0.5 cm] (65cm,50cm) rectangle (109.5cm,82.2cm);
        \end{tikzpicture}}\label{fig:p90Mesh}}
        \hspace{3pt}
    \subfloat[Zoom on \ref{fig:p90Mesh}.]{
			\resizebox*{4.1cm}{!}
            {\begin{tikzpicture}
            \node[inner sep=0] (image) {\includegraphics[draft=\draftmode,trim={65cm 50cm 35cm 100cm},clip=true]{\imgfolder/2aStent/p90MeshPV}};
            \draw[red, dashed, dash pattern = on 3cm off 3cm, line width=0.5cm] (image.south west) rectangle (image.north east);
            \end{tikzpicture}}
            \label{fig:p90MeshZoom}}
	\subfloat[75\%.]{
			\resizebox*{2.4cm}{!}{\begin{tikzpicture}
            \node[anchor=south west]{\includegraphics[draft=\draftmode]{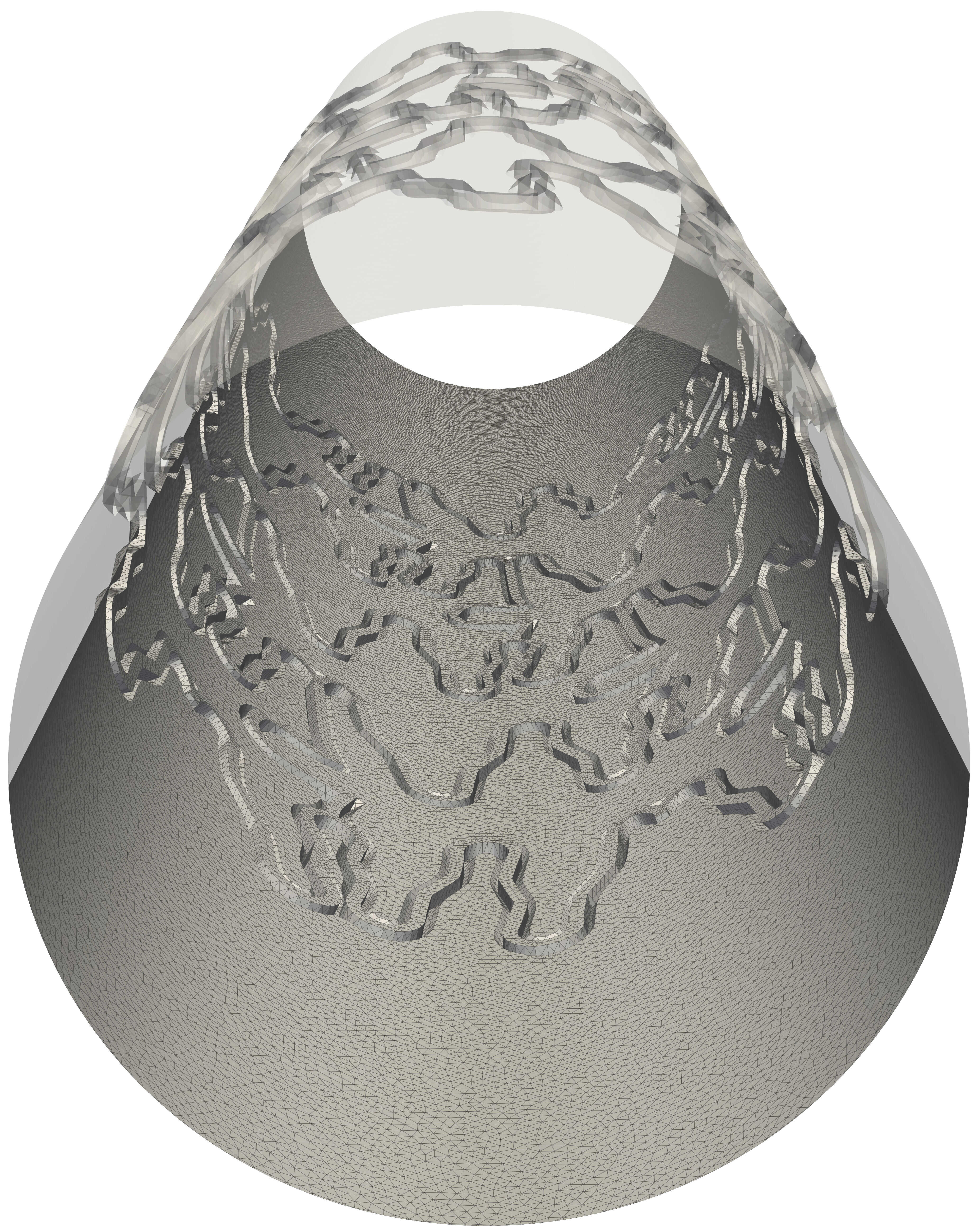}};
            \draw[red, dashed, dash pattern = on 3cm off 3cm, line width = 0.5 cm] (65cm,50cm) rectangle (109.5cm,82.2cm);
            \end{tikzpicture}}\label{fig:p25Mesh}}
        \hspace{3pt}
    \subfloat[Zoom on \ref{fig:p25Mesh}.]{
			\resizebox*{4.1cm}{!}
            {\begin{tikzpicture}
            \node[inner sep=0] (image) {\includegraphics[draft=\draftmode,trim={65cm 50cm 35cm 100cm},clip=true]{\imgfolder/2aStent/p25MeshPV}};
            \draw[red, dashed, dash pattern = on 3cm off 3cm, line width=0.5cm] (image.south west) rectangle (image.north east);
            \end{tikzpicture}}
            \label{fig:p25MeshZoom}}\\
	\subfloat[Uniform.]{
		\resizebox*{4.5cm}{!}{\includegraphics[draft=\draftmode,trim={75cm 46cm 36.3cm 8.4cm},clip=true]{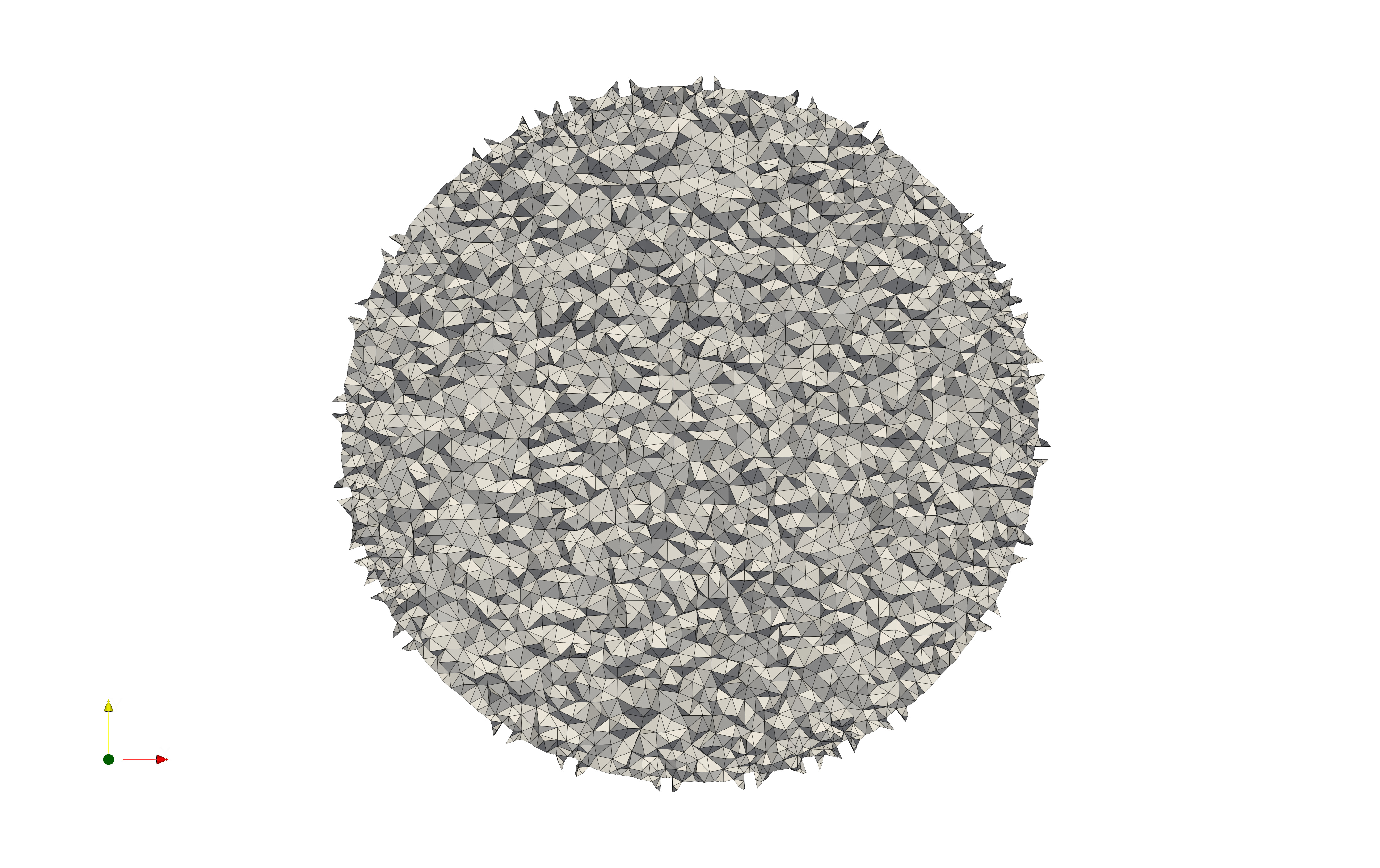}}\label{meshCoarse}}
    \hspace{3pt}
	\subfloat[TMR, 75\%.]{
		\resizebox*{4.5cm}{!}{\includegraphics[draft=\draftmode,trim={75cm 46cm 36.3cm 8.4cm},clip=true]{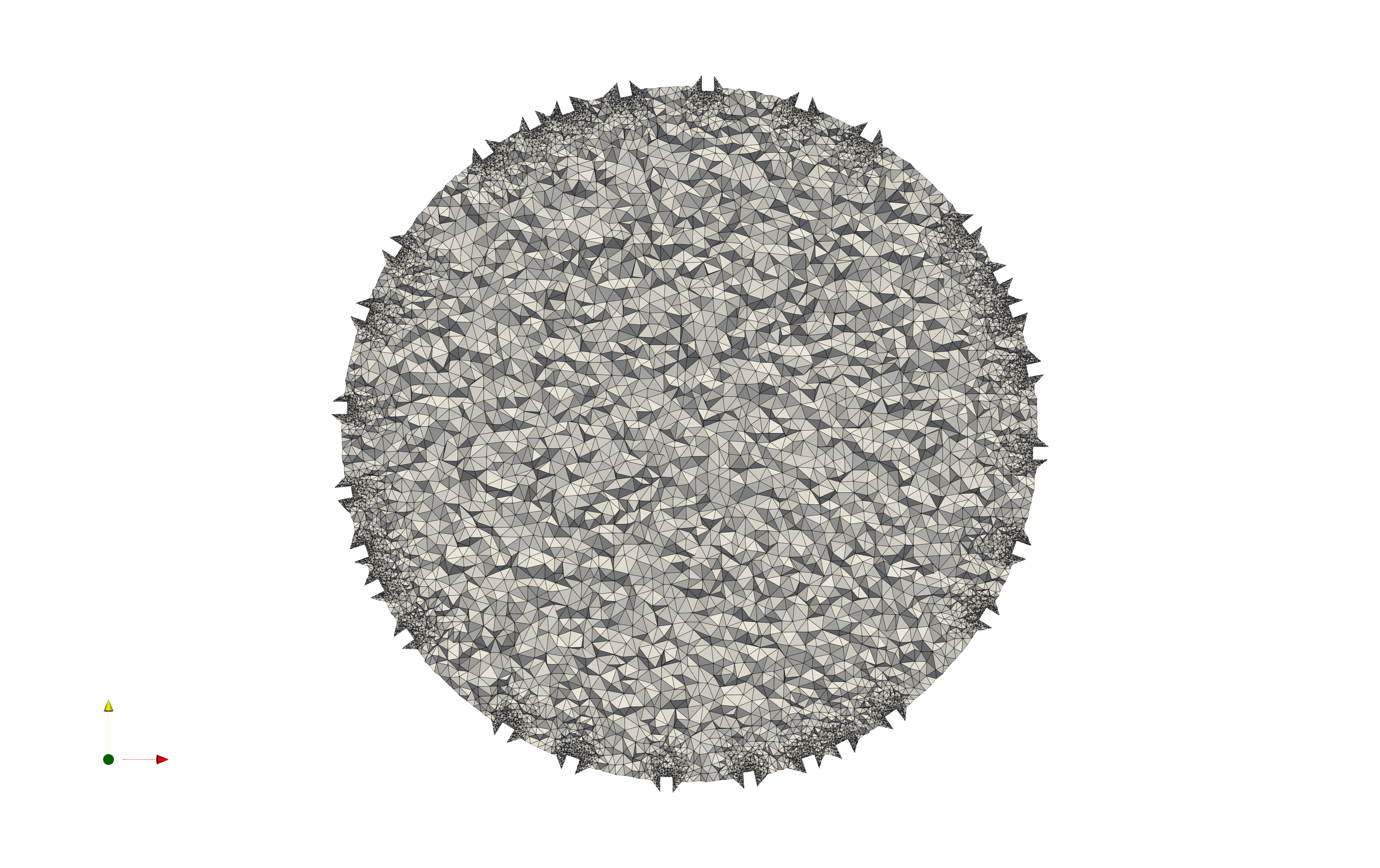}}\label{meshFine}}
    \hspace{3pt}
	\subfloat[Reference.]{
		\resizebox*{4.5cm}{!}{\includegraphics[draft=\draftmode,trim={75cm 46cm 36.3cm 8.4cm},clip=true]{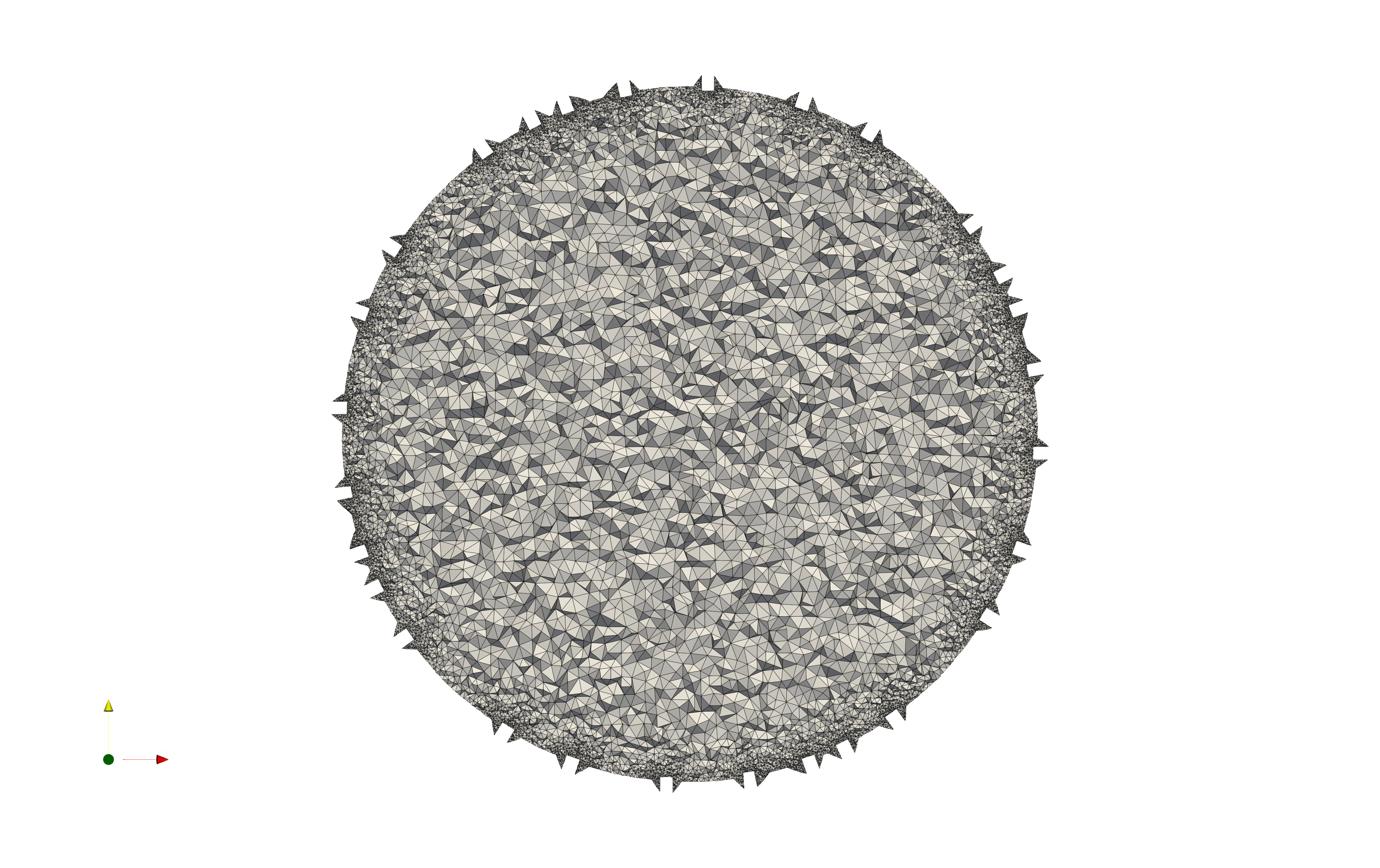}}\label{meshFinest}}
  \caption{Geometry and mesh of artery lumen with \textit{Xience V} stent in cross-section (\ref{indentation10}-\ref{indentation75}, \ref{fig:p90Mesh}-\ref{fig:p25MeshZoom}) and longitudinal view for different indentation percentages. Figures \ref{meshCoarse}-\ref{meshFinest} show the different mesh refinements for 75\% indentation.}\label{fig:geom}
\end{figure}

We proceed to intersect the artery wall with the 3D model of the \textit{Xience V} stent. If the stent is only lightly pushed against the artery wall, we observe a 10\% indentation percentage. High levels of indentation, e.g., 75\%, imply that high pressure was applied during the surgical procedure \cite{Cornelissen2023Development}. Common indentation levels are between 10\% and 55\%, but in the scope of this work, we aim at exploring all in silico possibilities. Hence, the stent is uniformly expanded to obtain four levels of indentation: 10\%, 25\%, 50\% and 75\%, see Figures \ref{indentation10}-\ref{indentation75}. The artery diameter is kept at 3.6 mm and the stent is adjusted such that its external diameter is $d_S  = d_A(1+In)$ where $d_A$ is the diameter of the lumen before the stent implantation and $In$ is the indentation percentage. During the implantation procedure the stent is pushed further into the artery wall, therefore bending it close to the stent struts. To this extent, we define a transition area between the stent contact surface and the artery wall, highlighted in purple in Figures \ref{fig:p90Geom} and \ref{fig:p25Geom}. The transition area is estimated based on ex-vivo microscopy imaging \cite{Cornelissen2023Development}. In patient-specific geometries, different indentation percentages are observed for single struts and depending on the artery configuration.\\

For blood flow simulations, the domain we are interested in is only the artery lumen. The lumen boundaries are obtained intersecting the inner artery wall $\Gamma_w$ (cylindrical surface and transition areas) and the inner-most surface of the struts $\Gamma_S$ (see pink regions in  Figures \ref{fig:p90Geom} and \ref{fig:p25Geom}). The external surface of the stent and the struts volume are not part of the computational domain and therefore not meshed. Figures \ref{fig:p90Mesh}-\ref{fig:p25MeshZoom} show the mesh resolution of the intersection surface near the transition area for two indentation levels. After meshing the intersection surface, we obtain the 3D mesh for the artery lumen. For each indentation we evaluate three mesh refinements, as shown in Figures \ref{meshCoarse}-\ref{meshFinest}. The uniform mesh has a mesh size of circa 0.1 mm. The reason for this is to keep the mesh elements comparable in size to the struts \cite{manjunatha2024silico}. The targeted mesh refinement (TMR) case has a varying mesh size between 0.01 and 0.1 mm, where the smallest elements surround the stent and the transition areas. In the reference case, the boundary elements (artery wall, stent, and transition areas) are all kept at 0.01 mm mesh size and the volume elements transition from 0.01 mm to 0.1 mm mesh size from the boundary towards the center of the lumen.

%% file: sections/2b-method.tex
\subsection{Blood flow modeling and hemodynamic indicators}
\label{2bMethod}
\begin{figure}
    \centering
    \subfloat[Lumen boundaries: $\Gamma_{in}$ in yellow, $\Gamma_w$ in grey, $\Gamma_{stent}$ in pink and $\Gamma_{out}$ in blue.]{
	\resizebox*{8.4cm}{!}{\includegraphics[draft=\draftmode]{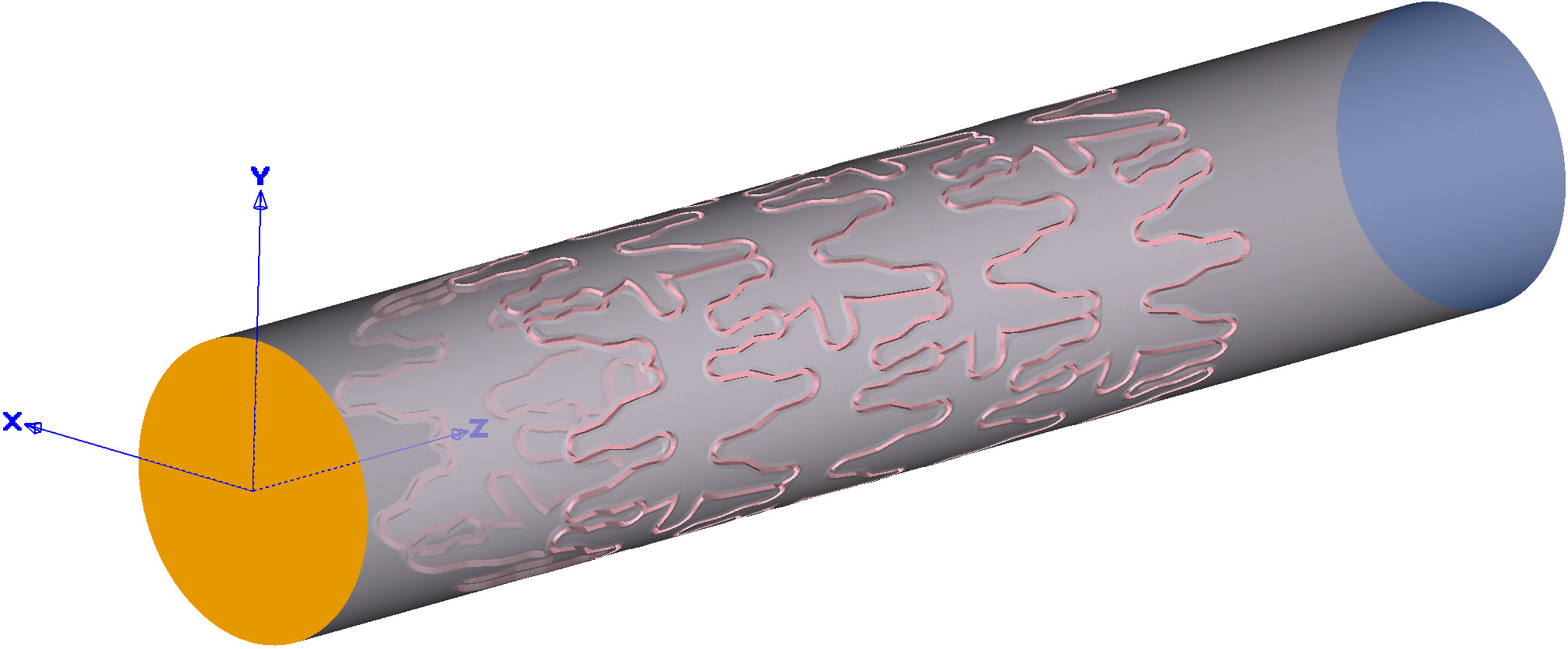}}\label{fig:boundaries}}\hfill
	\hspace{3pt}
    \subfloat[Flow rate over one heart beat imposed at inflow $\Gamma_{in}$.]{
	\resizebox*{5.4cm}{!}{\includegraphics[draft=\draftmode]{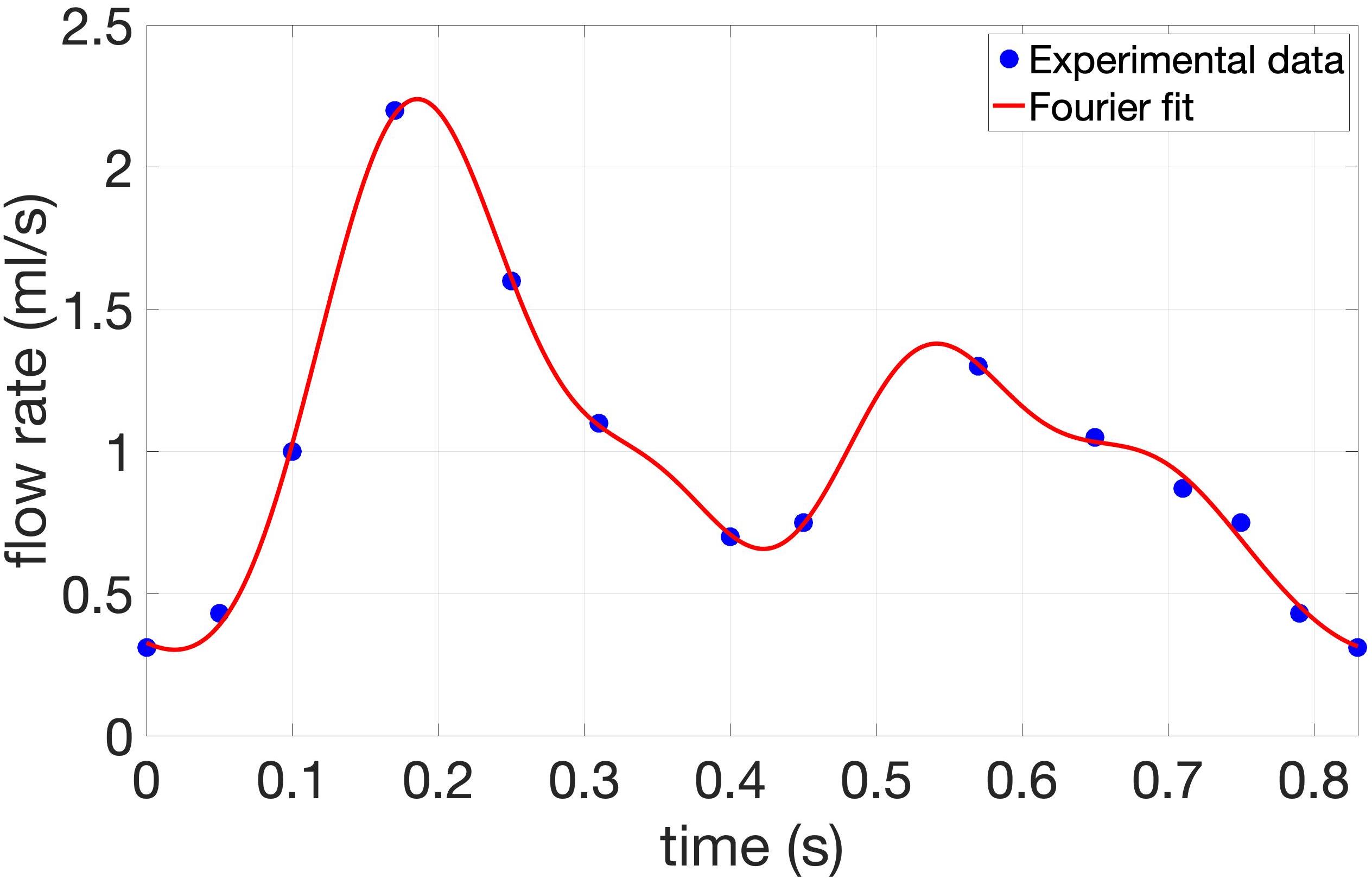}}\label{fig:flowRate}}
	\caption{Definition of $\Gamma$ and inflow boundary conditions.}\label{fig:BC}
\end{figure}

Incompressible Navier-Stokes equations in the Eulerian configuration describe the blood flow. The system of equations \eqref{NS-first}-\eqref{NS-last} shows the momentum and mass conservation with boundary and initial conditions:

\begin{align}
	\rho \left( \mathbf{u}_t+\mathbf{u}\cdot \nabla \mathbf{u}\right) -\nabla \cdot \bm{\sigma} & = \mathbf{0}, & & \text { in } \Omega \times(0, T],\label{NS-first}\\
	\nabla \cdot \mathbf{u} & =0, & & \text { in } \Omega \times(0, T],\\
	\mathbf{u} & =\mathbf{g}, & & \text { on } \partial \Omega,\\
	\mathbf{u} & =\mathbf{u}_0, & & \text { in } \Omega \text{ at } t =0,\label{NS-last}
\end{align}

where $\Omega$ is the spatial domain, $(0, T]$ denotes the time horizon, $\mathbf{u}=\mathbf{u}(\mathbf{x},t)$ is the velocity vector, $p=p(\mathbf{x},t)$ is the pressure and $\rho$ the blood density. The stress tensor for incompressible and viscous fluids is defined as

\begin{equation}
	\bm{\sigma} = -p \mathbf{I} + 2 \mu \mathbf{E},
	\label{sigma}
\end{equation}

where $\mathbf{E}(\mathbf{u}) = \frac{1}{2}\left( \nabla \mathbf{u} + \nabla \mathbf{u}^\top\right)$ is the rate-of-strain tensor and $\mu$ is the dynamic viscosity which is assumed to be constant. Given the artery diameter of circa 3.6 mm and the idealized geometry of this work, we assume a Newtonian constitutive model for the blood \cite{leuprecht2001computer}. Common choices for density and viscosity are $\rho=1056$ kg/m$^3$ and $\mu = 3.5 \times 10^{-3}$ Pa\hspace{1mm}s for 45\% hematocrit and 37\degree Celsius. The artery wall is assumed to be non-moving in the presence of rigid metal stents \cite{moore2002fluid}. Fig. \ref{fig:boundaries} shows the subsets of the boundary: on the artery wall $\Gamma_{w}$ and on the stent inner surface $\Gamma_{stent}$, \textit{no-slip} boundary conditions are imposed; on $\Gamma_{out}$, we impose for the velocity to be perfectly orthogonal to the outflow surface; and $\Gamma_{in}$ has a parabolic profile. The inflow velocity magnitude is varying in time to mimic the periodic pulsatile regime over one heart beat lasting 0.83 s \cite{manjunatha2024silico}. Figure \ref{fig:flowRate} shows the Fourier interpolation of experimental data of the flow rate $Q(t)$ in the right coronary artery \cite{bertolotti2001numerical, hsiao2012hemodynamic}. During systole, the flow increases from a minimum flow rate of $0.31$ ml/s to a maximum of $2.2$ ml/s. The diastolic phase shows another increase, but the peak is much lower compared to the systolic one. Initial conditions $\mathbf{u}_0 = \mathbf{g}(\mathbf{x},0)$ are imposed in $\Omega$ based on the velocity obtained from a steady simulation with flow rate equal to $Q(0)$.\\

We discretize system \eqref{NS-first}-\eqref{NS-last} in space by means of stabilized FEM and we choose $\mathbb{P}_1-\mathbb{P}_1$ element pair. Given the geometry and the inflow velocity, we do not expect to be in a turbulent regime, but we expect to see vortices close to the stent struts. The maximum Reynolds number is $Re=300$, thus GLS stabilization \cite{pauli2017stabilized, donea2003finite} for the momentum equation and for the incompressibility constraint is sufficient. For time discretization, we choose the BDF2 multi-step method \cite{forti2015semi}. We linearize the convective term by means of Newton-Raphson method and solve the resulting linear system with a GMRES solver \cite{saad1986gmres} and ILUT preconditioning \cite{saad2003iterative}.\\

The WSS tensor $\bm{\tau}$ is computed as \cite{john2017influence}:

\begin{equation}
	\bm{\tau} = \bm{\sigma}\mathbf{n} - [(\bm{\sigma}\mathbf{n})\cdot \mathbf{n}]\mathbf{n} = 2 \mu (\mathbf{E}\mathbf{n} - [(\mathbf{E}\mathbf{n})\cdot \mathbf{n}]\mathbf{n}),
\end{equation}

and its magnitude is defined as $\wss = |\bm{\tau}|$. To evaluate the WSS over a certain time span, a well-known quantity is TAWSS:
\begin{equation}
	\tawss= \frac{1}{T}\int_{0}^{T} |\bm{\tau} (t)|dt.
	\label{TAWSS}
\end{equation}

Taking $(0,T)$ as one heart beat cycle, TAWSS shows which areas have persistently low values of WSS. This choice is justified if the periodic flow is reached, and therefore one heart beat is representative of any heart beat. Averaging over multiple heart beats would be computationally more expensive but would not add any more information. OSI is an indicator of the WSS deflection from the main flow direction and highlights recirculation near the artery wall:

\begin{equation}
	\osi = \frac{1}{2}\left(1-\frac{|\int_{0}^{T} \bm{\tau} (t)dt|}{\int_{0}^{T} |\bm{\tau} (t)|dt}\right).
	\label{OSI}
\end{equation}

It spans between 0 where shear stresses completely align with the main flow direction and 0.5 which indicates possible vortices and stagnation.
RRT in \eqref{RRT} merges the information of TAWSS and OSI in one quantity:

\begin{equation}
\rrt = \frac{1}{(1-2\hspace{2mm}\osi)\tawss}.
\label{RRT}
\end{equation}

High values of RRT correspond to low values of TAWSS and recirculation areas with high OSI values.

%% file: sections/3-results.tex
\begin{table}
    \begin{tabular}{lcccccccc} \toprule
     & $n_{el}$ & $n_n$ & $\Bar{h}$ & $h_{min}$ & $n_n$ on $\Gamma$ & $n_n$ on $\Gamma_w$ & \# cores & wall-clock time \\ \midrule
    Uniform
    & $\sim$ 4 M & $\sim$ 0.7 M & 0.1 mm & 0.1 mm & $\sim$ 110 K & $\sim$ 80 K & 240 & $\sim$ 3.5 h\\
    TMR & $\sim$ 27 M & $\sim$ 5 M & 0.1 mm &  0.01 mm & $\sim$ 1,25 M & $\sim$ 310 K & 480 & $\sim$ 16 h\\
    Reference & $\sim$ 35 M & $\sim$ 6.5 M & 0.1 mm & 0.01 mm & $\sim$ 2 M & $\sim$ 900 K & 960 & $\sim$ 13 h\\ \bottomrule
    \end{tabular}
    \caption{Average mesh parameters for any indentation.}
    \label{meshShort}
\end{table}
In this section, we analyze the numerical results obtained for the hemodynamic indicators introduced in Section \ref{2bMethod} in stented arteries with a \textit{Xience V} stent at different indentations. Table \ref{meshShort} shows an overview of mesh parameters for different refinements. The solutions obtained on the finest mesh are taken as reference. For each mesh, $n_{el}$ is the total number of elements, $h$ denotes the mesh size, and the total number of nodes is $n_n$. Meshes vary from 4 to 35 million elements for any indentation, depending on the refinement level. The simulation of one cycle $T$ for a fully resolved stented artery runs for 16--17 hours on 480 cores with targeted mesh refinement, while the reference case requires much more computational resources for a comparable run-time. The results are obtained after three heart beats, with time step $\Delta t = 0.005$ s. The periodic regime is reached after one full heart beat. Thus, hemodynamic quantities are obtained via post-processing of the second cycle. All simulations are obtained with the highly parallelizable in-house code XNS \cite{pauli2017stabilized} and are performed on the supercomputers JURECA at Forschungszentrum J\"ulich \cite{krause2018jureca} and CLAIX 2018 at RWTH Aachen University.

\subsection{WSS and TAWSS}
\label{3aWSSandTAWSS}
We qualitatively analyze WSS and TAWSS, in particular on the reference mesh for 75\% and 10\% indentation.\\

\begin{figure}
	\centering
    \subfloat[75\%.]{
        \resizebox*{7cm}{!}{\includegraphics[draft=\draftmode]{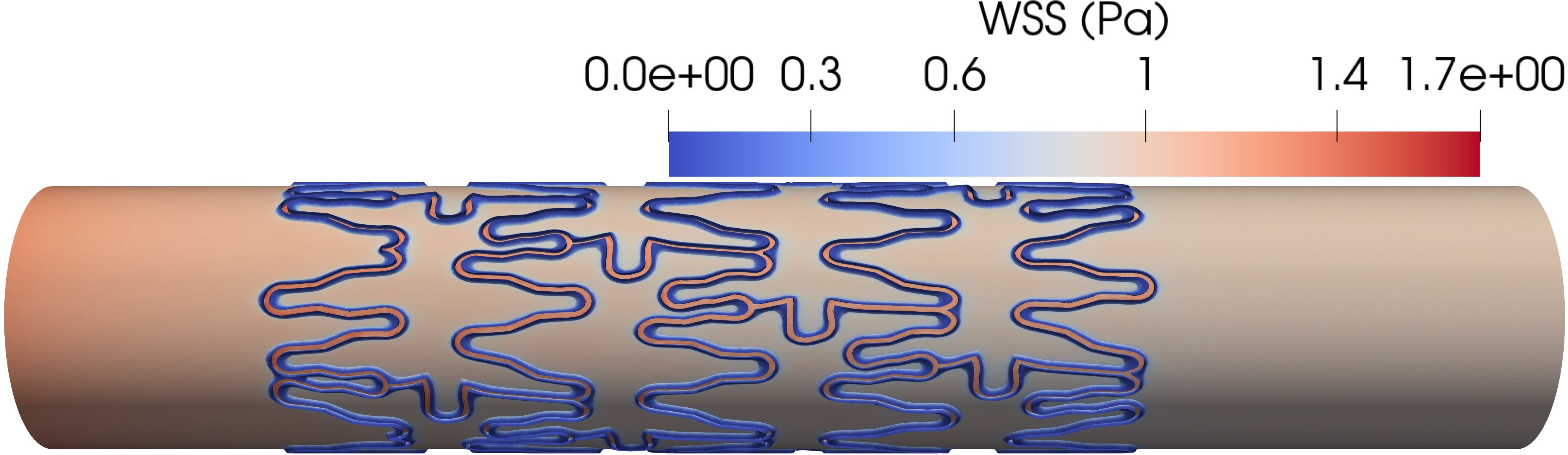}}\label{fig:WSS75}}
    \subfloat[10\%.]{
        \resizebox*{7cm}{!}{\includegraphics[draft=\draftmode]{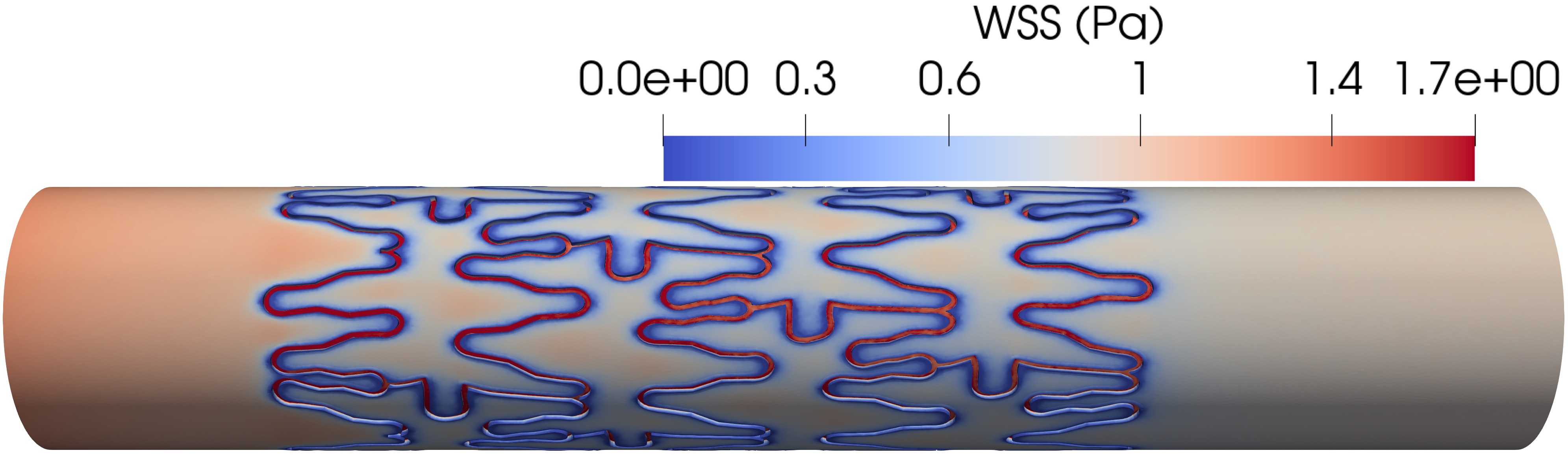}}\label{fig:WSS10}}\\
    \subfloat[Areas of artery wall at $t = 0.4$ s with WSS $<$ 0.4 Pa.]{%
		\resizebox*{7.8cm}{!}{\includegraphics[draft=\draftmode]{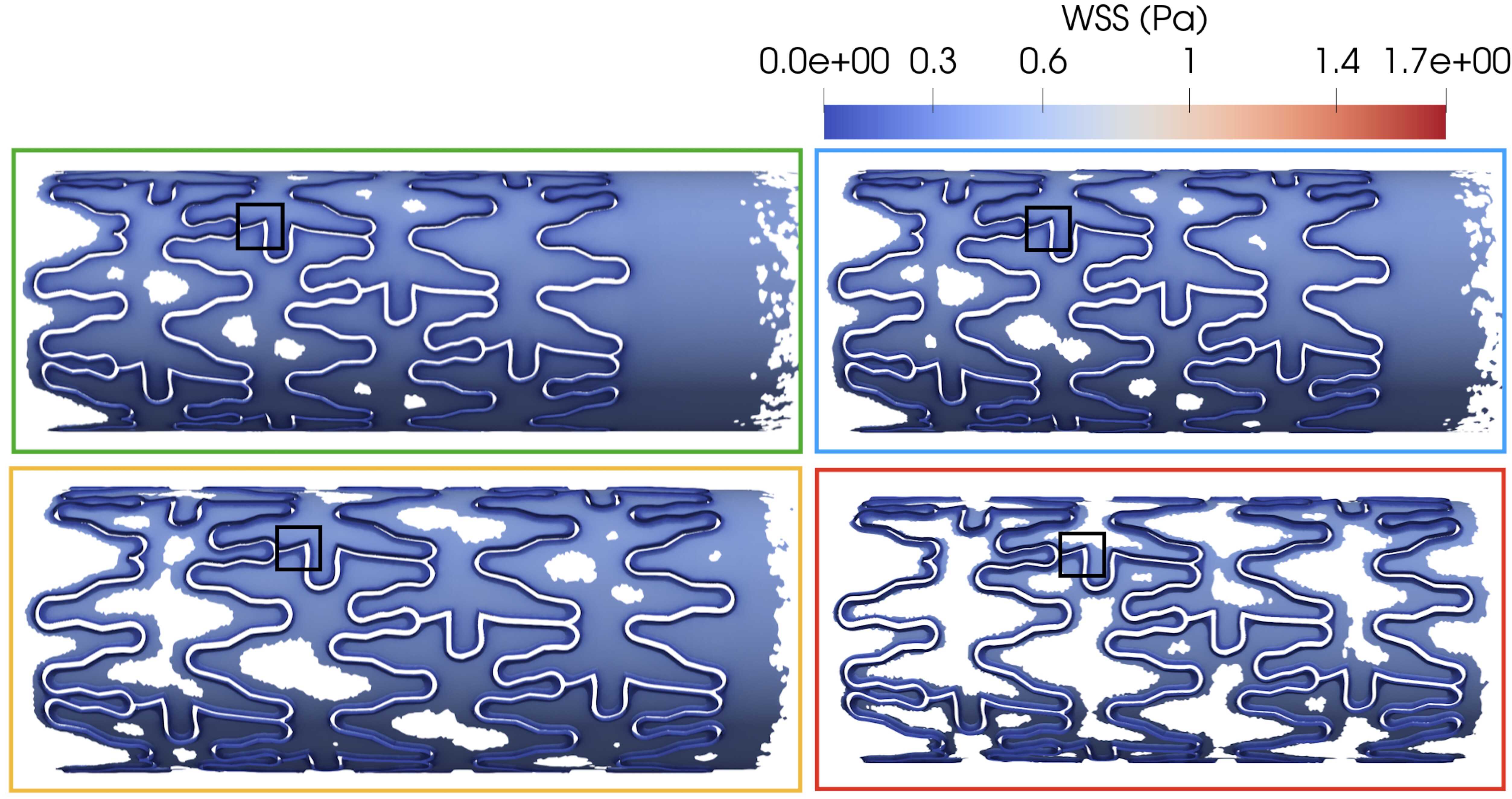}}\label{fig:WSS-low0.4}}\hspace{1pt}
	\subfloat[WSS values near stent over one cycle. Zoom on square in Fig. \ref{fig:WSS-low0.4}.]{%
		\resizebox*{6.2cm}{!}{\includegraphics[draft=\draftmode]{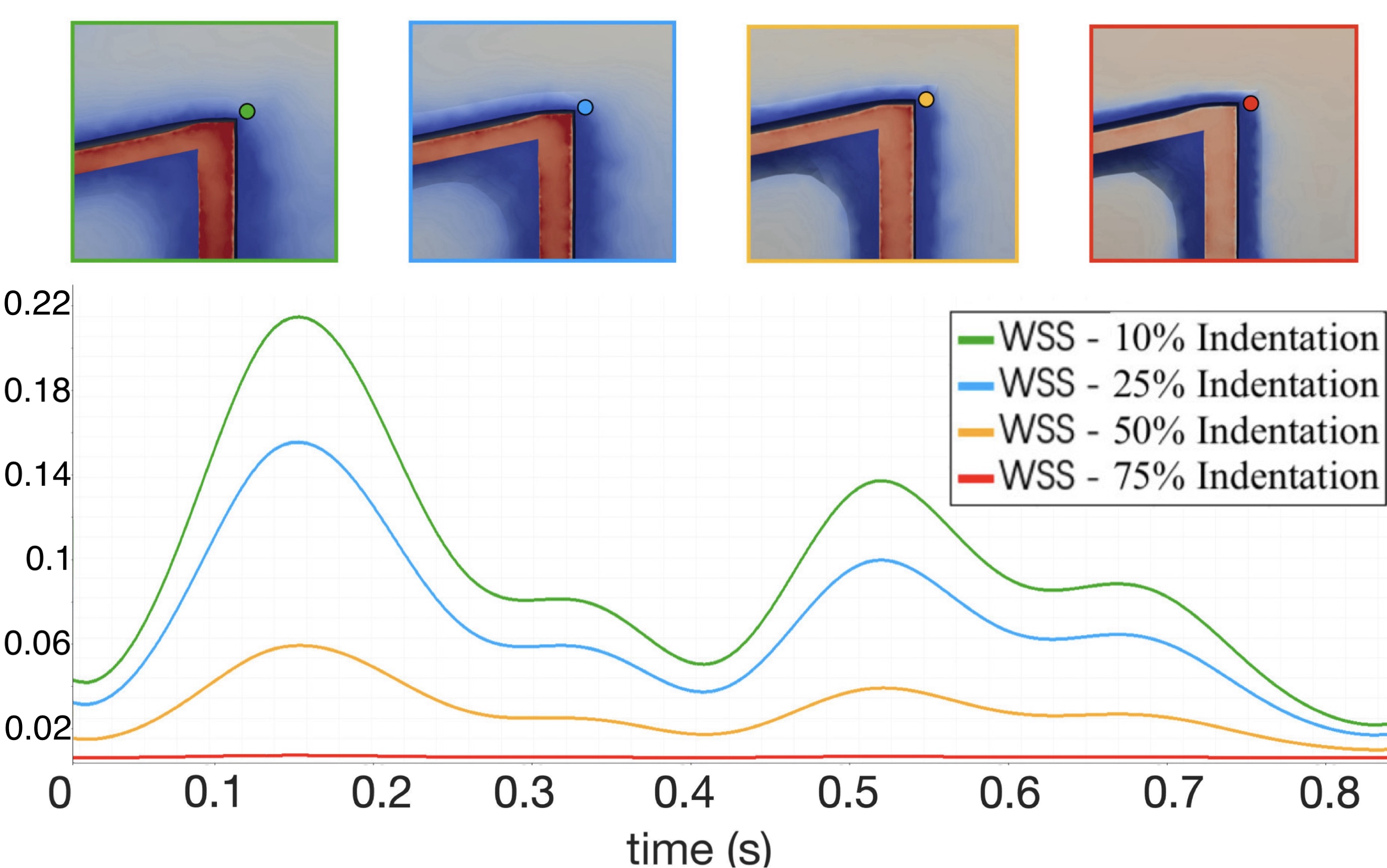}}\label{fig:WSS-lowZoom}}
	\caption{Reference WSS at time $t = 0.25$ s for two indentation percentages (\ref{fig:WSS75} and \ref{fig:WSS10}). Plots of critical WSS for all indentations (\ref{fig:WSS-low0.4} and \ref{fig:WSS-lowZoom}). Plots are color-coded to indentation percentages: 10\% (green), 25\% (blue), 50\% (yellow), 75\% (red).}\label{fig:WSS}
\end{figure}

A qualitative overview of the reference WSS on the artery wall and on the stent inner surface is shown in Figures \ref{fig:WSS75} and \ref{fig:WSS10}. In both cases, peaks of WSS near $1.7$ Pa are observed on the stent surface, which is more exposed to the laminar flow in the lumen. In line with \cite{manjunatha2024silico}, the WSS values on the artery wall far away from the stent and transition areas are very similar for both indentations. We define a critical threshold for values of WSS \textless $0.4$ Pa, which are especially problematic according to \cite{benard2006computational, chiastra2013computational}. Fig. \ref{fig:WSS-low0.4} highlights areas of critical WSS at time $t = 0.4$ s. We observe larger areas of critical WSS for 10\% indentation, which decrease for higher indentations. However, if we inspect the transition areas adjacent to the stent struts as shown in Fig. \ref{fig:WSS-lowZoom}, the case of 75\% indentation displays the lowest values of WSS over the whole cycle. On the other hand, WSS values over one cycle systematically increase for decreasing indentation percentage.\\

\begin{figure}
	\centering
    \subfloat[Reference, 75\%.]{
        \resizebox*{6.9cm}{!}{
        \begin{tikzpicture}
        \node[anchor=south west]{\includegraphics[draft=\draftmode]{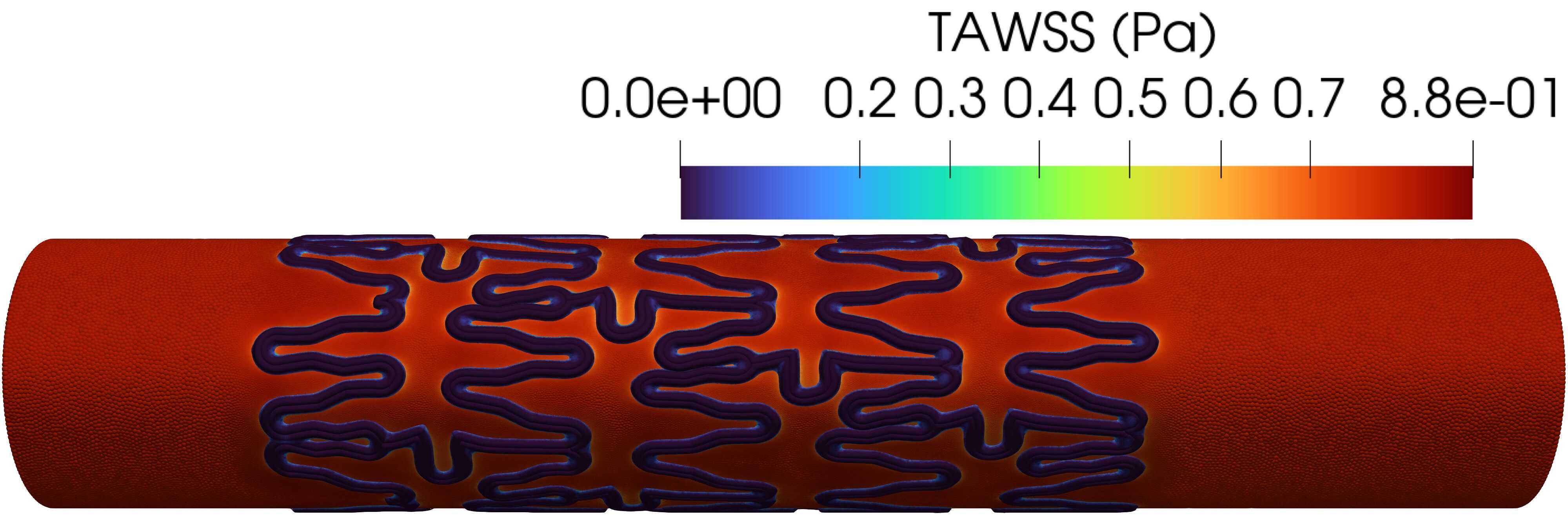}};
        \draw[black, dashed, dash pattern = on 3cm off 3cm, line width = 0.5 cm] (22cm,0cm) rectangle (105cm,26cm);
        \end{tikzpicture}}
        \label{fig:TAWSSRef25}}
     \subfloat[Reference, 10\%.]{
        \resizebox*{6.9cm}{!}{\begin{tikzpicture}
        \node[anchor=south west]{\includegraphics[draft=\draftmode]{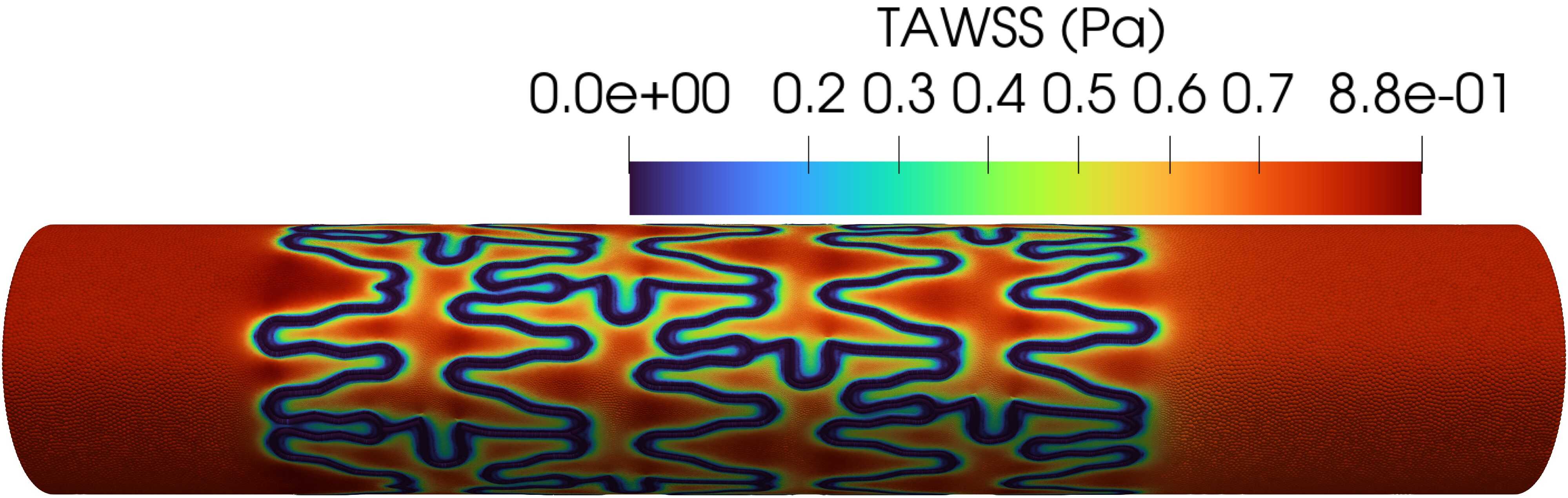}};
        \draw[black, dashed, dash pattern = on 3cm off 3cm, line width = 0.5 cm] (22cm,0cm) rectangle (105cm,25.2cm);
        \end{tikzpicture}}
        \label{fig:TAWSSRef90}}\\
    \subfloat[Zoom on \ref{fig:TAWSSRef25}.]{
        \resizebox*{6.9cm}{!}{
        \begin{tikzpicture}
            \node[inner sep=0] (image) {\includegraphics[draft=\draftmode,trim={22cm 0 36cm 0},clip=true]{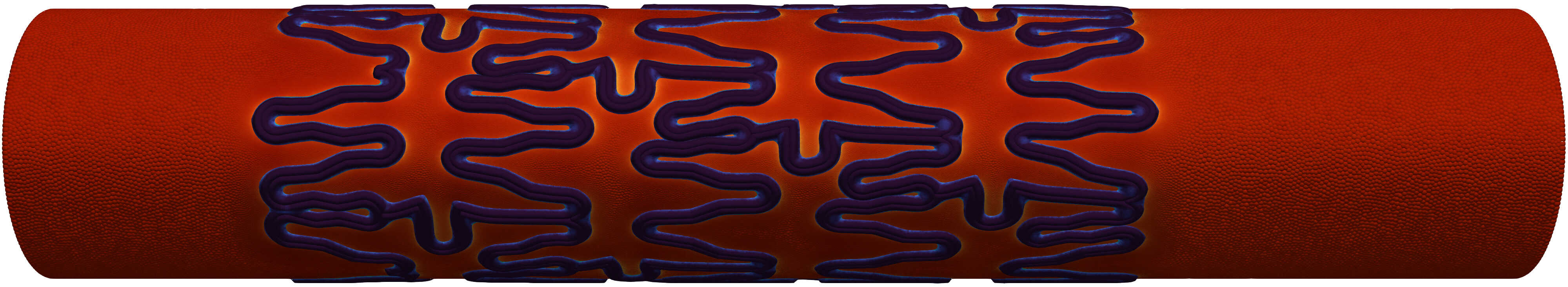}};
            \draw[yellow, dashed, dash pattern = on 1cm off 1cm, line width=0.5cm] (-9cm,-5cm) rectangle (23.5cm,5cm);
            \draw[black, dashed, dash pattern = on 3cm off 3cm, line width = 0.5 cm] (image.south west) rectangle (image.north east);
            \end{tikzpicture}}\label{fig:TAWSSRef25Zoom}}
    \subfloat[Zoom on \ref{fig:TAWSSRef90}.]{
        \resizebox*{6.9cm}{!}{\begin{tikzpicture}
            \node[inner sep=0] (image) {\includegraphics[draft=\draftmode,trim={22cm 0 36cm 0},clip=true]{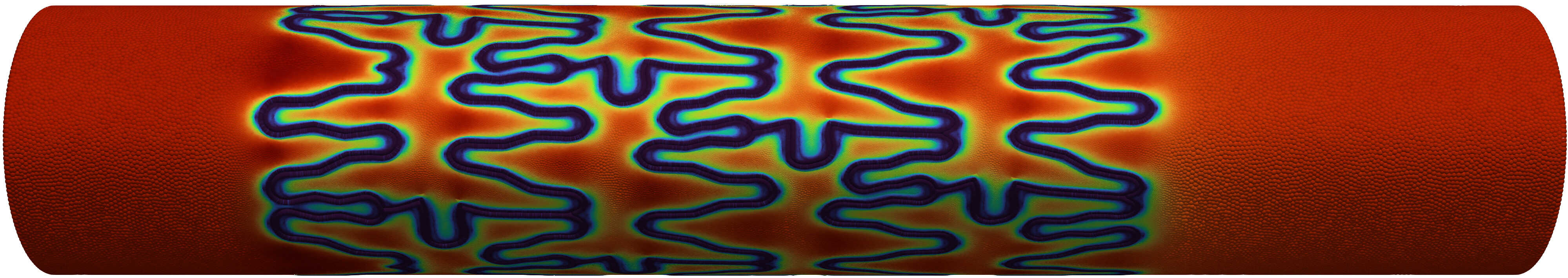}};
            \draw[black, dashed, dash pattern = on 3cm off 3cm, line width = 0.5 cm] (image.south west) rectangle (image.north east);
            \end{tikzpicture}}\label{fig:TAWSSRef90Zoom}}\\
    \subfloat[Zoom on \ref{fig:TAWSSRef25Zoom}.]{%
    	\resizebox*{4.65cm}{!}{\begin{tikzpicture}[baseline={($ (current bounding box.north) - (0,5pt) $)}]
            \node[inner sep=0] (image) {\scalebox{1}[-1]{\includegraphics[draft=\draftmode]{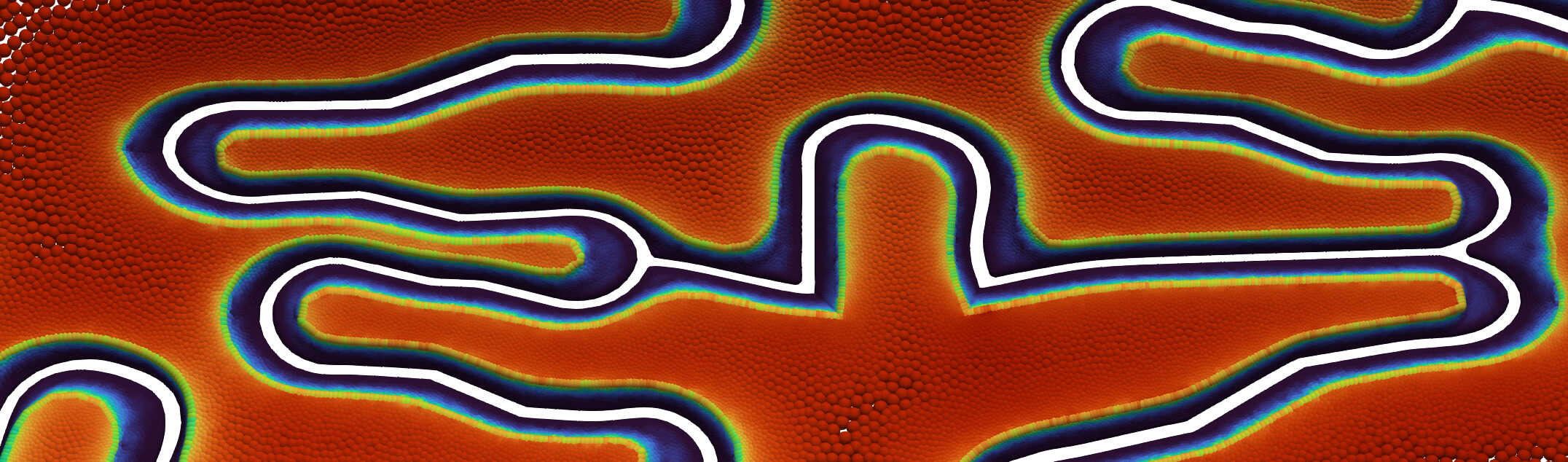}}};
            \draw[yellow, dashed, dash pattern = on 2cm off 2cm, line width = 1 cm] (image.south west) rectangle (image.north east);
            \end{tikzpicture}}\label{fig:ffTAWSS25Zoom}}
    \hspace{0.5 pt}
    \subfloat[TMR, 75\%.]{%
		\resizebox*{4.65cm}{!}{\scalebox{1}[-1]{\includegraphics[draft=\draftmode]{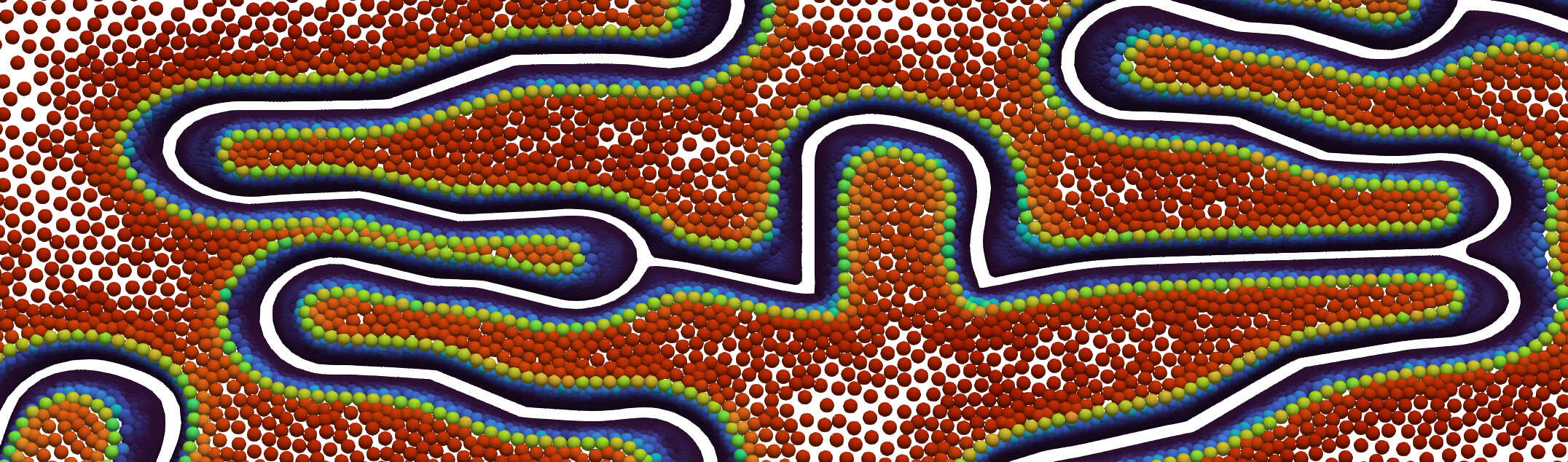}}}
        \label{fig:fTAWSS25Zoom}}
    \subfloat[Uniform, 75\%.]{%
		\resizebox*{4.65cm}{!}{\scalebox{1}[-1]{\includegraphics[draft=\draftmode]{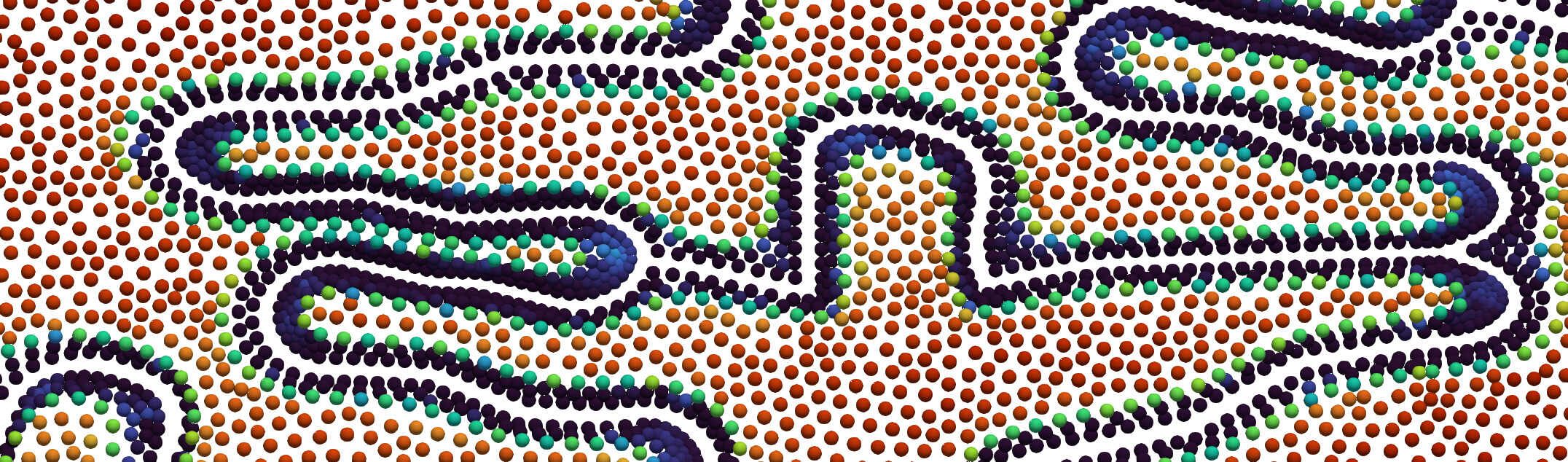}}}
        \label{fig:TAWSS25Zoom}}\\
    \subfloat[Reference, 50\%.]{%
    	\resizebox*{4.65cm}{!}{\scalebox{1}[-1]{\includegraphics[draft=\draftmode]{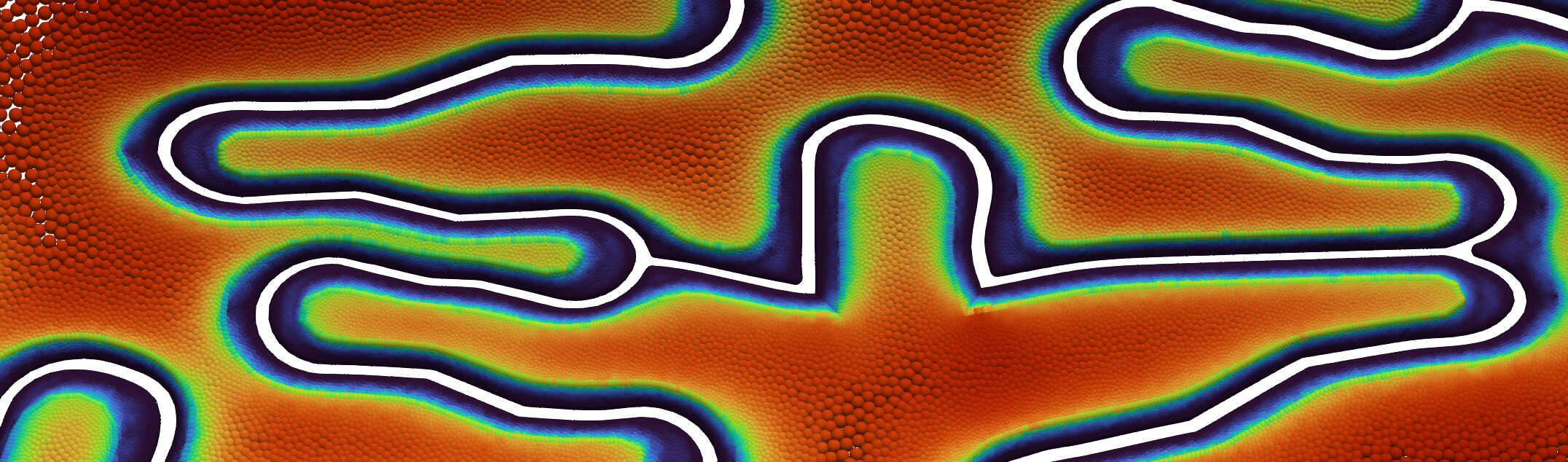}}}
        \label{fig:ffTAWSS50Zoom}}
    \subfloat[TMR, 50\%.]{%
		\resizebox*{4.65cm}{!}{\scalebox{1}[-1]{\includegraphics[draft=\draftmode]{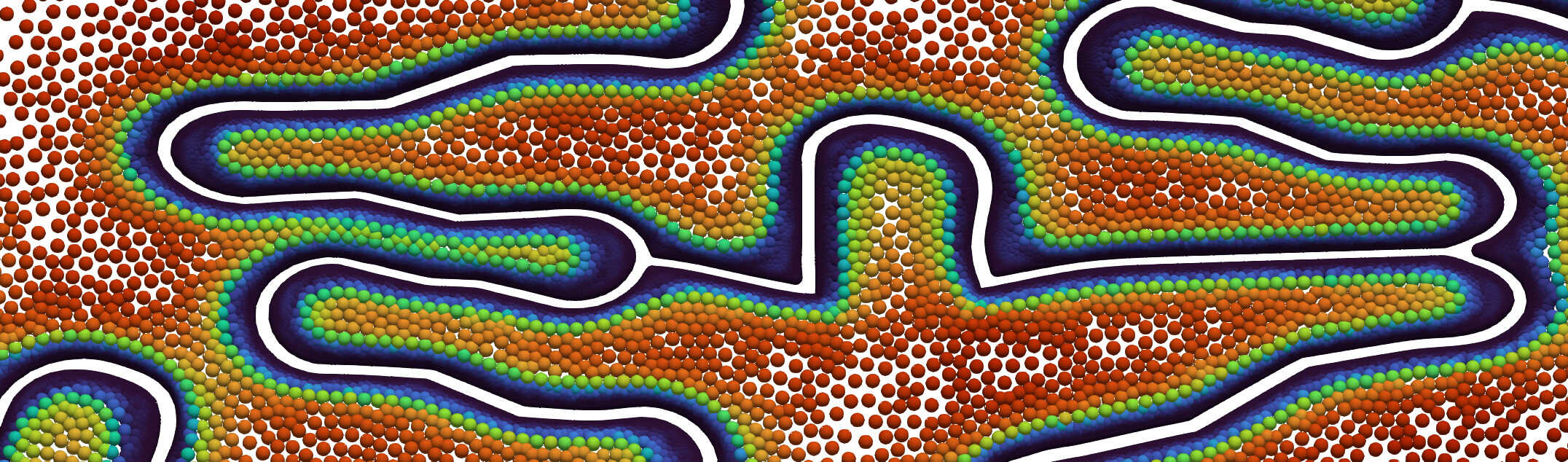}}}
        \label{fig:fTAWSS50Zoom}}
    \subfloat[Uniform, 50\%.]{%
		\resizebox*{4.65cm}{!}{\scalebox{1}[-1]{\includegraphics[draft=\draftmode]{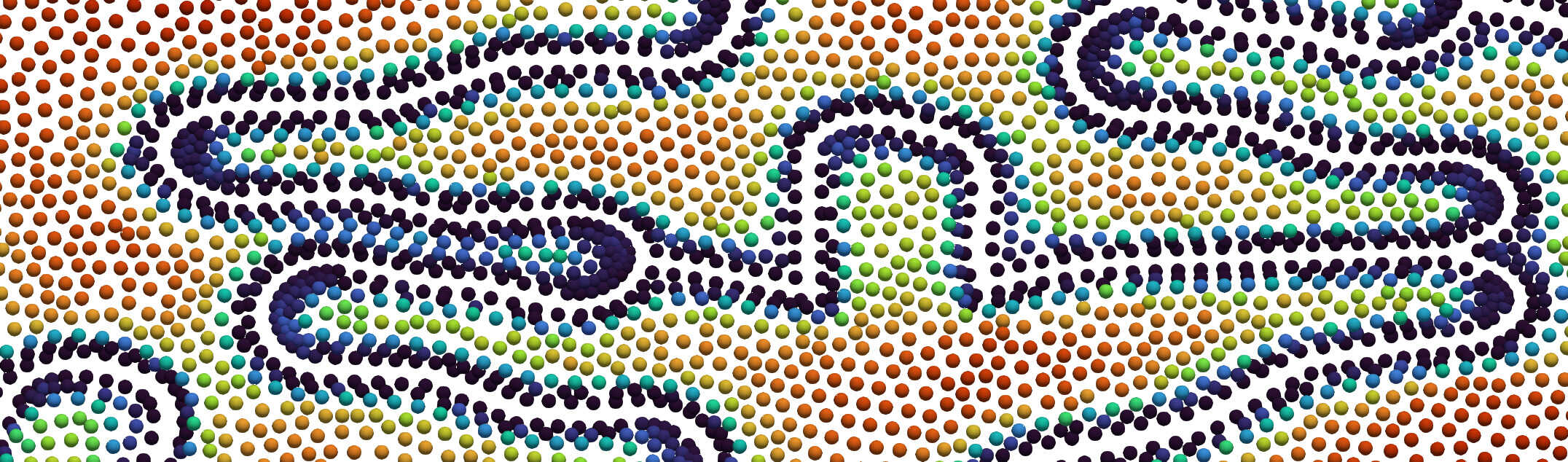}}}
        \label{fig:TAWSS50Zoom}}\\
    \subfloat[Reference, 25\%.]{%
    	\resizebox*{4.65cm}{!}{\scalebox{1}[-1]{\includegraphics[draft=\draftmode]{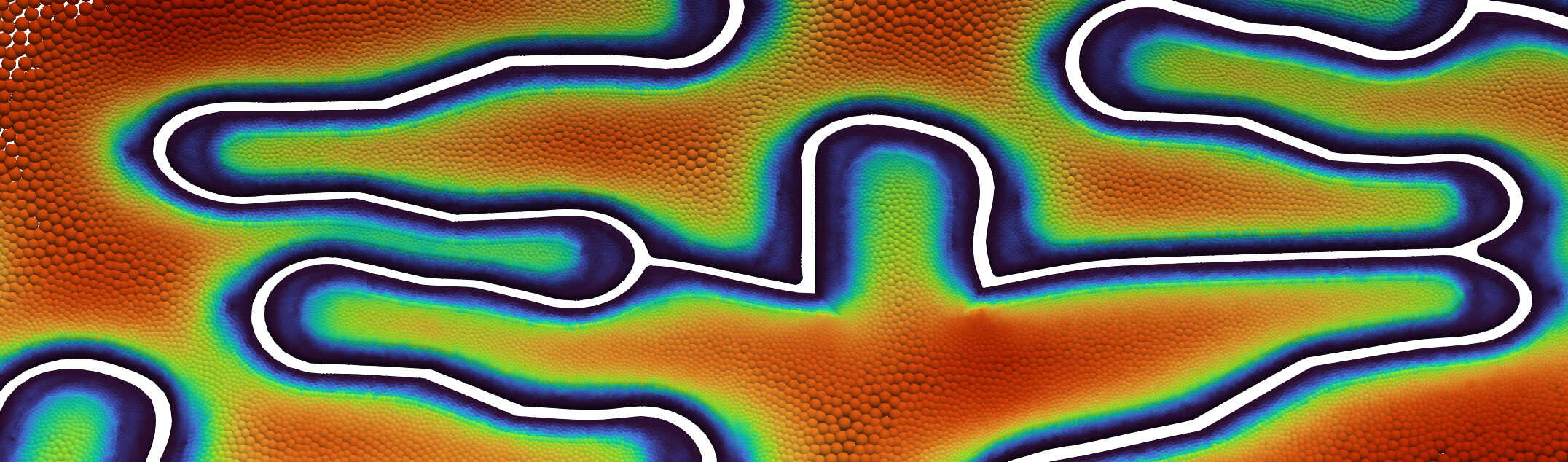}}}
        \label{fig:ffTAWSS75Zoom}}
    \subfloat[TMR, 25\%.]{%
		\resizebox*{4.65cm}{!}{\scalebox{1}[-1]{\includegraphics[draft=\draftmode]{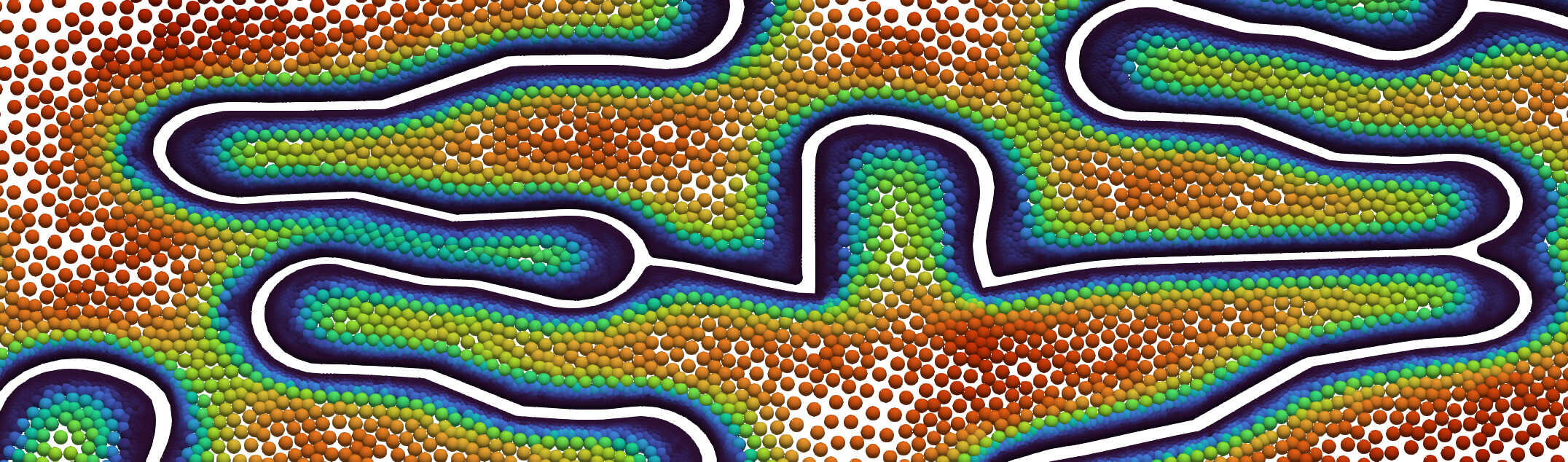}}}
        \label{fig:fTAWSS75Zoom}}
    \subfloat[Uniform, 25\%.]{%
		\resizebox*{4.65cm}{!}{\scalebox{1}[-1]{\includegraphics[draft=\draftmode]{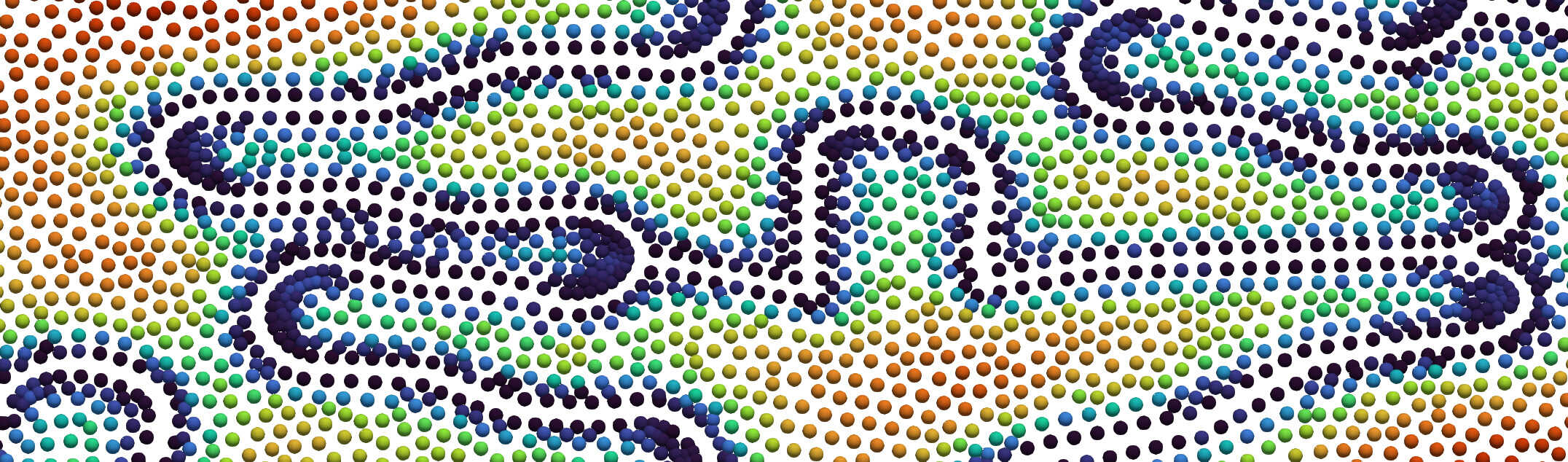}}}
        \label{fig:TAWSS75Zoom}}\\
    \subfloat[Reference, 10\%.]{%
    	\resizebox*{4.65cm}{!}{\scalebox{1}[-1]{\includegraphics[draft=\draftmode]{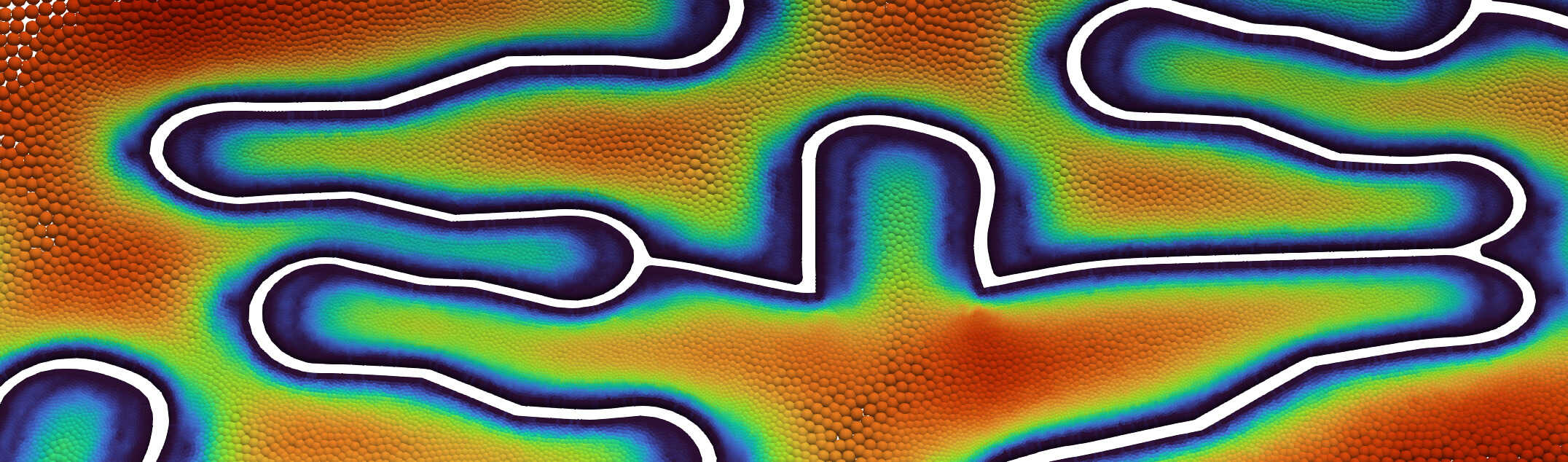}}}
        \label{fig:ffTAWSS90Zoom}}
	\subfloat[TMR, 10\%.]{%
		\resizebox*{4.65cm}{!}{\scalebox{1}[-1]{\includegraphics[draft=\draftmode]{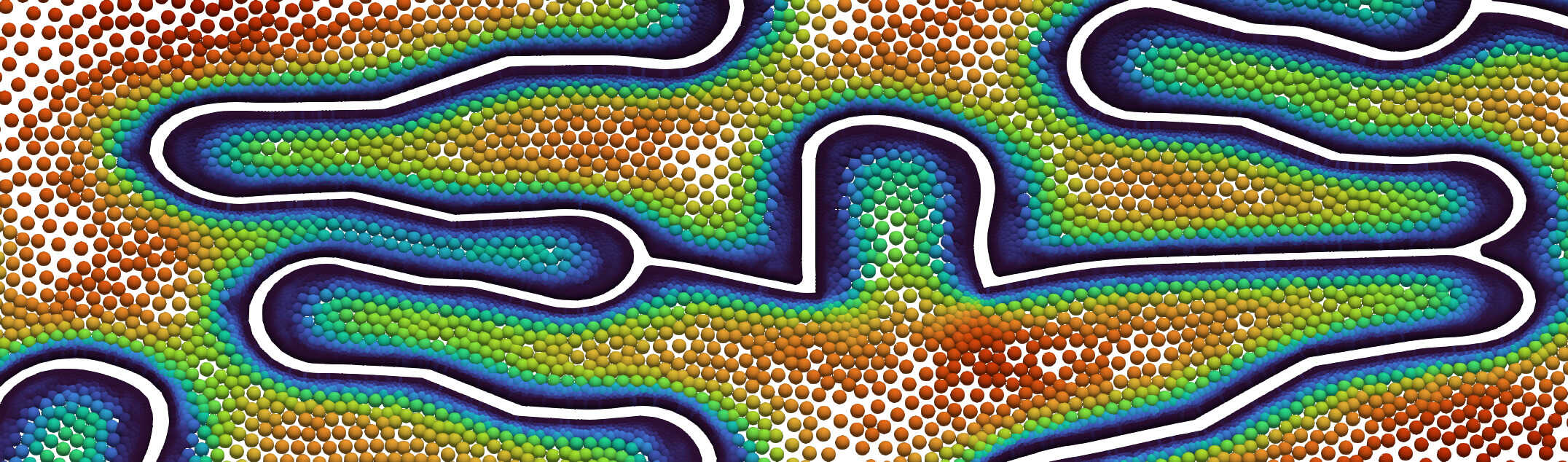}}}
        \label{fig:fTAWSS90Zoom}}
	\subfloat[Uniform, 10\%.]{%
		\resizebox*{4.65cm}{!}{\scalebox{1}[-1]{\includegraphics[draft=\draftmode]{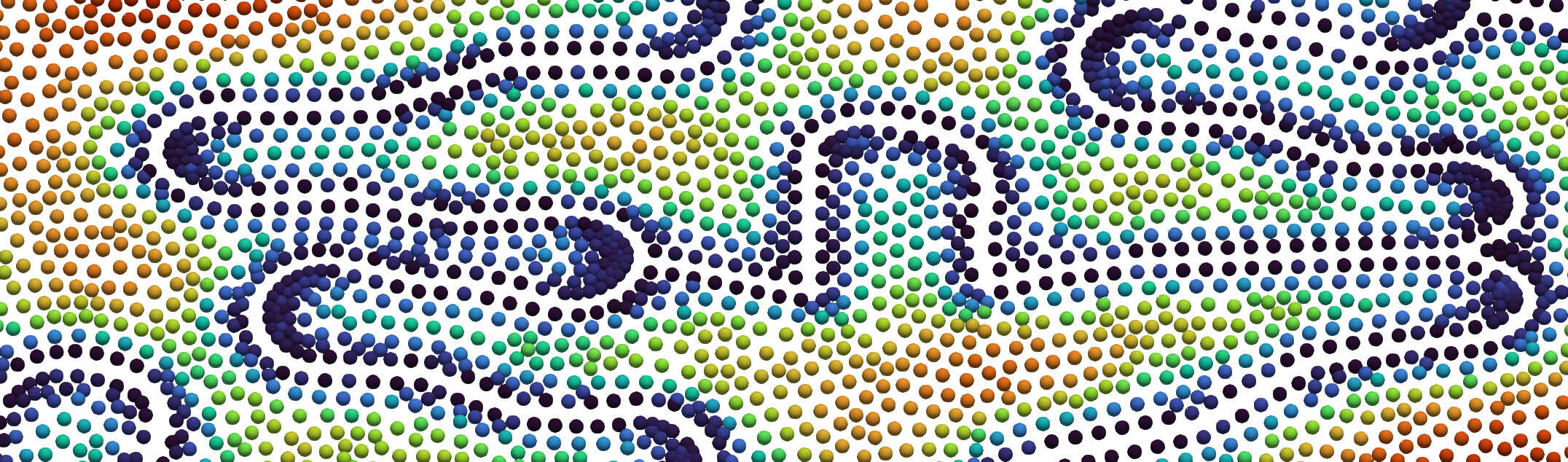}}}
        \label{fig:TAWSS90Zoom}}
	\caption{TAWSS for all mesh refinements and indentations.}\label{fig:TAWSS75-10}
\end{figure}
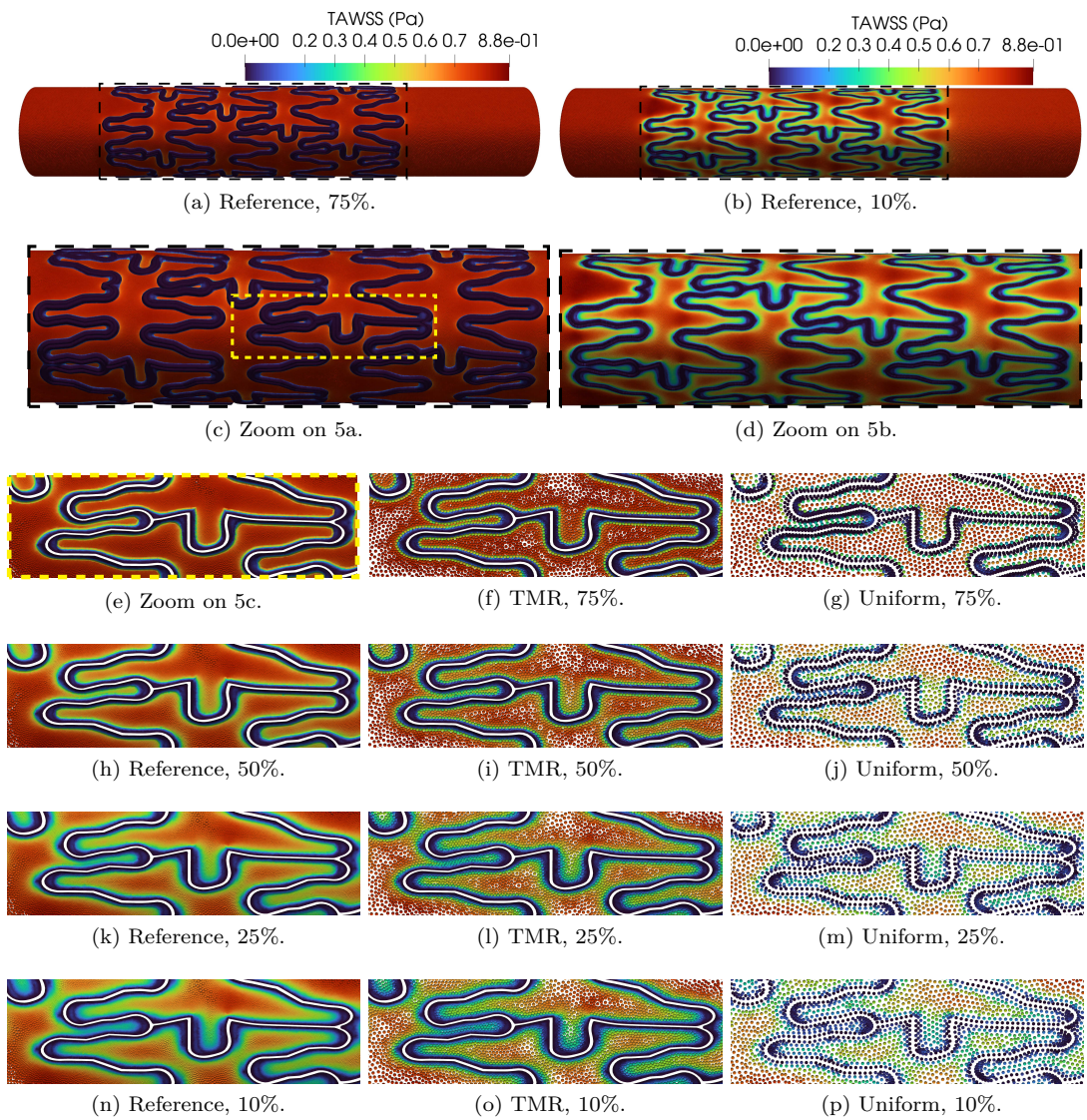

The reference TAWSS for the two extreme indentation percentages chosen in this paper is shown in Figures \ref{fig:TAWSSRef25}-\ref{fig:TAWSSRef90Zoom}. We can qualitatively observe that the TAWSS values range from 0 Pa (in blue) to 0.8 Pa (in red) and low values of TAWSS are detected in both cases close to the stent struts. However, for low indentation levels we observe a smoother transition from 0 values to physiological TAWSS. In the case of 75\% indentation the transition areas are much deeper and sharper, resulting in a sudden change of TAWSS from values close to 0 to physiological ones. If we focus on a particular set of struts on the reference mesh (see Figures \ref{fig:ffTAWSS25Zoom}, \ref{fig:ffTAWSS50Zoom}, \ref{fig:ffTAWSS75Zoom} and \ref{fig:ffTAWSS90Zoom}), high indentations show larger areas of extremely low TAWSS (close to 0) and sharper transition to $\tawss \approx 0.8$ Pa. Simulations with targeted mesh refinement (see Figures \ref{fig:fTAWSS25Zoom}, \ref{fig:fTAWSS50Zoom}, \ref{fig:fTAWSS75Zoom} and \ref{fig:fTAWSS90Zoom}) capture TAWSS values very accurately, particularly on transition areas, for all indentations. On the other hand, TAWSS values on uniform meshes (Figures \ref{fig:TAWSS25Zoom}, \ref{fig:TAWSS50Zoom}, \ref{fig:TAWSS75Zoom} and \ref{fig:TAWSS90Zoom}) roughly reproduce the indentation patterns observed for reference and targeted mesh refinement, but fail to capture the microdynamics near transition areas.

\subsection{Deviation from physiological TAWSS}
\label{3bTAWSSDev}

\begin{figure}
	\centering
    \subfloat[75\%.]{
        \resizebox*{6.9cm}{!}{\begin{tikzpicture}
            \node[inner sep=0] (image) {\includegraphics[draft=\draftmode]{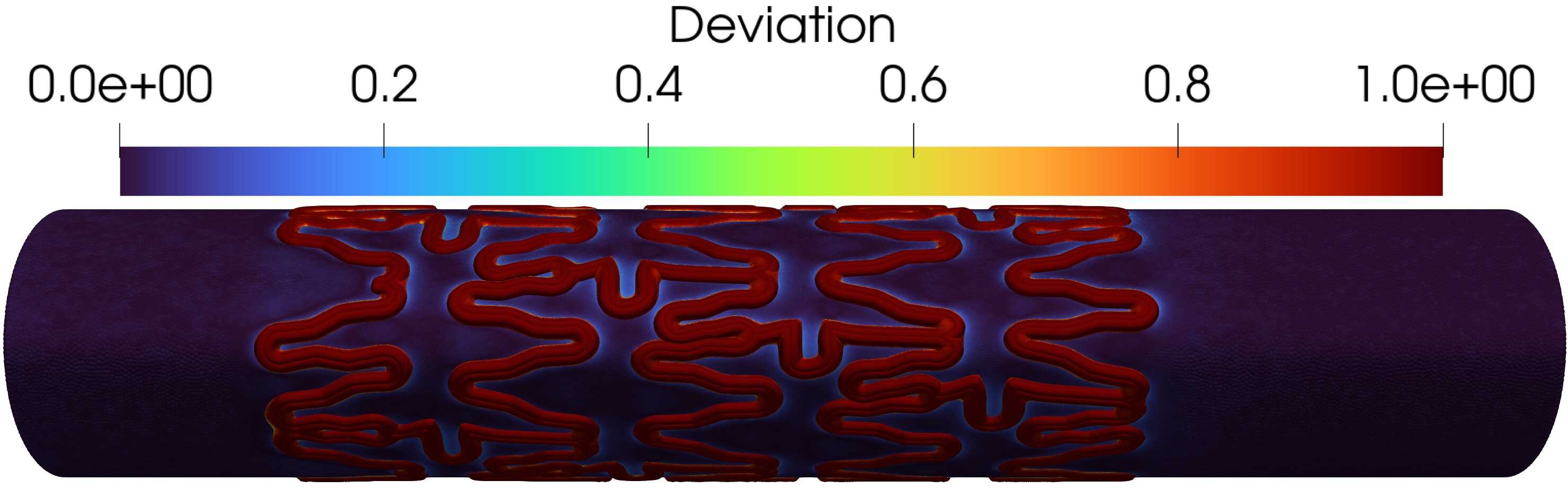}};
            \draw[yellow, dashed, dash pattern = on 1cm off 1cm, line width=0.3cm] (-12cm,-13cm) rectangle (12cm,0cm);
            \end{tikzpicture}}
        \label{fig:Dev75ff}}
     \subfloat[10\%.]{
        \resizebox*{6.9cm}{!}{\begin{tikzpicture}
            \node[inner sep=0] (image) {\includegraphics[draft=\draftmode]{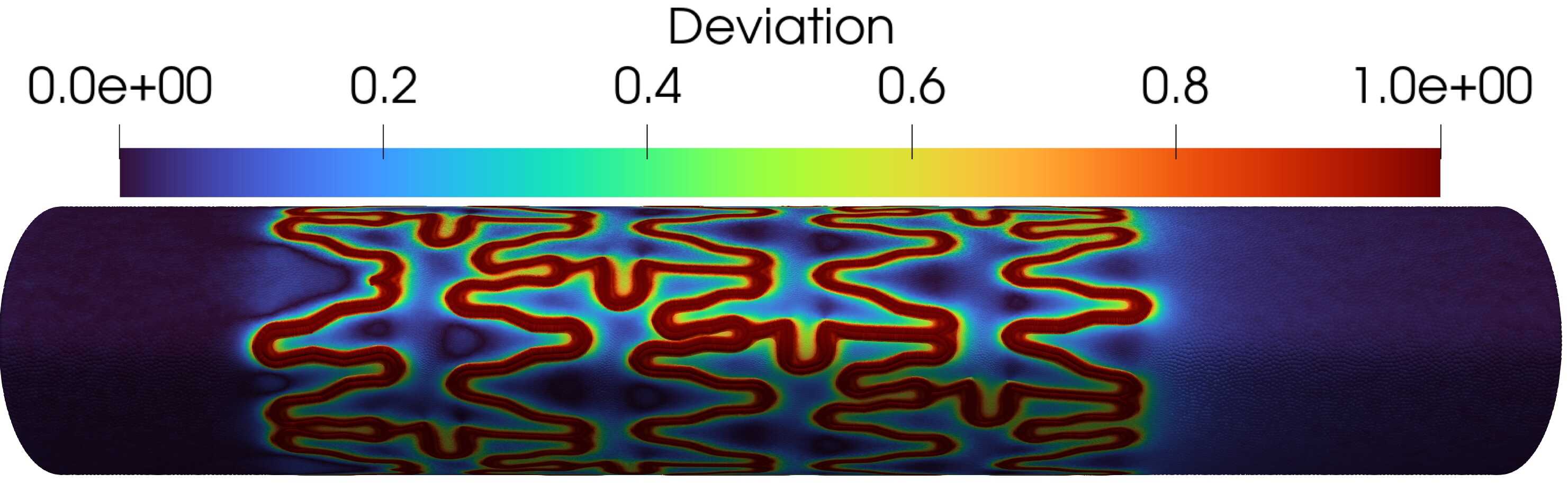}};
            \draw[yellow, dashed, dash pattern = on 1cm off 1cm, line width=0.3cm] (-12cm,-13cm) rectangle (12cm,0cm);
            \end{tikzpicture}}
        \label{fig:Dev10ff}}\\
    \subfloat[Zoom on \ref{fig:Dev75ff}.]{
        \resizebox*{6.9cm}{!}{
        \begin{tikzpicture}
            \node[inner sep=0] (image) {\includegraphics[draft=\draftmode]{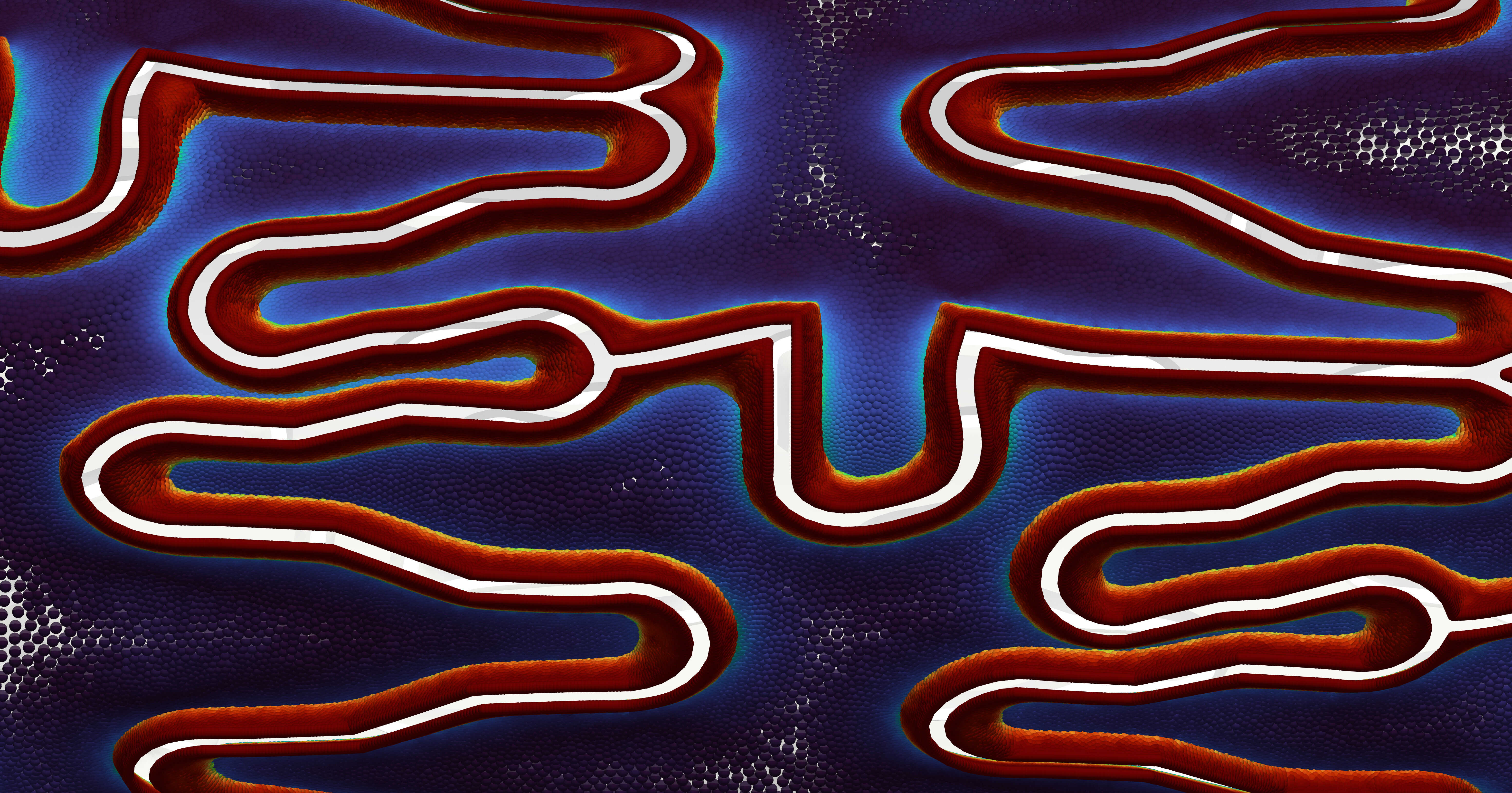}};
            \draw[yellow, dashed, dash pattern = on 4cm off 4cm, line width=2cm] (image.south west) rectangle (image.north east);
            \end{tikzpicture}} \label{fig:Dev75ffZoom}
    }
    \subfloat[Zoom on \ref{fig:Dev10ff}.]{
        \resizebox*{6.9cm}{!}{\begin{tikzpicture}
            \node[inner sep=0] (image) {\includegraphics[draft=\draftmode]{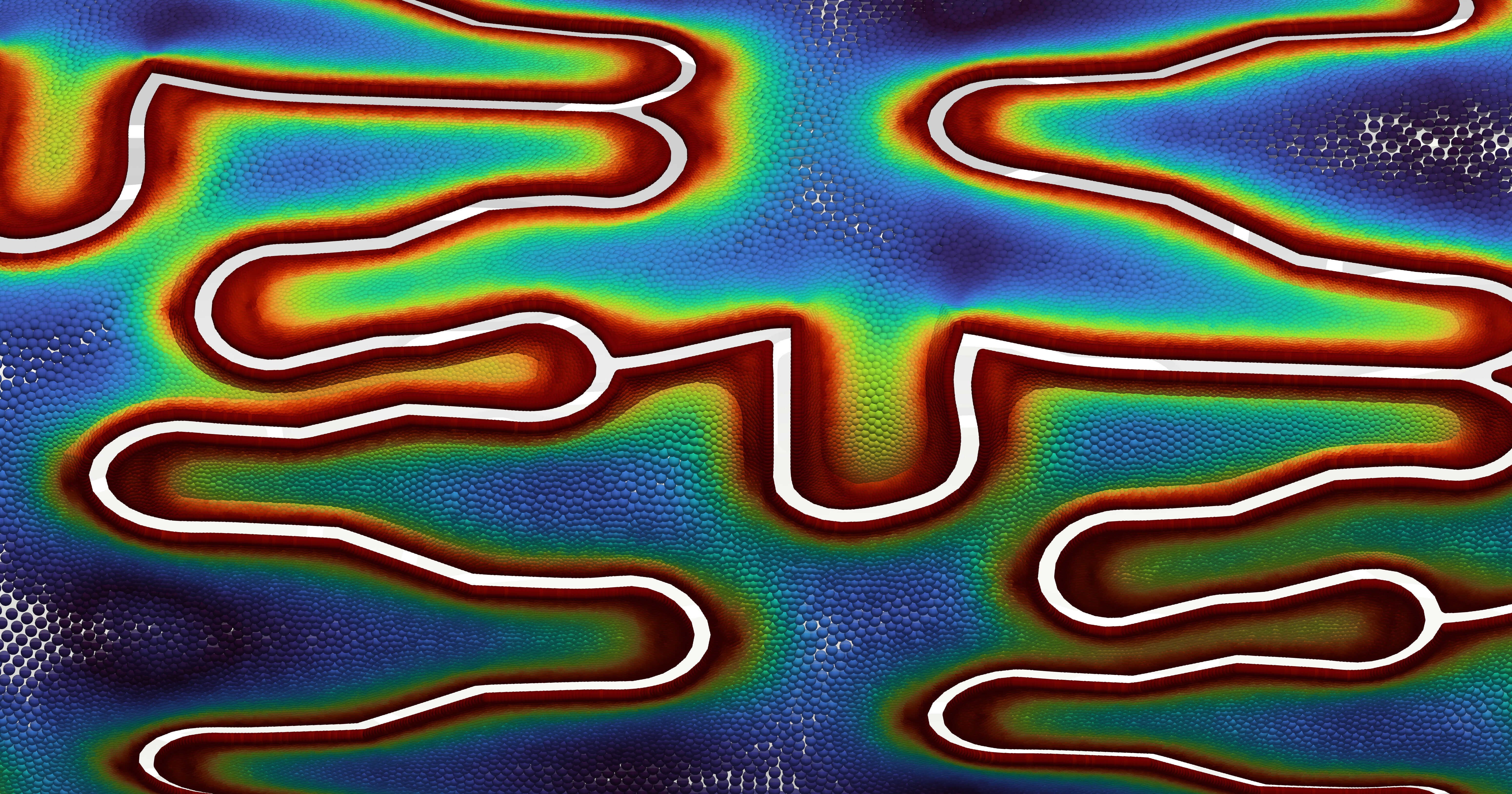}};
            \draw[yellow, dashed, dash pattern = on 4cm off 4cm, line width=2cm] (image.south west) rectangle (image.north east);
            \end{tikzpicture}} \label{fig:Dev10ffZoom}
    }
    \caption{Reference TAWSS deviation $\delta$ for two indentation percentages.}
    \label{fig:TAWSSDeviation75-10}
\end{figure}
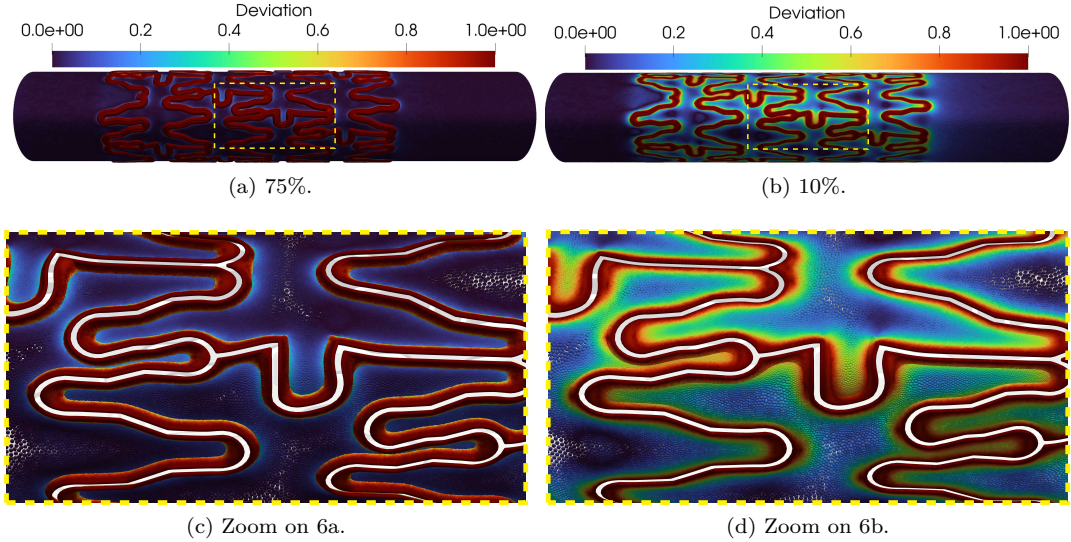

To better estimate the effects of the struts and indentation compared to healthy hemodynamics, we investigate the deviation $\delta$ from the physiological TAWSS value of 0.8 Pa \cite{manjunatha2024silico}, defined as:

\begin{equation}
    \delta = \frac{|\tawss - 0.8|}{0.8}.
\end{equation}

A deviation close to 0 corresponds to values within a healthy range, while a $\delta$ close to 1 highlights critical values of TAWSS. In Figures \ref{fig:Dev75ff} and \ref{fig:Dev10ff} we observe that the transition pattern for 10\% and 75\% is similar to the one observed in Figure \ref{fig:TAWSS75-10}: the case of 75\% indentation shows highly deviating values in the transition areas and 10\% indentation gradually transitions from deviating values in the stent vicinity to 0 deviation on the non-deformed artery wall; see strut details in Figures \ref{fig:Dev75ffZoom} and \ref{fig:Dev10ffZoom}.\\

\begin{figure}
    \centering
    \resizebox*{14cm}{!}{\includegraphics[draft=\draftmode]{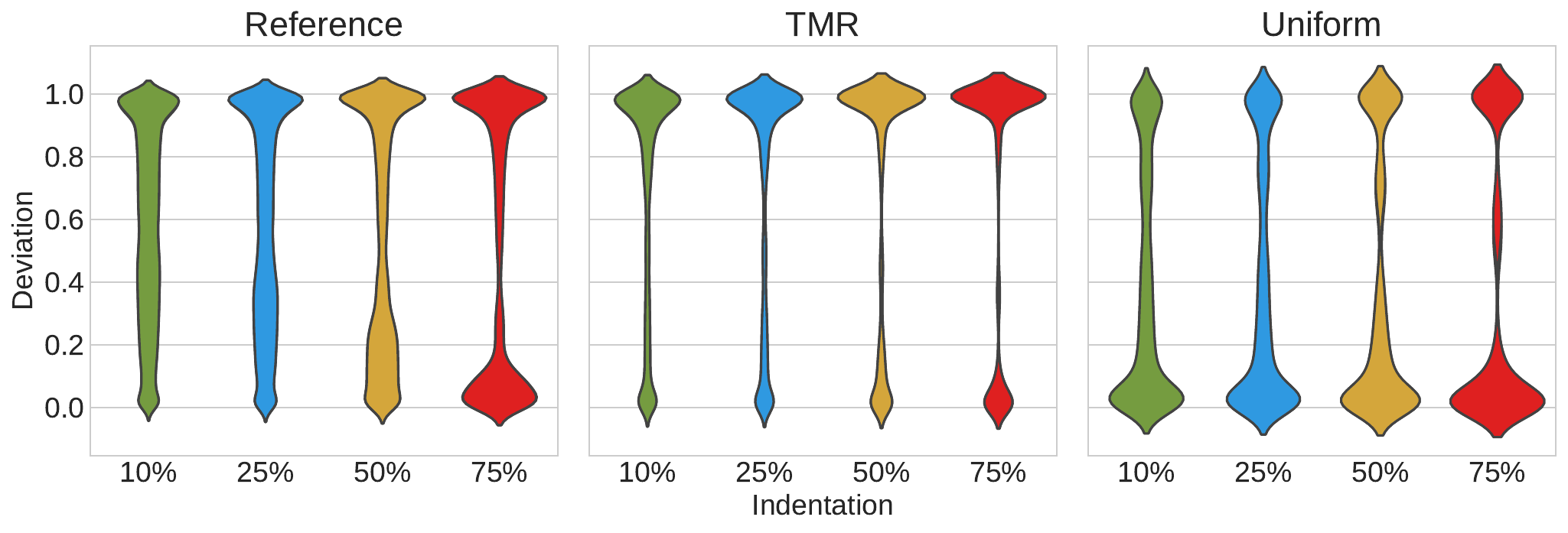}}
	\caption{Violin plots of $\delta$ density distribution for all mesh refinements and indentations.}\label{fig:TAWSSDeviation}
\end{figure}

Figure \ref{fig:TAWSSDeviation} provides an overview of indentation percentage effects on the deviation $\delta$ for all mesh refinements. The violin plots estimate the density distribution of $\delta$, based on the number of mesh nodes associated with a certain deviation value. In the reference case, we observe that low indentations have a more uniformly distributed deviation between 0 and 1, which becomes more hour-glass-shaped for higher indentations. In particular, 75\% indentation has the highest density of both $\delta \approx 1$ and $\delta \approx 0$. The simulations with targeted mesh refinement reproduce qualitatively accurate results for all indentation percentages, in particular for deviation $\delta \approx 1$ in the stent proximity. Physiological TAWSS values are located where the targeted refined mesh is coarser. Therefore all indentations have low density for $\delta < 0.6$. In the uniform case, the deviation trend for all indentations is similar for $\delta \approx 1$. However, the density distribution is much larger for $\delta \approx 0$, inverting the trend shown in the reference case.

\subsection{Critical values of TAWSS}
\label{3cCriticalTAWSS}

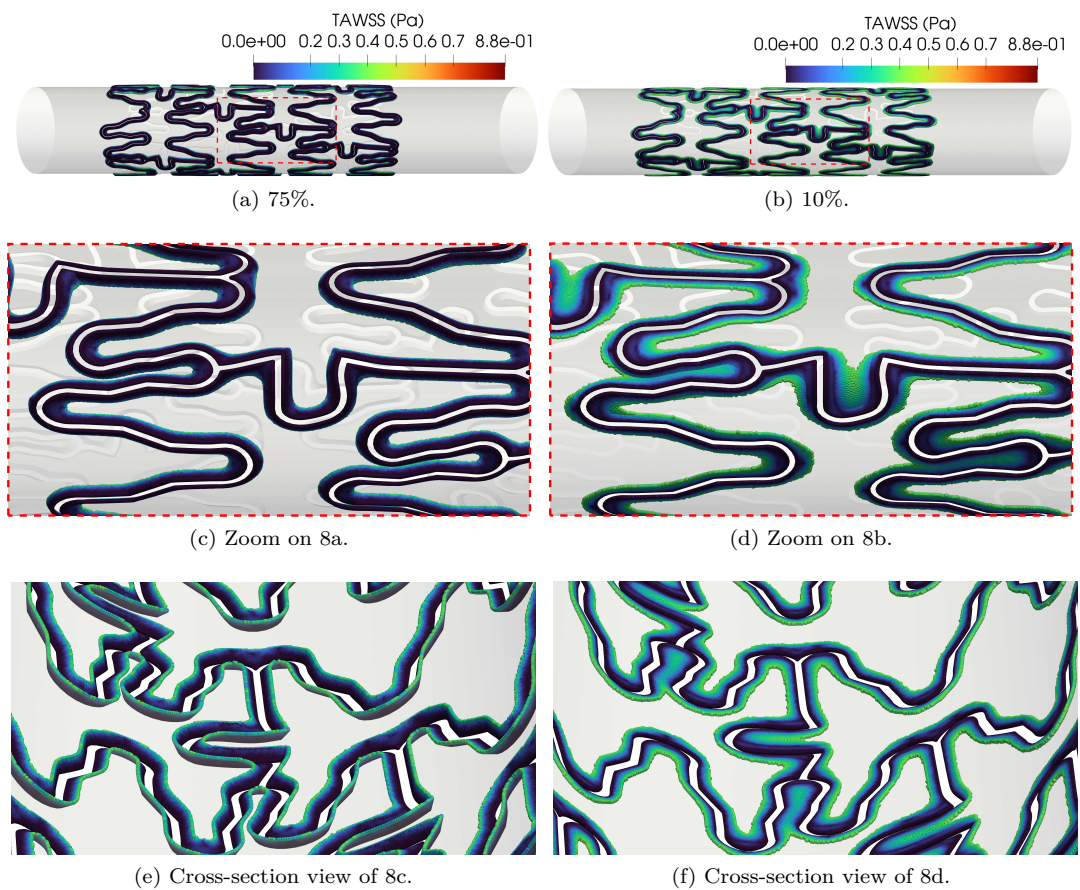
\begin{figure}
	\centering
    \subfloat[75\%.]{
        \resizebox*{6.9cm}{!}{
         \begin{tikzpicture}
            \node[inner sep=0] (image) {\includegraphics[draft=\draftmode]{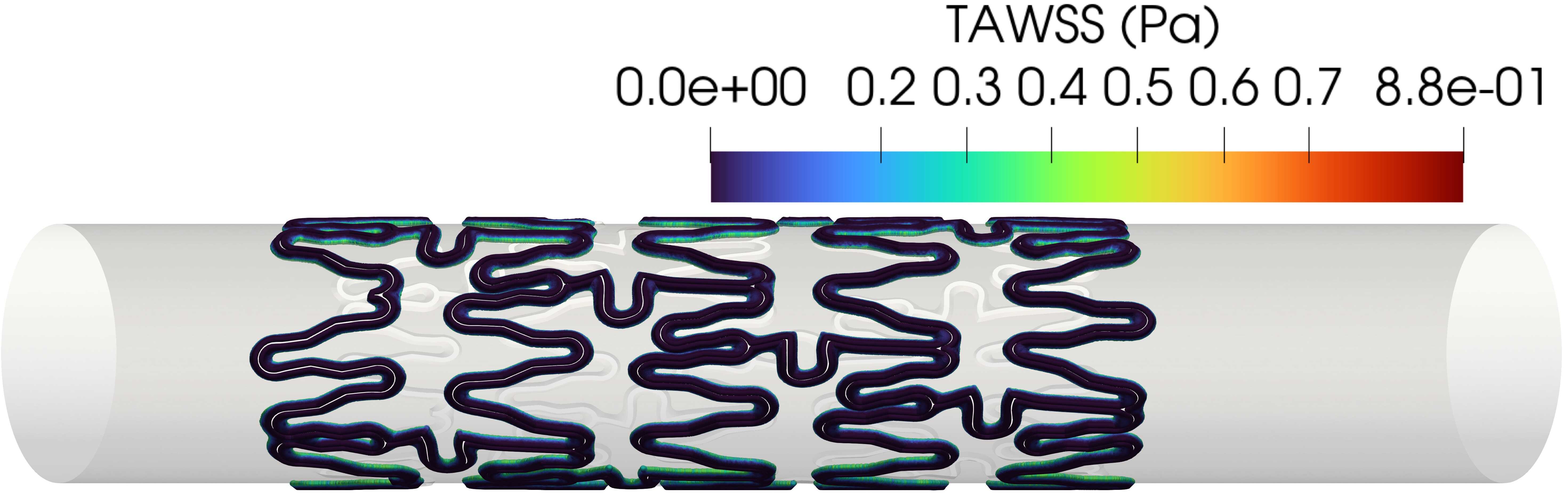}};
            \draw[red, dashed, dash pattern = on 2cm off 2cm, line width=0.5cm] (-24cm,-27cm) rectangle (23cm,-1cm);
            \end{tikzpicture}}
        \label{fig:TAWSS0.4-75}}
    \subfloat[10\%.]{
        \resizebox*{6.9cm}{!}{\begin{tikzpicture}
            \node[inner sep=0] (image) {\includegraphics[draft=\draftmode]{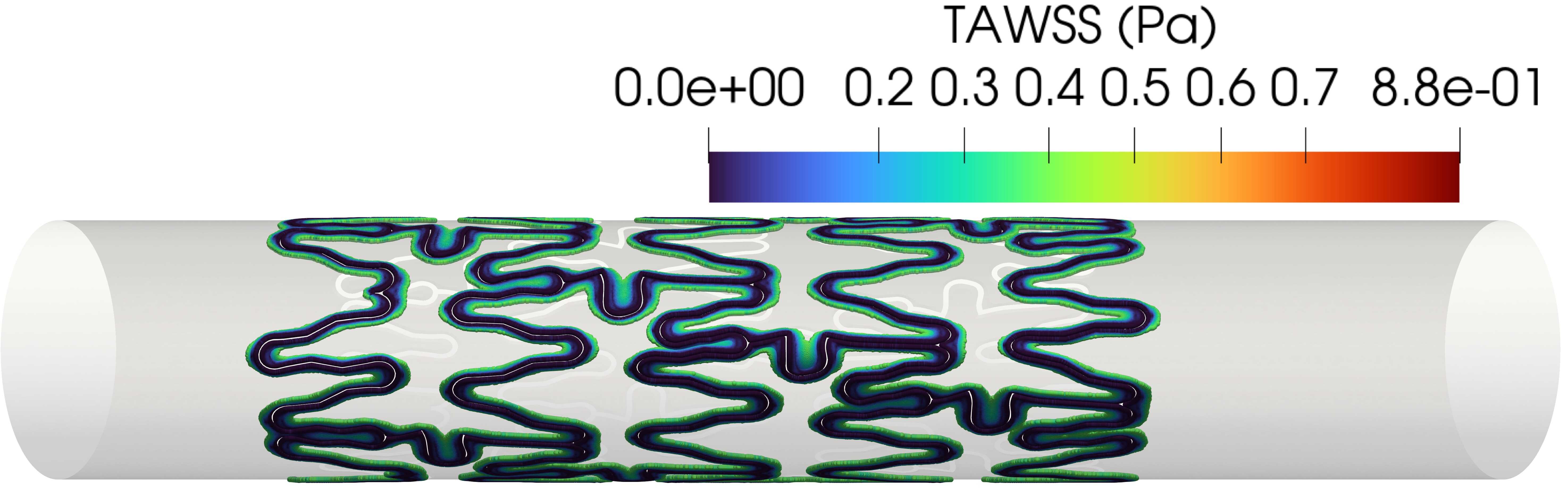}};
            \draw[red, dashed, dash pattern = on 2cm off 2cm, line width=0.5cm] (-24cm,-27cm) rectangle (23cm,-1cm);
            \end{tikzpicture}}
        \label{fig:TAWSS0.4-10}}\\
    \subfloat[Zoom on \ref{fig:TAWSS0.4-75}.]{
        \resizebox*{6.9cm}{!}{
        \begin{tikzpicture}
            \node[inner sep=0] (image) {\includegraphics[draft=\draftmode]{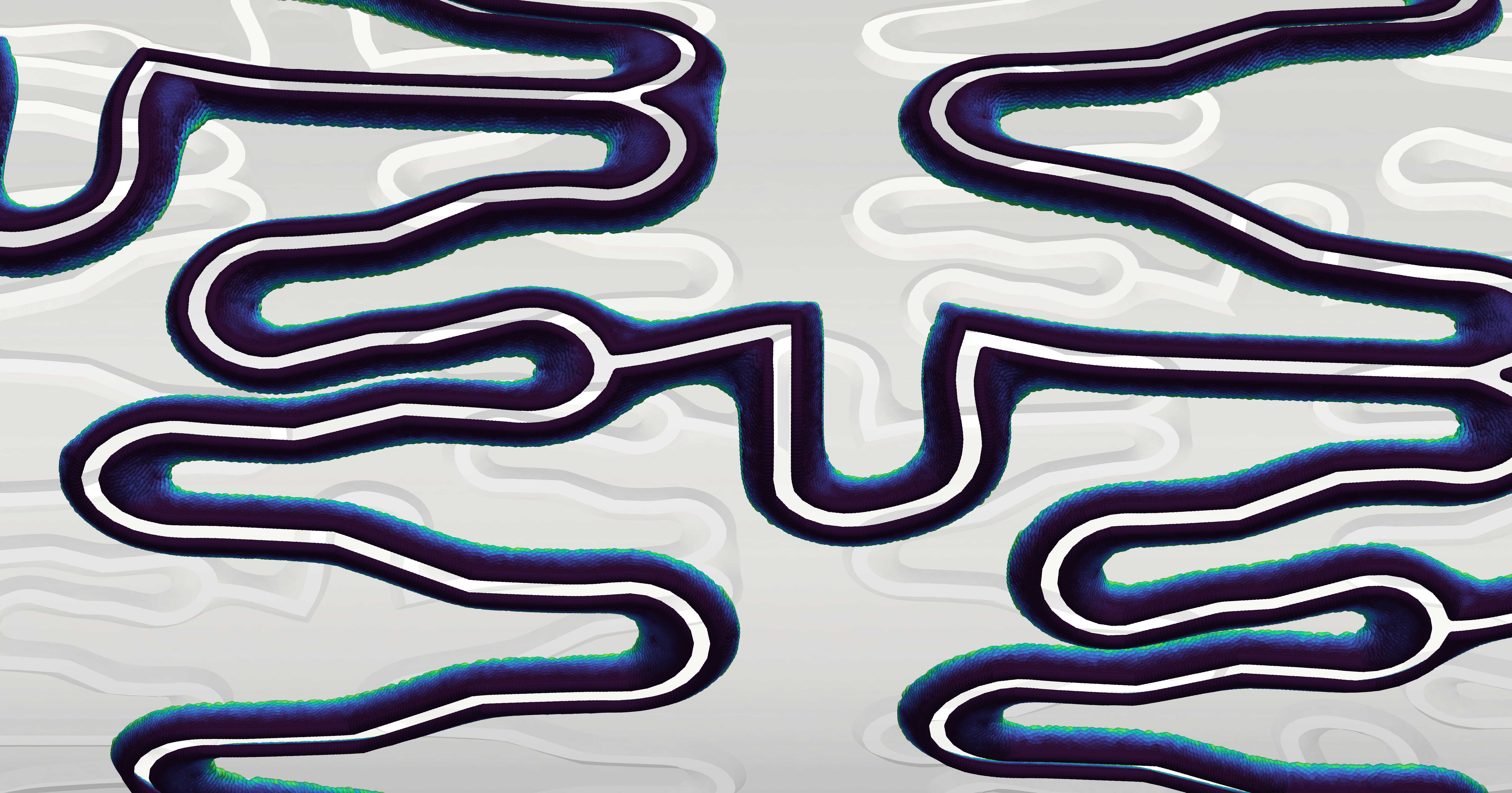}};
            \draw[red, dashed, dash pattern = on 3cm off 3cm, line width=1cm] (image.south west) rectangle (image.north east);
            \end{tikzpicture}}
        \label{fig:TAWSS0.4-75Zoom}}
    \subfloat[Zoom on \ref{fig:TAWSS0.4-10}.]{
        \resizebox*{6.9cm}{!}{\begin{tikzpicture}
            \node[inner sep=0] (image) {\includegraphics[draft=\draftmode]{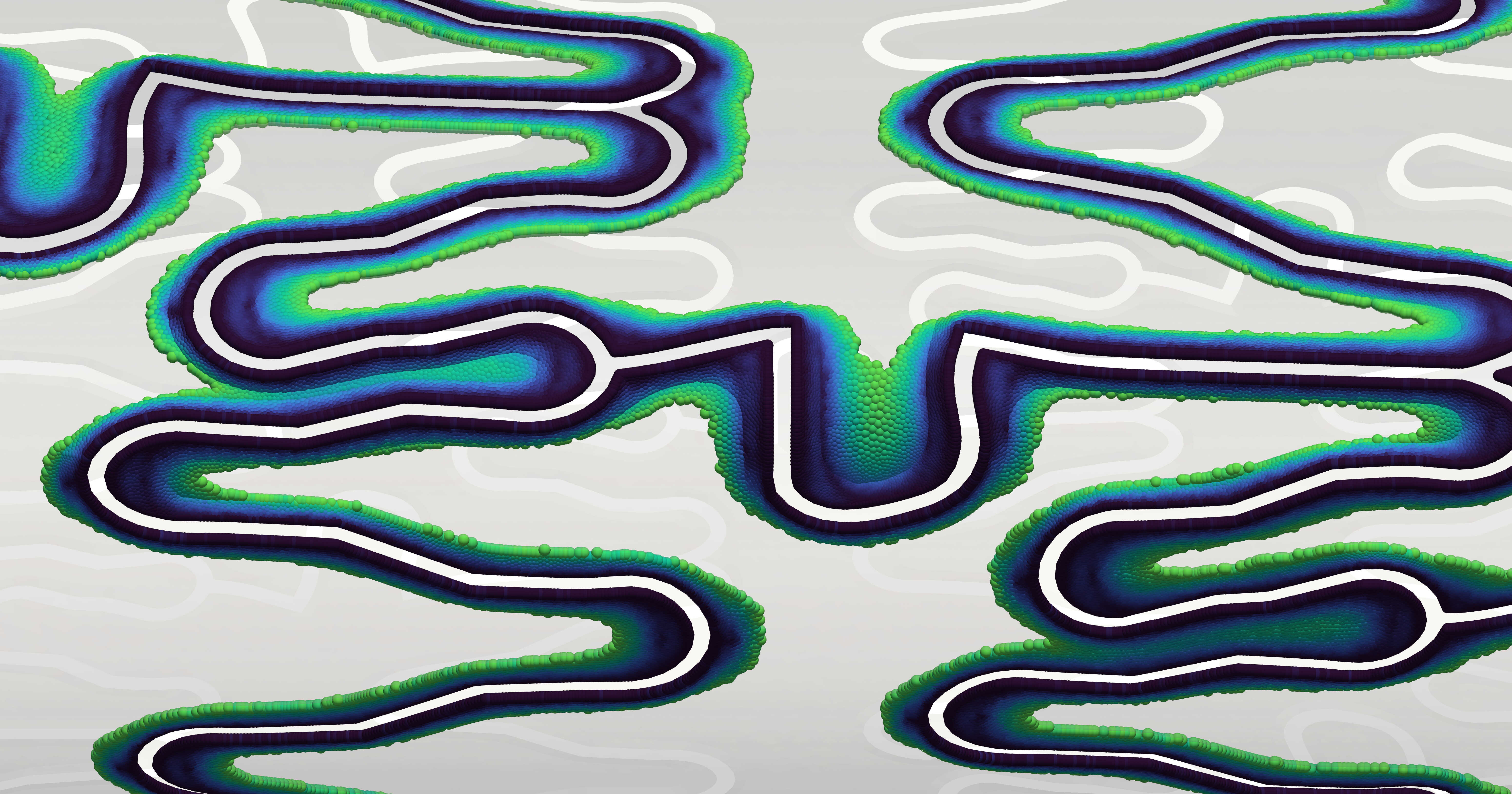}};
            \draw[red, dashed, dash pattern = on 3cm off 3cm, line width=1cm] (image.south west) rectangle (image.north east);
            \end{tikzpicture}}
        \label{fig:TAWSS0.4-10Zoom}}\\
    \subfloat[Cross-section view of \ref{fig:TAWSS0.4-75Zoom}.]{
    \resizebox*{6.9cm}{!}{\includegraphics[draft=\draftmode]{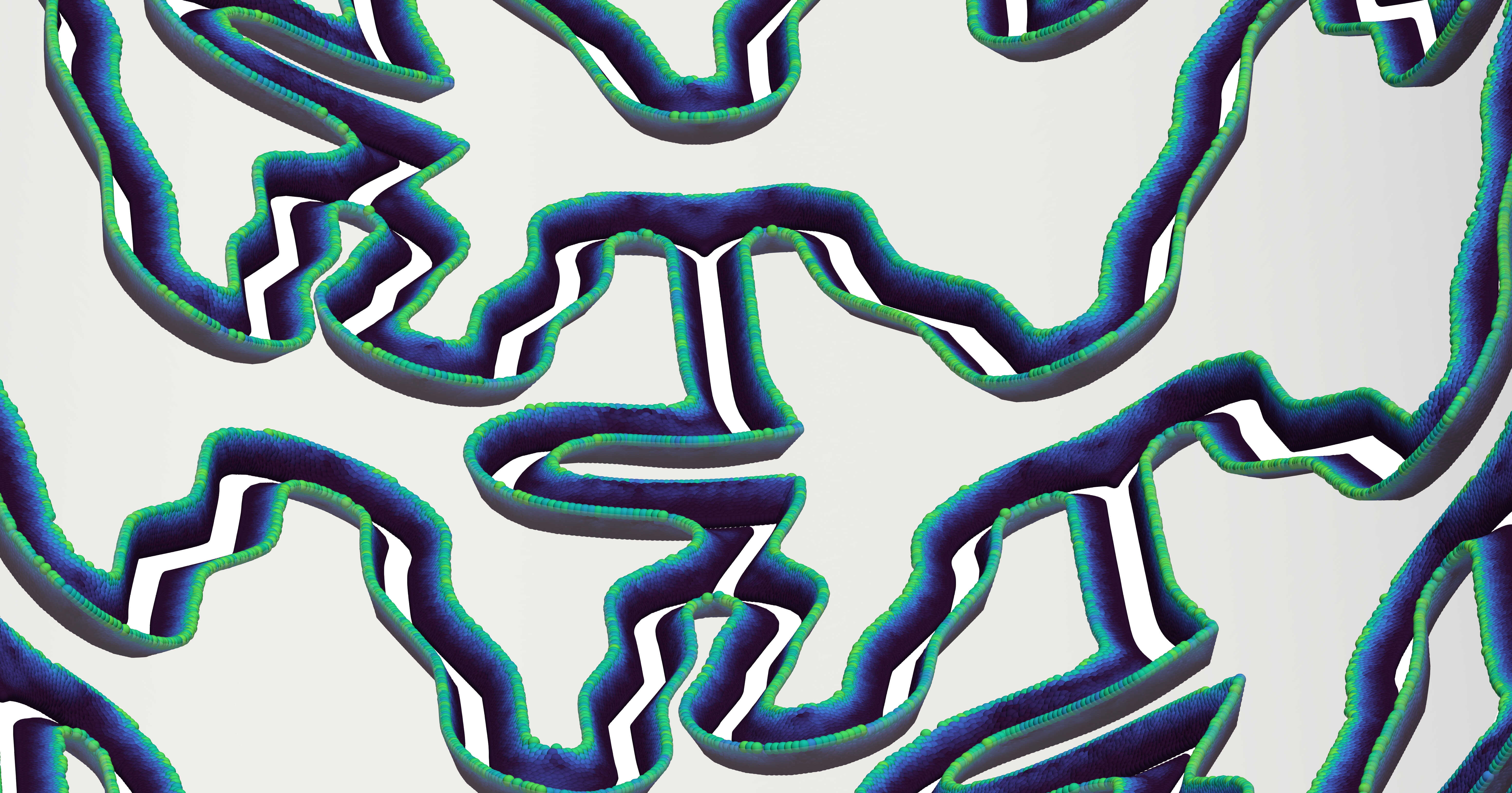}} \label{fig:TAWSS0.4-75ZoomCS}}
    \subfloat[Cross-section view of \ref{fig:TAWSS0.4-10Zoom}.]{
    \resizebox*{6.9cm}{!}{\includegraphics[draft=\draftmode]{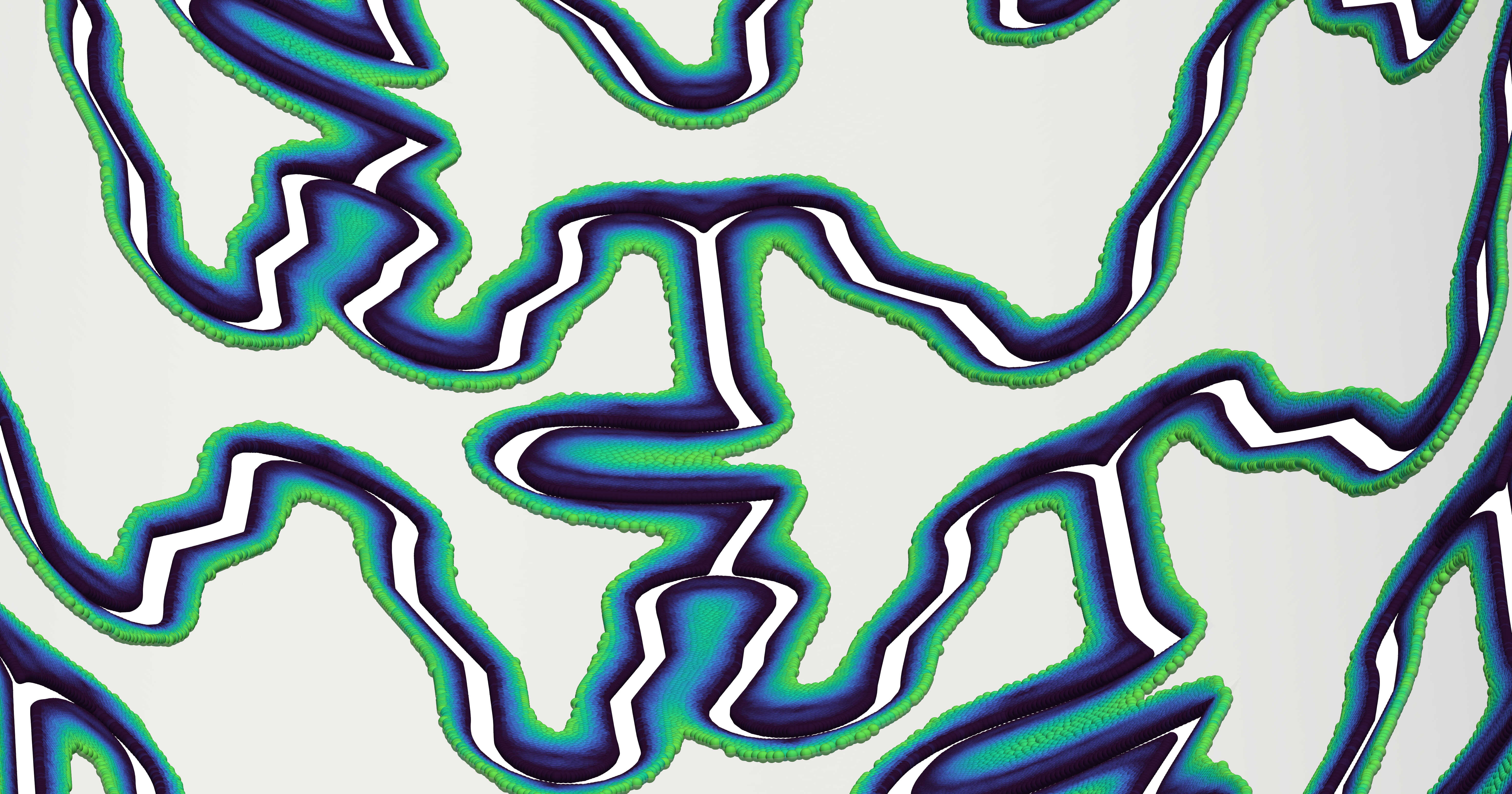}}\label{fig:TAWSS0.4-10ZoomCS}}
	\caption{Areas of artery wall with TAWSS $<$ 0.4 Pa on reference mesh for two indentation percentages.}\label{fig:TAWSS04}
\end{figure}

In this subsection, we analyze the critical sub-threshold TAWSS between 0 and 0.4 Pa \cite{benard2006computational, chiastra2013computational} both qualitatively and quantitatively.\\

Figure \ref{fig:TAWSS04} highlights where critical values are located, i.e., in the vicinity of the stent struts, almost completely overlapping with the transition areas for 75\% indentation, as shown in Fig. \ref{fig:TAWSS0.4-75ZoomCS}. The highlighted struts in Figure \ref{fig:TAWSS0.4-75Zoom} also showcase that for high indentation we observe small TAWSS variance and high density of values close to 0. On the other hand, low values of TAWSS for 10\% indentation are spread over a larger portion of artery wall and showcase higher density of TAWSS towards the upper end of the critical values range (see Figures \ref{fig:TAWSS0.4-10Zoom} and \ref{fig:TAWSS0.4-10ZoomCS}).\\

\begin{figure}
    \centering
    \subfloat[All mesh refinements: indentation comparison.]{
    \resizebox*{14cm}{!}{\includegraphics[draft=\draftmode]{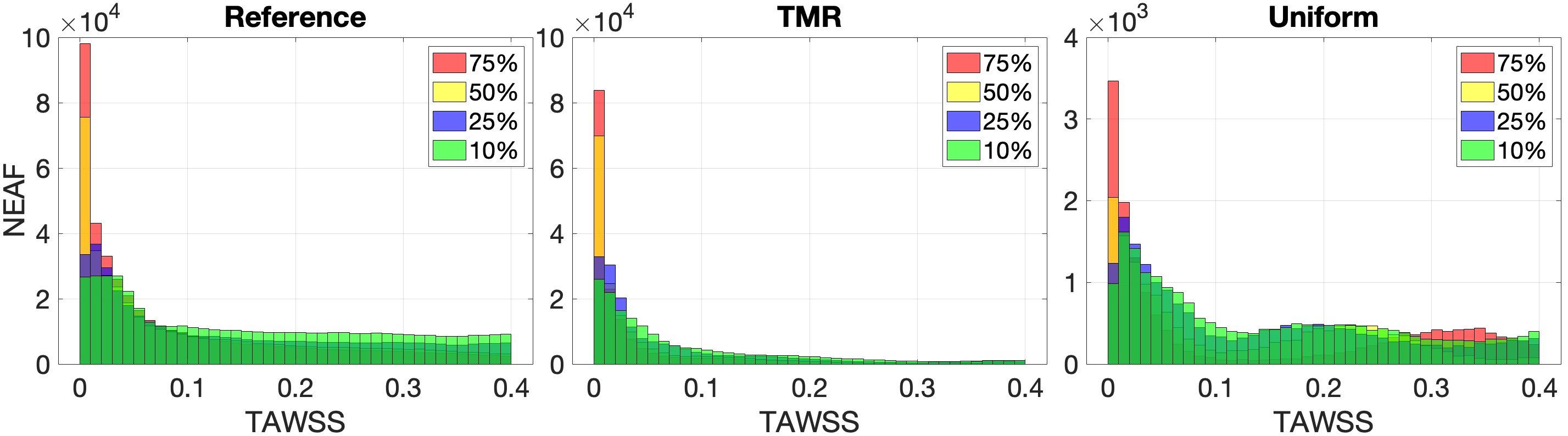}}\label{fig:TAWSS04histRefs}}\\
    \subfloat[All indentations: mesh refinement comparison.]{
    \resizebox*{14cm}{!}{\includegraphics[draft=\draftmode]{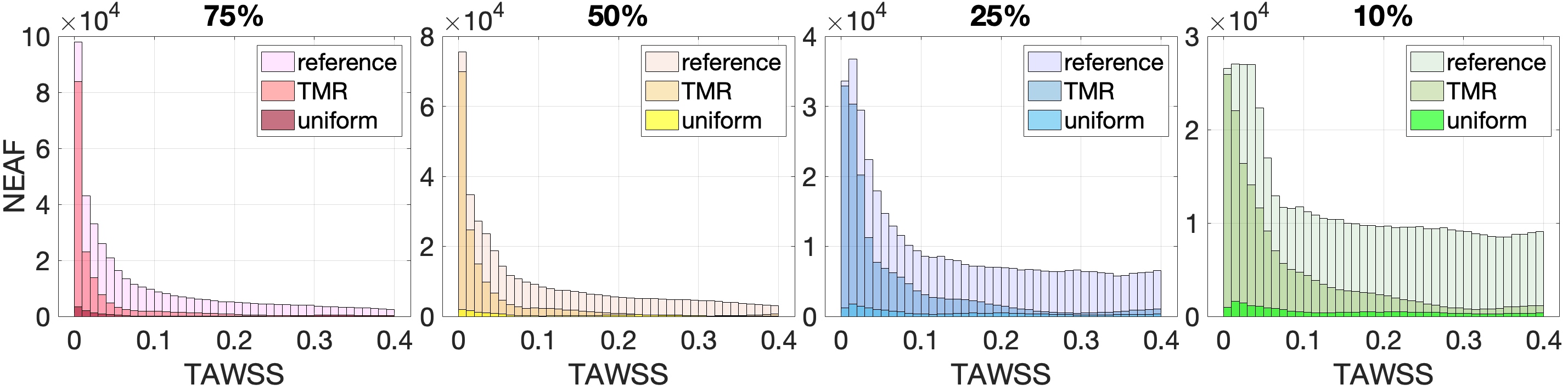}}\label{fig:TAWSS04histInds}}
    \caption{NEAF histograms of TAWSS \textless 0.4 Pa.}
    \label{fig:TAWSS04hist}
\end{figure}

A quantitative analysis is provided by plotting the nodal absolute empirical frequency (NEAF) of TAWSS values smaller than 0.4 Pa in the form of histograms. The NEAF is calculated by counting the number of mesh nodes with TAWSS within a certain range. Figures \ref{fig:TAWSS04histRefs} and \ref{fig:TAWSS04histInds} show the TAWSS histograms grouped by  mesh refinements and indentations respectively. We choose a uniform bin width of 0.01 Pa for a total of 40 bins per histogram. For the reference mesh (see left-most subplot in Fig. \ref{fig:TAWSS04histRefs}), 75\% indentation has the highest NEAF of TAWSS values between 0 and 0.01 Pa and the NEAF of the first bin decreases with indentation percentage. On the other hand, 10\% indentation has the highest NEAF of values between 0.39 and 0.4 Pa. From Figure \ref{fig:TAWSS04histInds}, we observe that for each indentation, NEAFs show a similarly decaying trend from $\tawss = 0$ to $\tawss = 0.4$ Pa in the case of targeted mesh refinement and reference mesh. This pattern is not present for the uniform meshes, where the NEAF variance is much higher between bins (see right-most subplot in Fig. \ref{fig:TAWSS04histRefs}).\\

\begin{figure}
    \centering
    \subfloat[Semilog plot of NEAF histogram data for all mesh refinements: indentation comparison.]{
        \resizebox*{14cm}{!}{\includegraphics[draft=\draftmode]{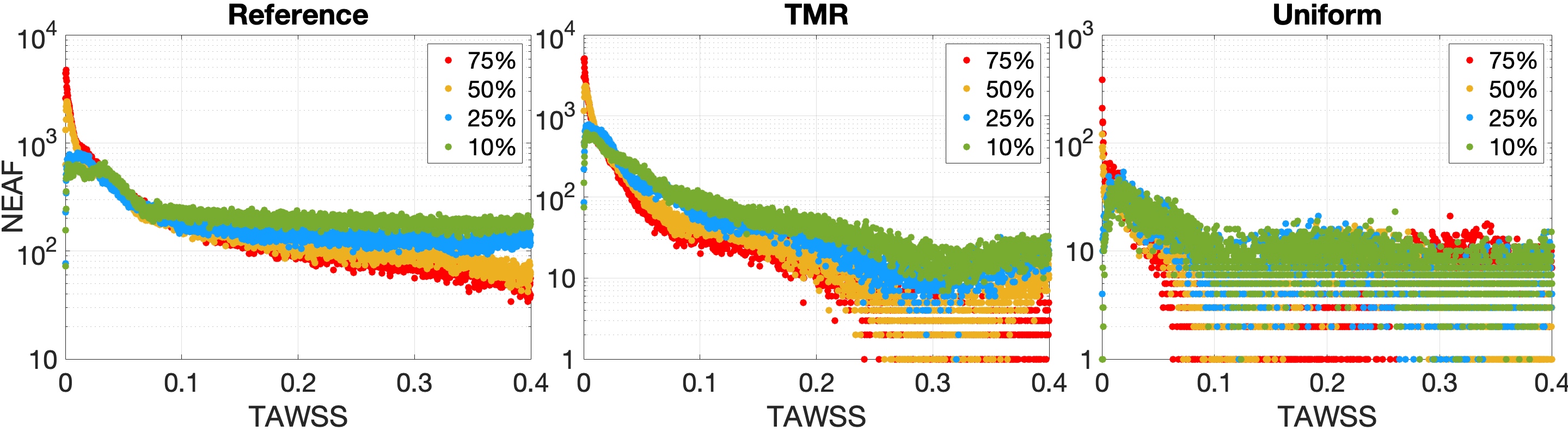}}\label{fig:semilogData}}\\
    \subfloat[Fitted smoothing splines and histogram data of NEAF for all mesh refinements and indentations.]{
        \resizebox*{14cm}{!}{\includegraphics[draft=\draftmode]{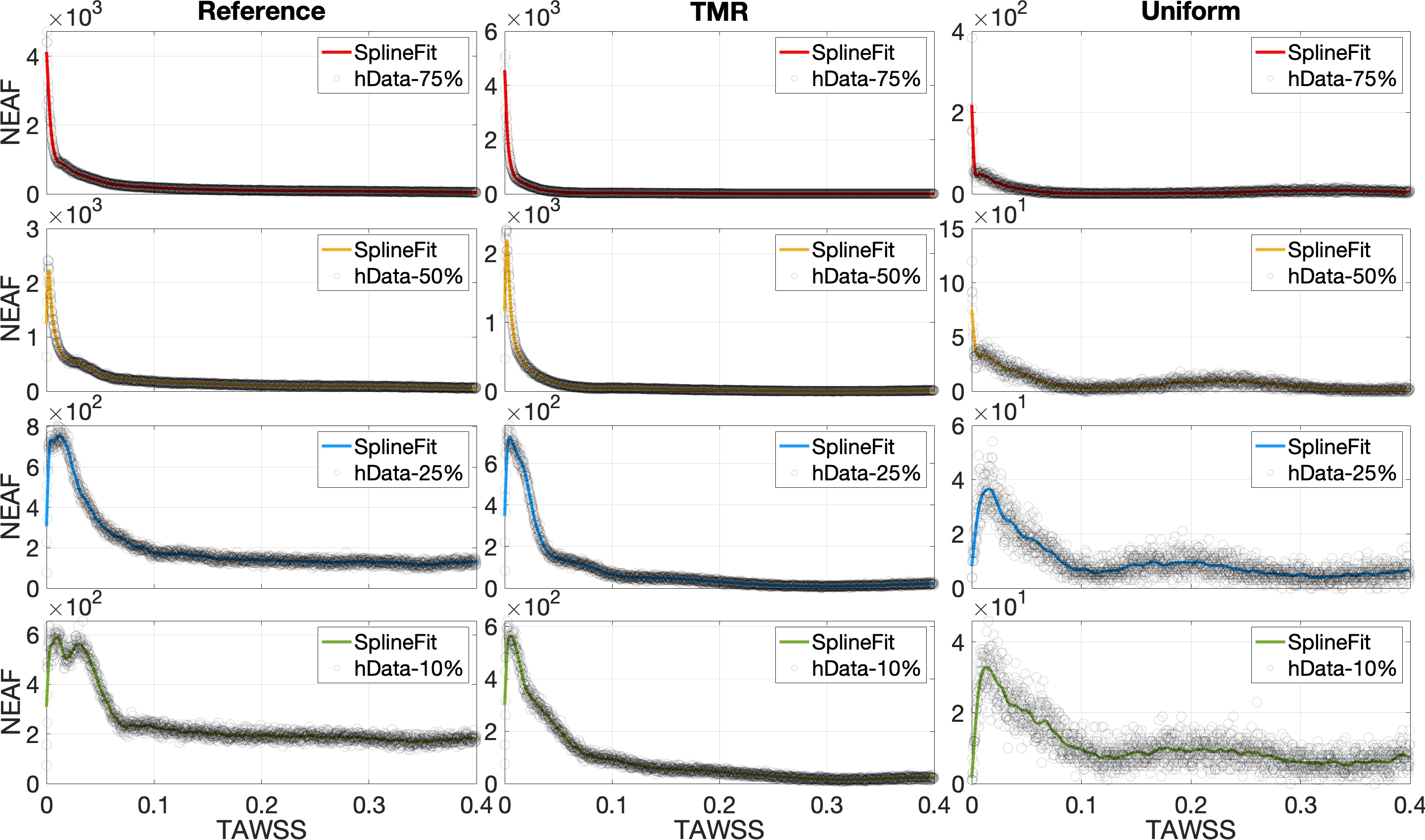}}\label{fig:semilogCurves}}\\
    \subfloat[Semilog plot of NEAF fitted splines for all mesh refinements: indentation comparison.]{
        \resizebox*{14cm}{!}{\includegraphics[draft=\draftmode]{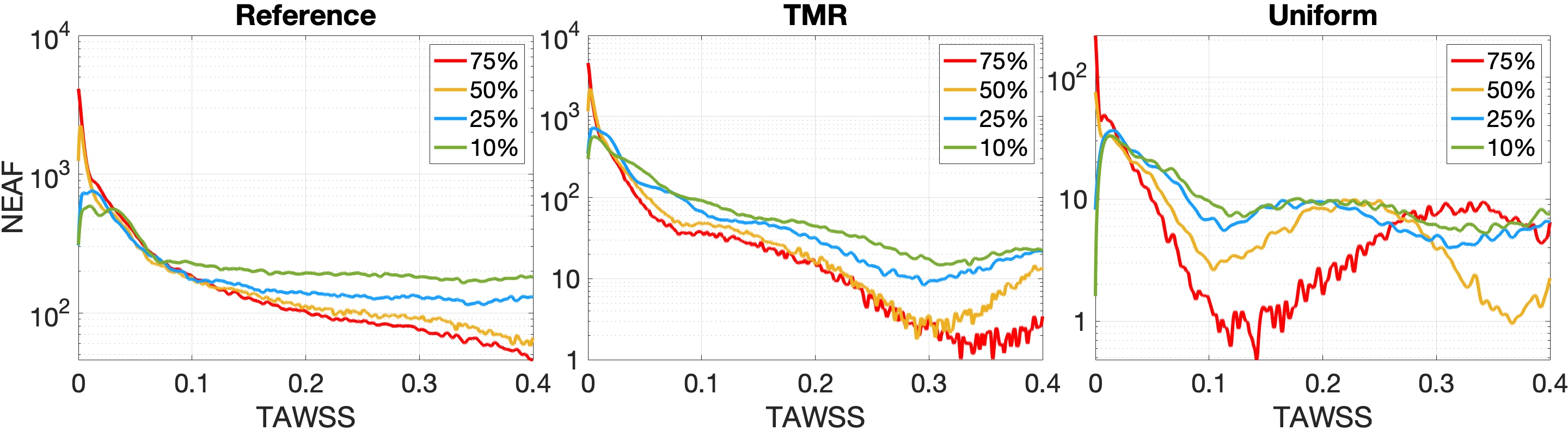}}\label{fig:semilogSplines}}
    \caption{NEAF plots of TAWSS \textless 0.4 Pa.}\label{fig:TAWSS04histAccurate}
\end{figure}

To prevent errors of NEAF estimation, we refine the histogram bin width to 0.0002 Pa and plot the histogram data in Fig. \ref{fig:semilogData}. Looking at the reference mesh (left-most subplot) the highest NEAF is accumulated close to zero for 75\% indentation and decreases (in order) for 50\%, 25\% and 10\% indentation. The NEAF decay is very fast for high indentations and slower for low indentations. This entails that the reference NEAF order is inverted for values close to 0.4 Pa. The highest NEAF in the upper end range is given by 10\% indentation and decreases with increasing indentation, i.e., (in order) 25\%, 50\% and 75\%. Low values of TAWSS have very close NEAF for each indentation in targeted mesh refinement and reference cases, especially for values smaller than 0.25 Pa. NEAF values with uniform meshes and targeted mesh refinement for $\tawss > 0.25$ Pa are hard to evaluate, since the refined bin size introduces a lot of noise.\\

To exclude additional noise added by the small histogram bin size, we interpolate the histogram values with splines \cite{eilers2010splines} using Matlab Curve Fitting Tool \cite{matlabcftool} for visual convenience. We ensure that the goodness of fit is satisfactory, by imposing $R^2 > 0.9$. The curves obtained are compared to the raw histogram data for each indentation and mesh refinement in Fig. \ref{fig:semilogCurves}. Figure \ref{fig:semilogSplines} groups the fitted curves for each mesh refinement. Comparing targeted mesh refinement and reference case, we observe a similar decay for each indentation percentage and nearly identical NEAF for $\tawss = 0$. Furthermore, the same NEAF decaying order is retrieved with targeted mesh refinement for values of TAWSS close to 0.4 Pa. This NEAF pattern is not reproduced with uniform meshes, in particular for high indentations (see right-most subplots in Fig. \ref{fig:semilogSplines}).

\subsection{OSI, RRT and limit values}
\label{3dOSIandRRT}
\begin{figure}
	\centering
    \subfloat[Reference, 75\%.]{
        \resizebox*{6.9cm}{!}{\begin{tikzpicture}
            \node[inner sep=0] (image) {\includegraphics[draft=\draftmode]{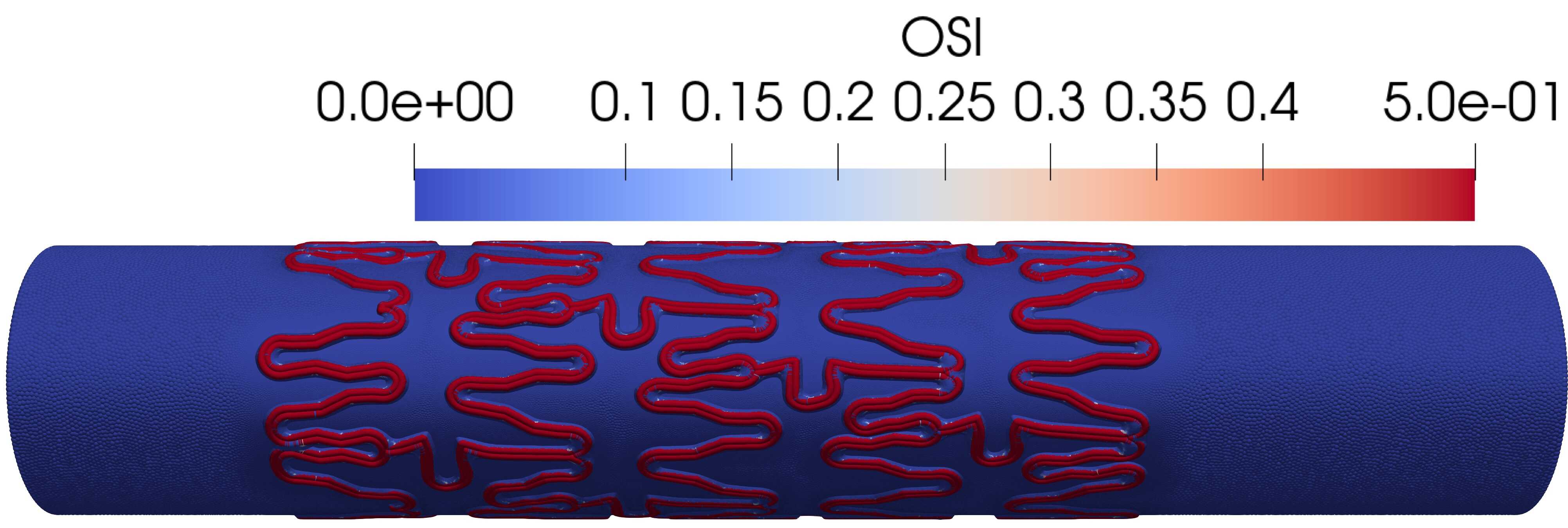}};
            \draw[gray, dashed, dash pattern = on 1cm off 1cm, line width=0.3cm] (-49cm,-23.5cm) rectangle (35cm,2.5cm);
            \end{tikzpicture}}\label{fig:OSI75}
    }
    \subfloat[Reference, 10\%.]{
        \resizebox*{6.9cm}{!}{\begin{tikzpicture}
            \node[inner sep=0] (image) {\includegraphics[draft=\draftmode]{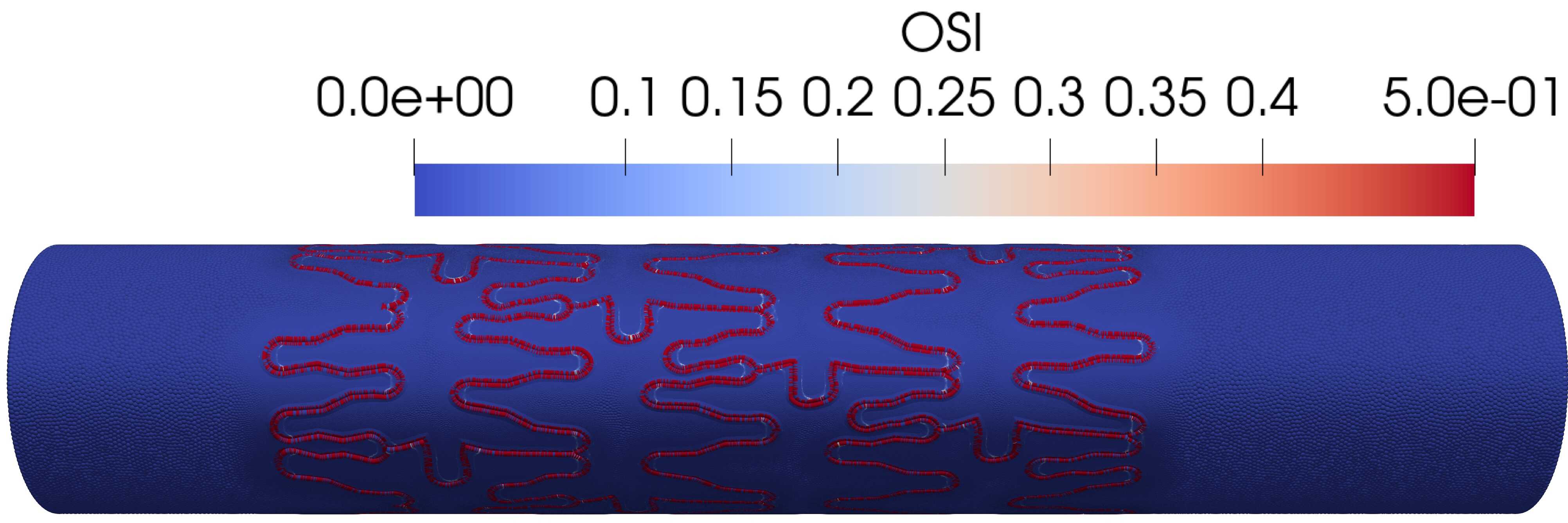}};
            \draw[gray, dashed, dash pattern = on 1cm off 1cm, line width=0.3cm] (-49cm,-23.5cm) rectangle (35cm,2.5cm);
            \end{tikzpicture}}\label{fig:OSI10}
        }\\
    \subfloat[Zoom on \ref{fig:OSI75}.]{
        \resizebox*{4.4cm}{!}{\begin{tikzpicture}
            \node[inner sep=0] (image){\includegraphics[draft=\draftmode,trim={22cm 0 36cm 0},clip=true]{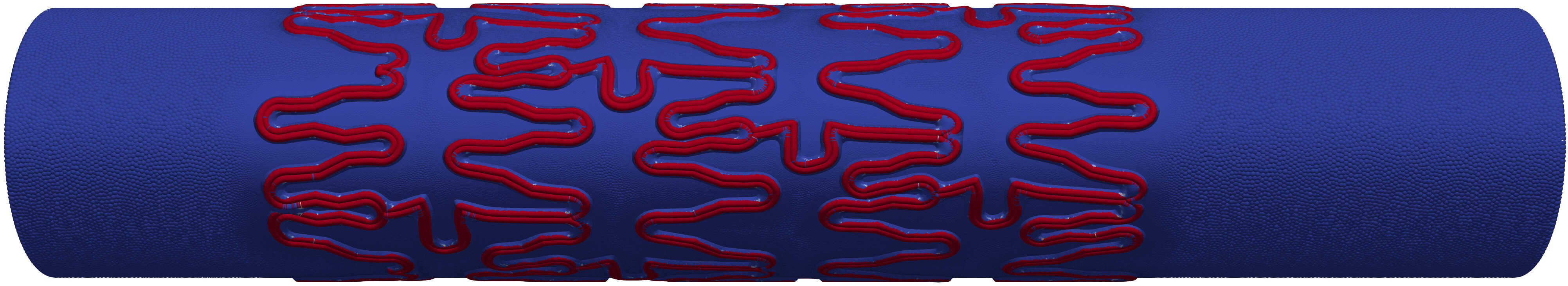}};
            \draw[gray, dashed, dash pattern = on 2cm off 2cm, line width=0.5cm] (image.south west) rectangle (image.north east);
        \end{tikzpicture}}\label{fig:OSI75Zoom}
    }
    \subfloat[TMR, 75\%.]{
        \resizebox*{4.4cm}{!}{\includegraphics[draft=\draftmode,trim={22cm 0 36cm 0},clip=true]{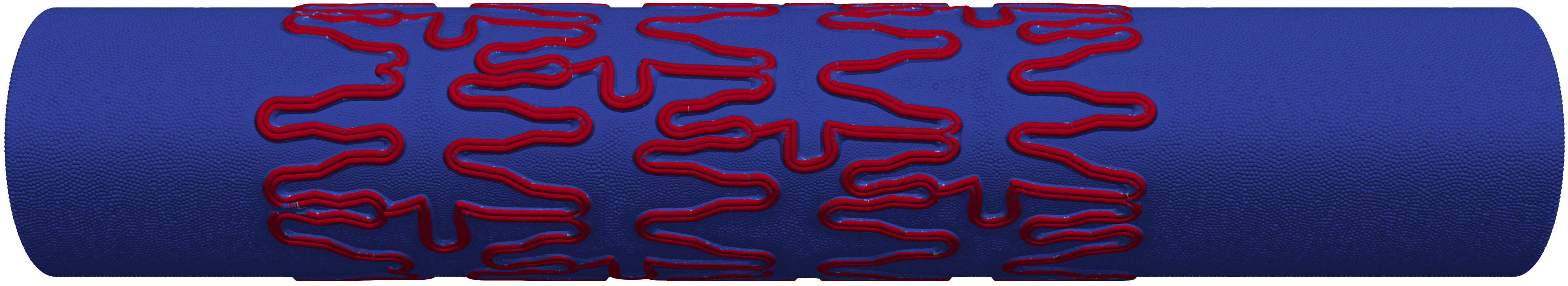}}\label{fig:OSI75ZoomTMR}
        }
    \subfloat[Uniform, 75\%.]{
        \resizebox*{4.4cm}{!}{\includegraphics[draft=\draftmode,trim={22cm 0 36cm 0},clip=true]{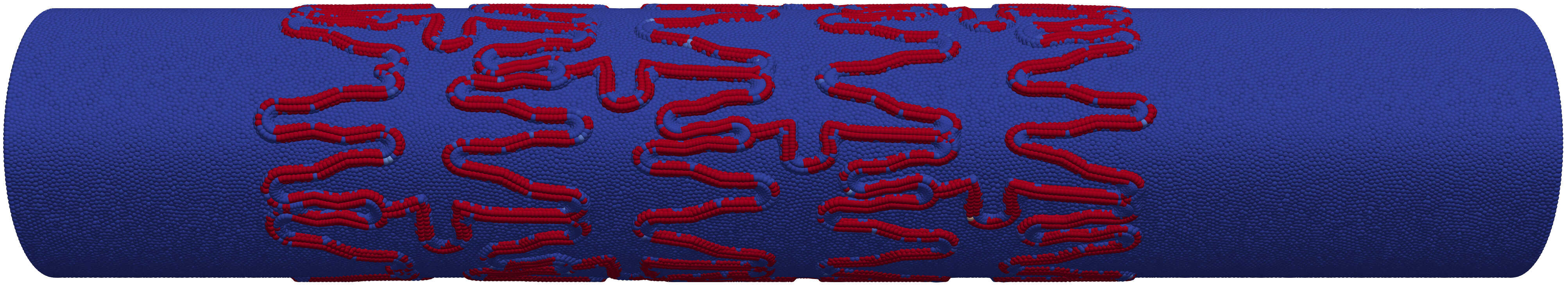}}\label{fig:OSI75ZoomUni}
        }\\
     \subfloat[Zoom on \ref{fig:OSI10}.]{
        \resizebox*{4.4cm}{!}{\begin{tikzpicture}
            \node[inner sep=0] (image){\includegraphics[draft=\draftmode,trim={22cm 0 36cm 0},clip=true]{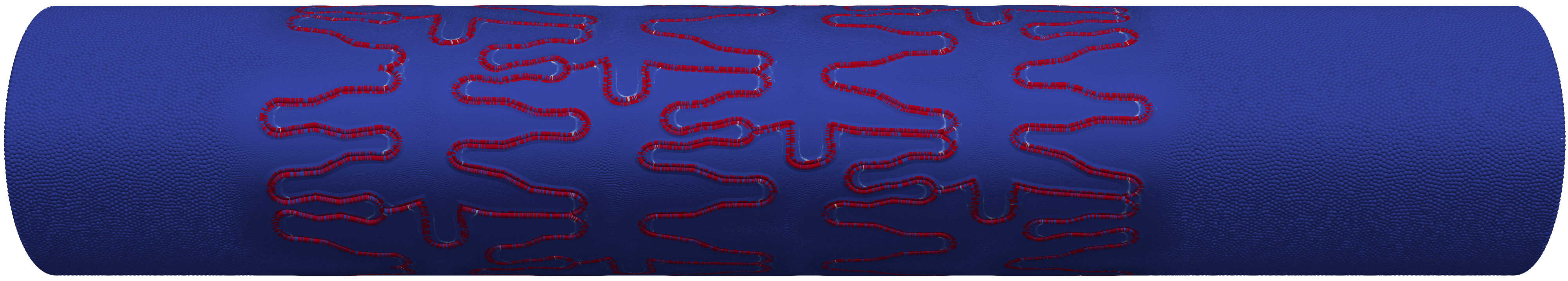}};
            \draw[gray, dashed, dash pattern = on 2cm off 2cm, line width=0.5cm] (image.south west) rectangle (image.north east);
        \end{tikzpicture}}\label{fig:OSI10Zoom}
        }
     \subfloat[TMR, 10\%.]{
        \resizebox*{4.4cm}{!}{\includegraphics[draft=\draftmode,trim={22cm 0 36cm 0},clip=true]{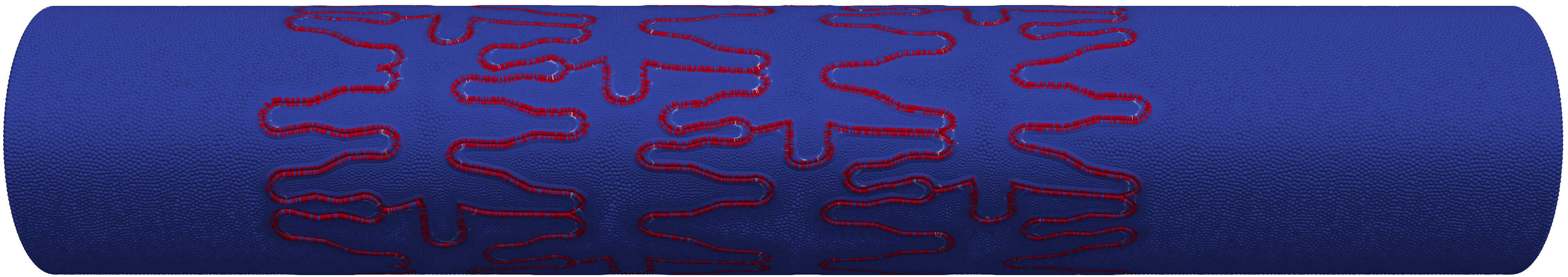}}\label{fig:OSI10ZoomTMR}
    }
    \subfloat[Uniform, 10\%.]{
        \resizebox*{4.4cm}{!}{\includegraphics[draft=\draftmode,trim={22cm 0 36cm 0},clip=true]{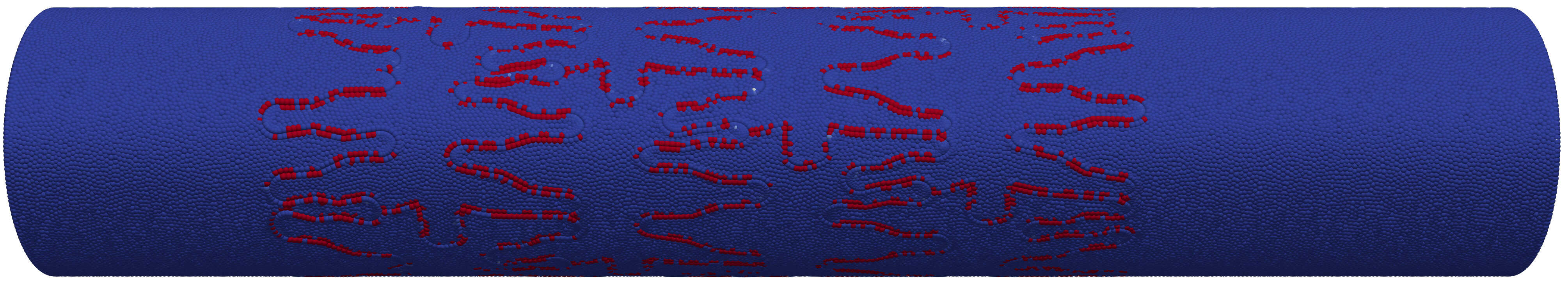}}\label{fig:OSI10ZoomUni}
        }
	\caption{OSI for all mesh refinements and two indentation percentages.}\label{fig:OSI75-10}
\end{figure}
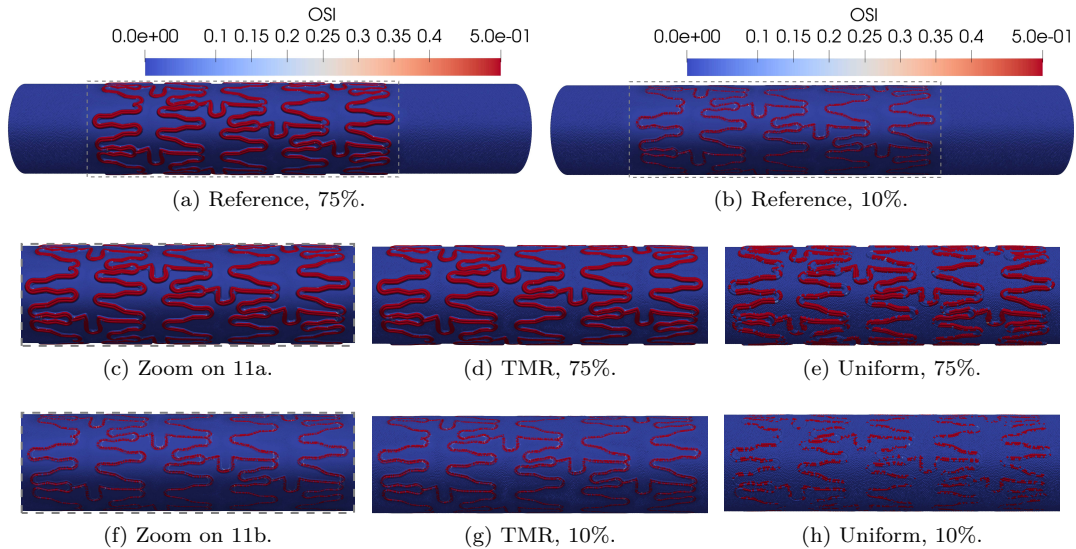
Effects of indentation on OSI and RRT are analyzed in this section.\\

\begin{figure}
    \centering
    \resizebox*{14cm}{!}{\includegraphics[draft=\draftmode]{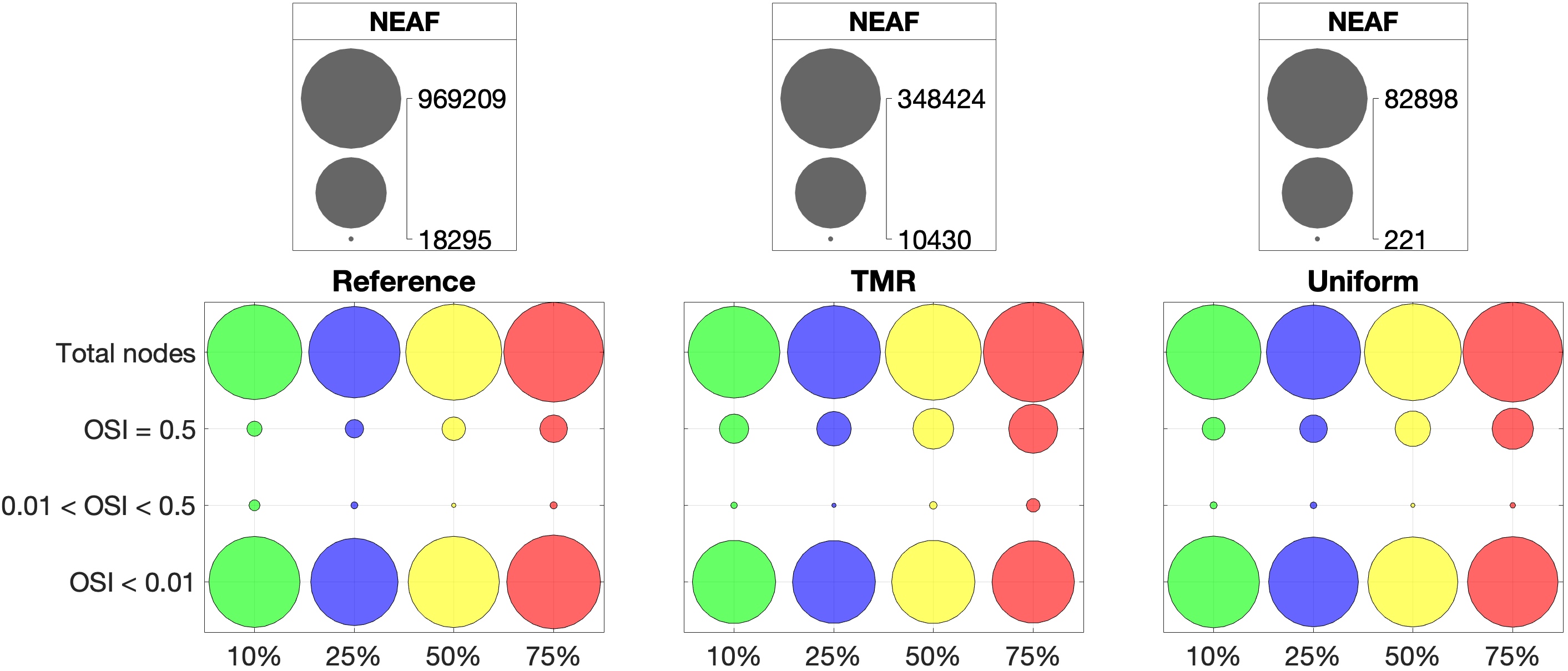}}
    \caption{NEAF bubble chart of OSI for all mesh refinements and indentations.}
    \label{fig:OSIBubble}
\end{figure}

\begin{figure}
    \centering
    \resizebox*{7.3cm}{!}{\includegraphics[draft=\draftmode]{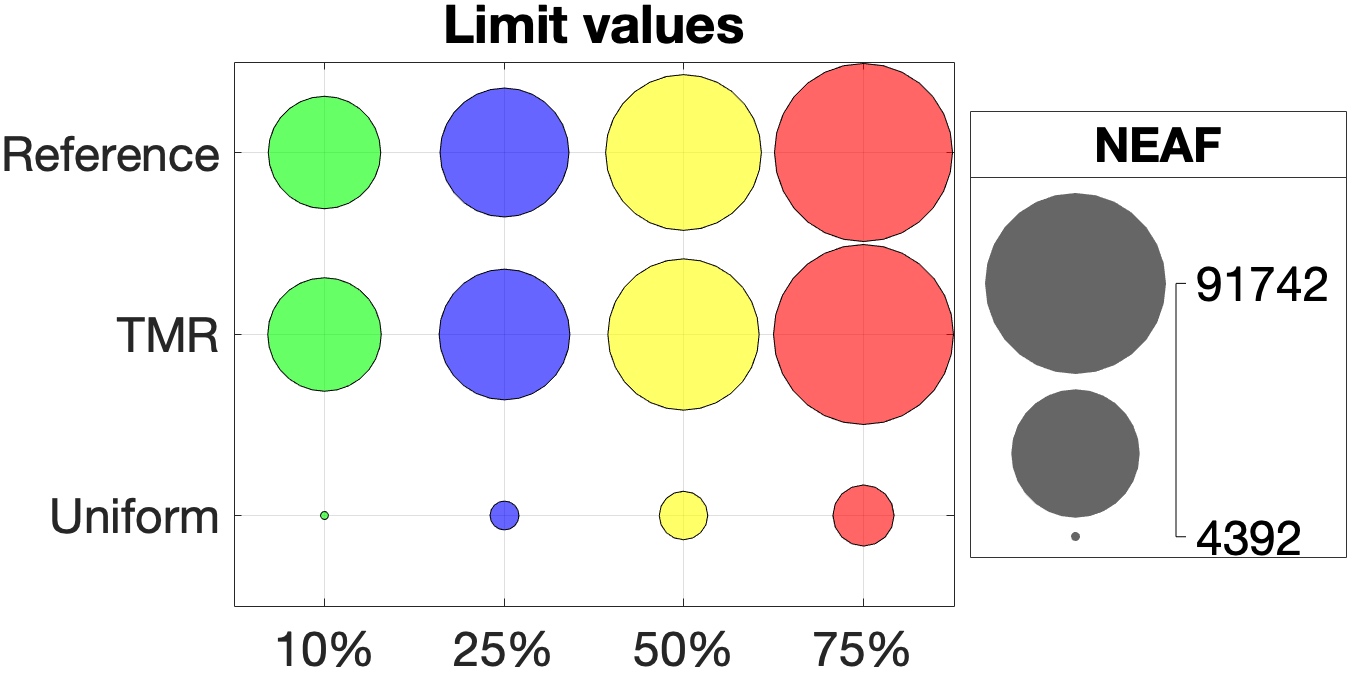}}
    \caption{NEAF bubble chart of hemodynamic indicators at limit values.}\label{fig:LimitBubble}
\end{figure}

Figure \ref{fig:OSI75Zoom} shows that 75\% indentation results in increased areas with OSI = 0.5, i.e., more stagnation around the stent struts, compared to 10\% indentation (see Fig. \ref{fig:OSI10Zoom}). The targeted mesh refinement is able to accurately capture the critical areas of OSI = 0.5 for both indentation percentages, as shown in Figures \ref{fig:OSI75ZoomTMR} and \ref{fig:OSI10ZoomTMR}. The uniform meshes underestimate the OSI value in several areas around the stent (Figures \ref{fig:OSI75ZoomUni} and \ref{fig:OSI10ZoomUni}). Most values of OSI are either very close to 0 or very close to 0.5. This pattern is detected for all indentation and refinements in Fig. \ref{fig:OSIBubble}. The bubble plot shows the NEAF for OSI $<$ 0.01, 0.01 $<$ OSI $<$ 0.5 and OSI = 0.5. The comparison between the total number of nodes and the NEAF for OSI in different value ranges show that the majority of mesh nodes detect $\osi < 0.01$. Only a small fraction of total nodes show limit values of OSI, which we compare in more detail in Fig. \ref{fig:LimitBubble}.\\

\begin{figure}
	\centering
    \subfloat[Reference, 75\%.]{
         \resizebox*{6cm}{!}{
         \begin{tikzpicture}[]
            \node[inner sep=0] (image) {\includegraphics[draft=\draftmode]{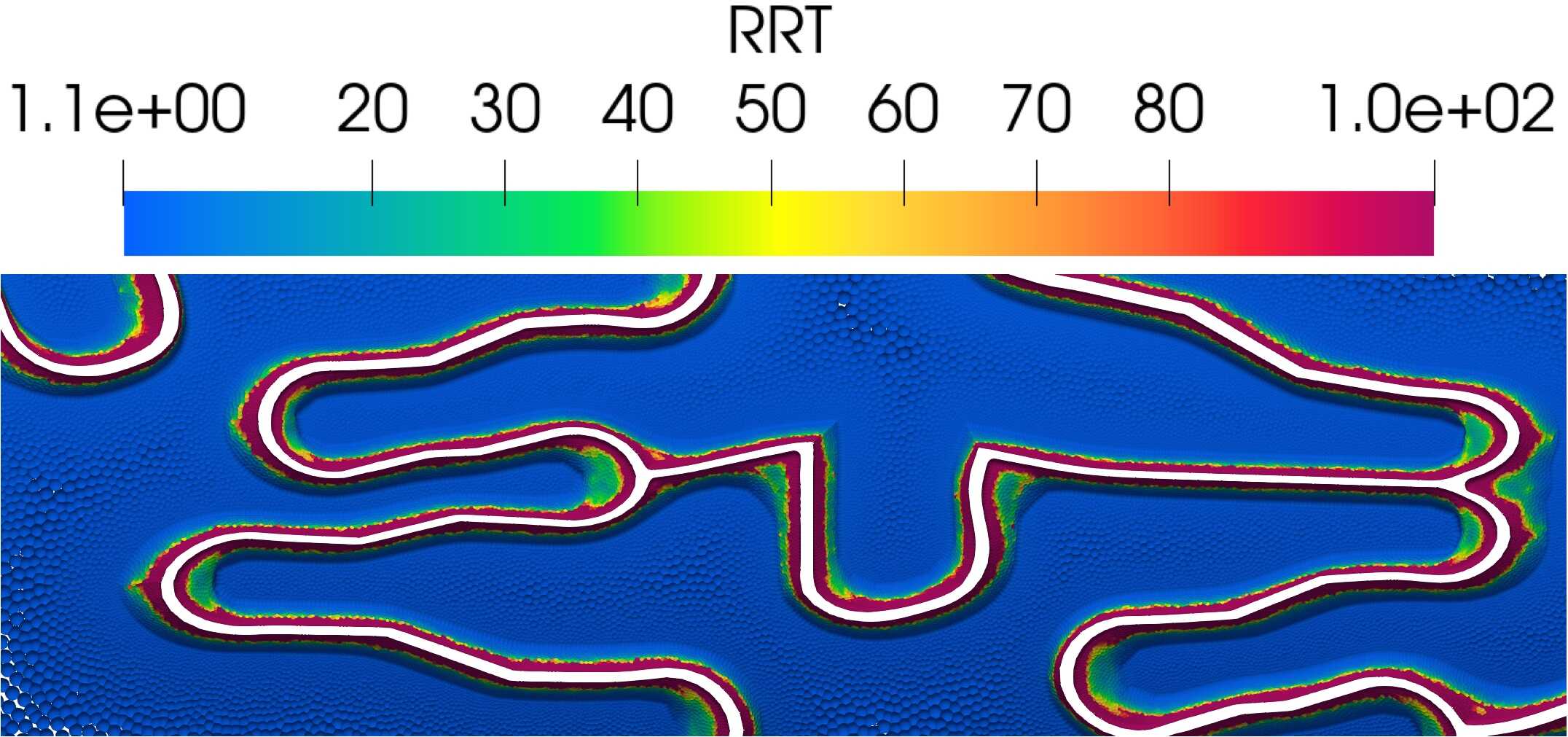}};
            \draw[yellow, dashed, dash pattern = on 2cm off 2cm, line width = 1 cm] (image.south west) rectangle (37.97cm,4.5cm);
            \draw[red, dashed, dash pattern = on 1cm off 1cm, line width = 0.3 cm] (-12cm,-13.5cm) rectangle (13cm,-2cm);
            \end{tikzpicture}}\label{fig:RRT75}}
    \subfloat[Reference, 10\%.]{
        \resizebox*{6cm}{!}{
         \begin{tikzpicture}[]
            \node[inner sep=0] (image) {\includegraphics[draft=\draftmode]{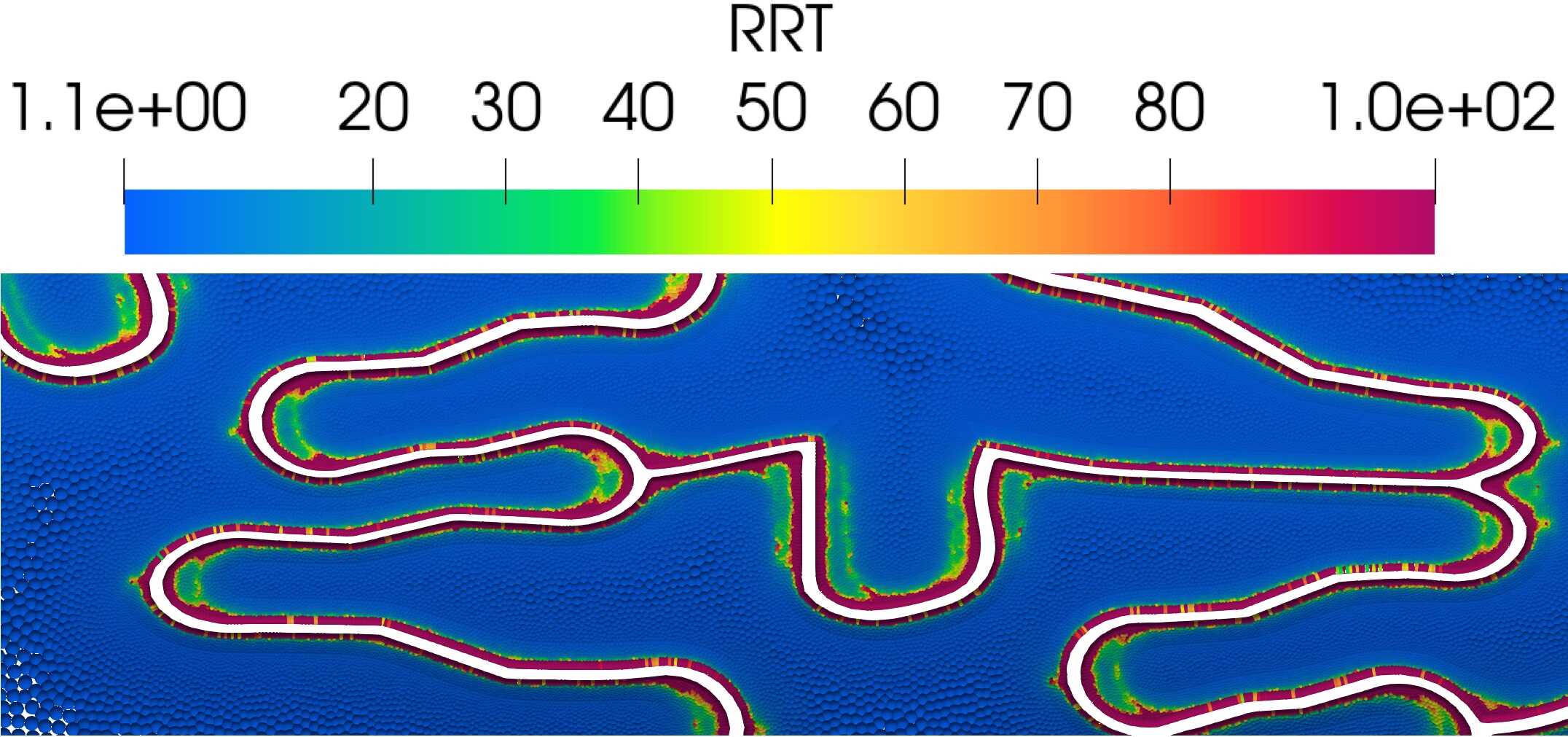}};
            \draw[yellow, dashed, dash pattern = on 2cm off 2cm, line width = 1 cm] (image.south west) rectangle (37.97cm,4.5cm);
            \draw[red, dashed, dash pattern = on 1cm off 1cm, line width = 0.3 cm] (-12cm,-13.5cm) rectangle (13cm,-2cm);
            \end{tikzpicture}}\label{fig:RRT10}}\\
    \subfloat[Zoom on \ref{fig:RRT75}.]{
            \resizebox*{4.5cm}{!}{
            \begin{tikzpicture}[baseline={($ (current bounding box.north) - (0,5pt) $)}]
                \node[inner sep=0] (image){\scalebox{1}[-1]{\includegraphics[draft=\draftmode,trim={25cm 6cm 25cm 4cm},clip=true]{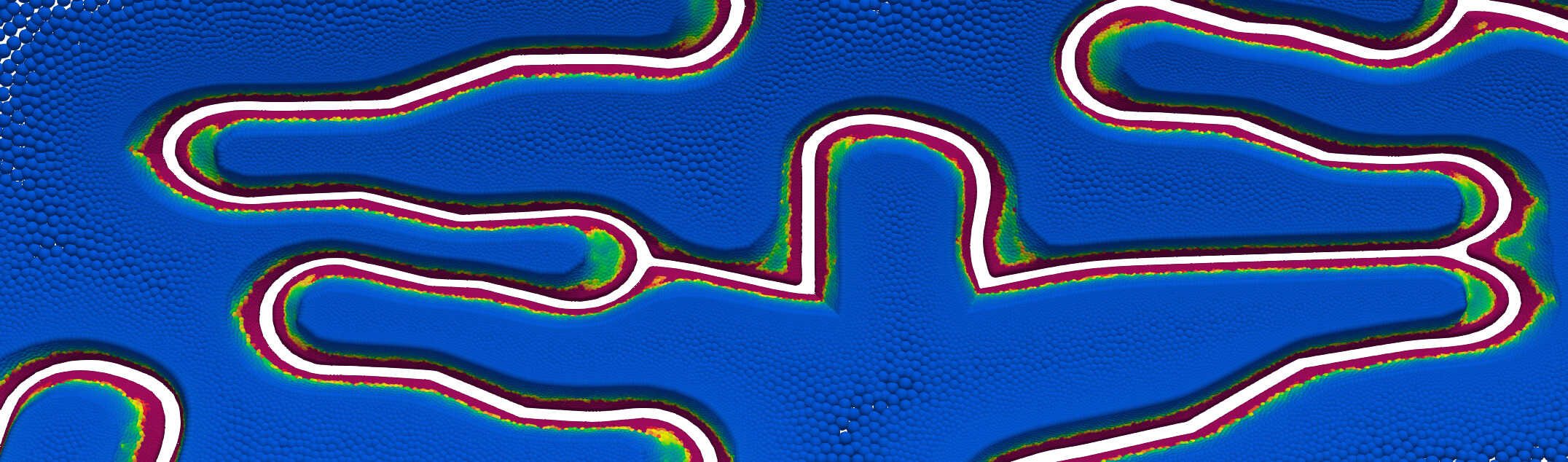}}};
                \draw[red, dashed, dash pattern = on 1cm off 1cm, line width = 0.3cm] (image.south west) rectangle (image.north east);
            \end{tikzpicture}}\label{fig:RRT75Zoom}
            }
    \subfloat[TMR, 75\%.]{
         \resizebox*{4.5cm}{!}{\scalebox{1}[-1]{\includegraphics[draft=\draftmode,trim={25cm 6cm 25cm 4cm},clip=true]{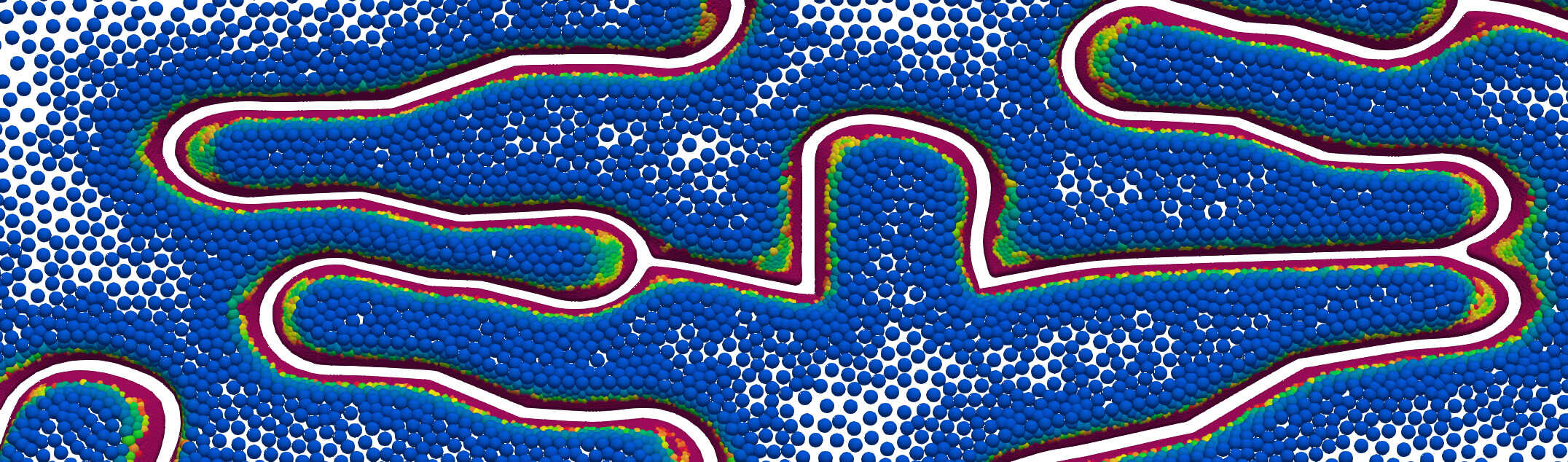}}}\label{fig:RRT75ZoomTMR}
         }
    \subfloat[Uniform, 75\%.]{
         \resizebox*{4.5cm}{!}{\scalebox{1}[-1]{\includegraphics[draft=\draftmode,trim={25cm 6cm 25cm 4cm},clip=true]{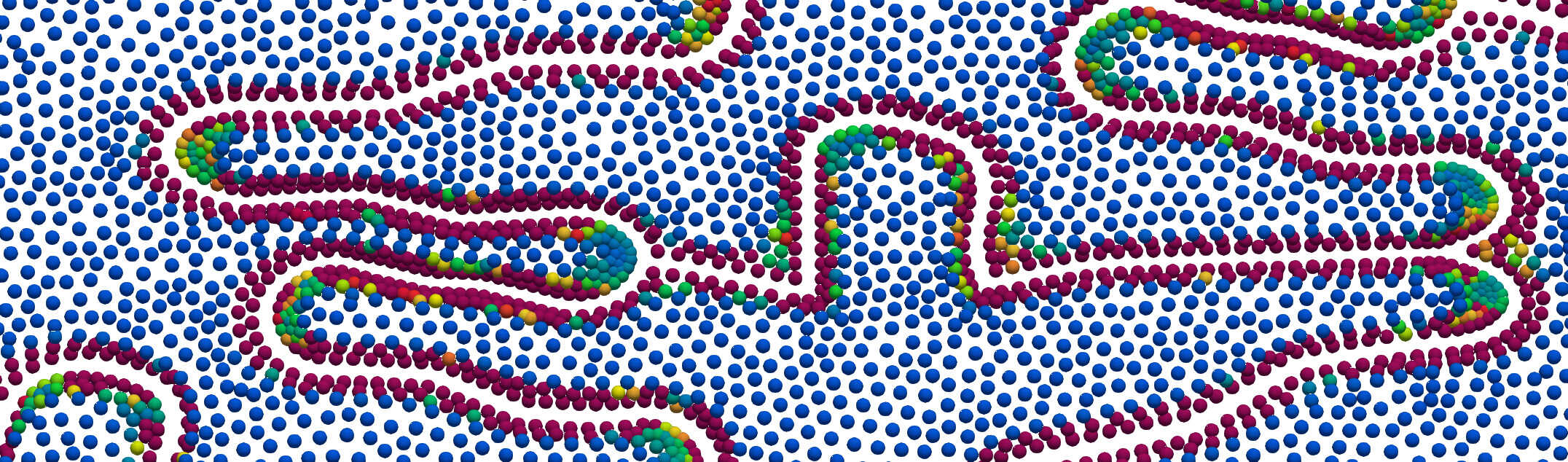}}}\label{fig:RRT75ZoomUni}
         }\\
    \subfloat[Zoom on \ref{fig:RRT10}.]{
        \resizebox*{4.5cm}{!}{\begin{tikzpicture}[baseline={($ (current bounding box.north) - (0,5pt) $)}]
                \node[inner sep=0] (image){\scalebox{1}[-1]{\includegraphics[draft=\draftmode,trim={25cm 6cm 25cm 4cm},clip=true]{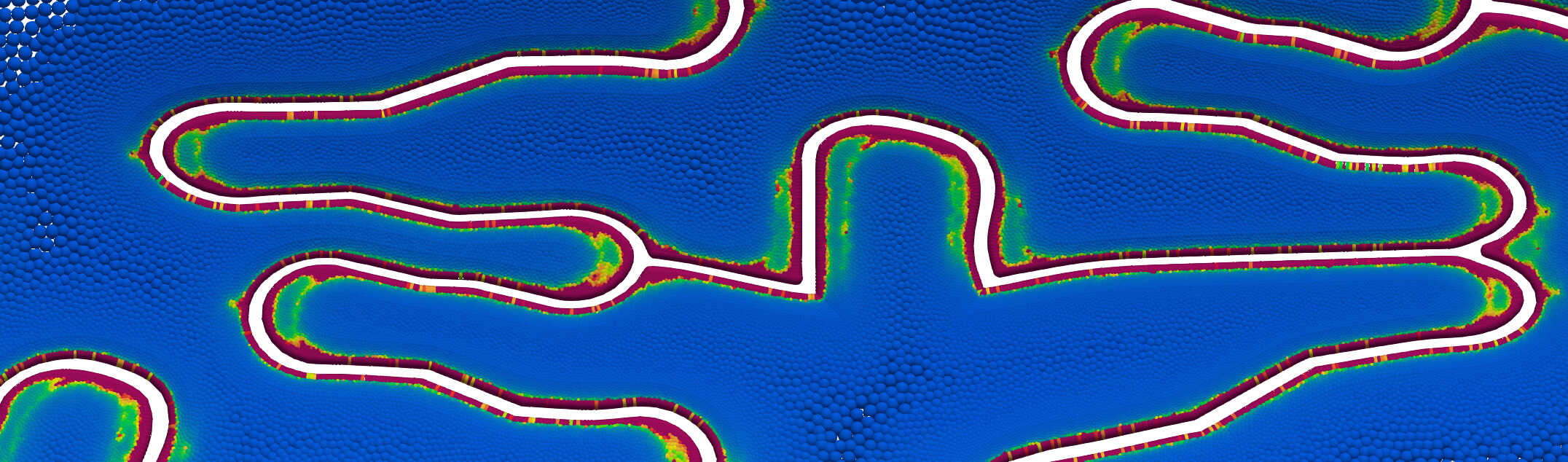}}};
                \draw[red, dashed, dash pattern = on 1cm off 1cm, line width = 0.3cm] (image.south west) rectangle (image.north east);
            \end{tikzpicture}}\label{fig:RRT10Zoom}
            }
    \subfloat[TMR, 10\%.]{
         \resizebox*{4.5cm}{!}{\scalebox{1}[-1]{\includegraphics[draft=\draftmode,trim={25cm 6cm 25cm 4cm},clip=true]{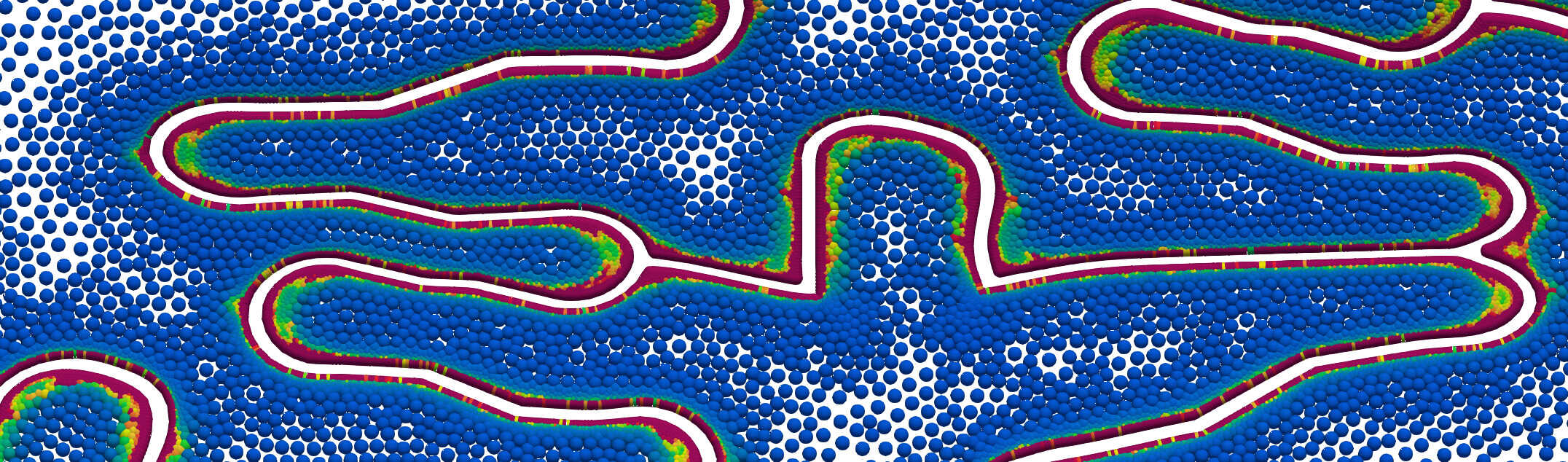}}}\label{fig:RRT10ZoomTMR}
         }
    \subfloat[Uniform, 10\%.]{
         \resizebox*{4.5cm}{!}{\scalebox{1}[-1]{\includegraphics[draft=\draftmode,trim={25cm 6cm 25cm 4cm},clip=true]{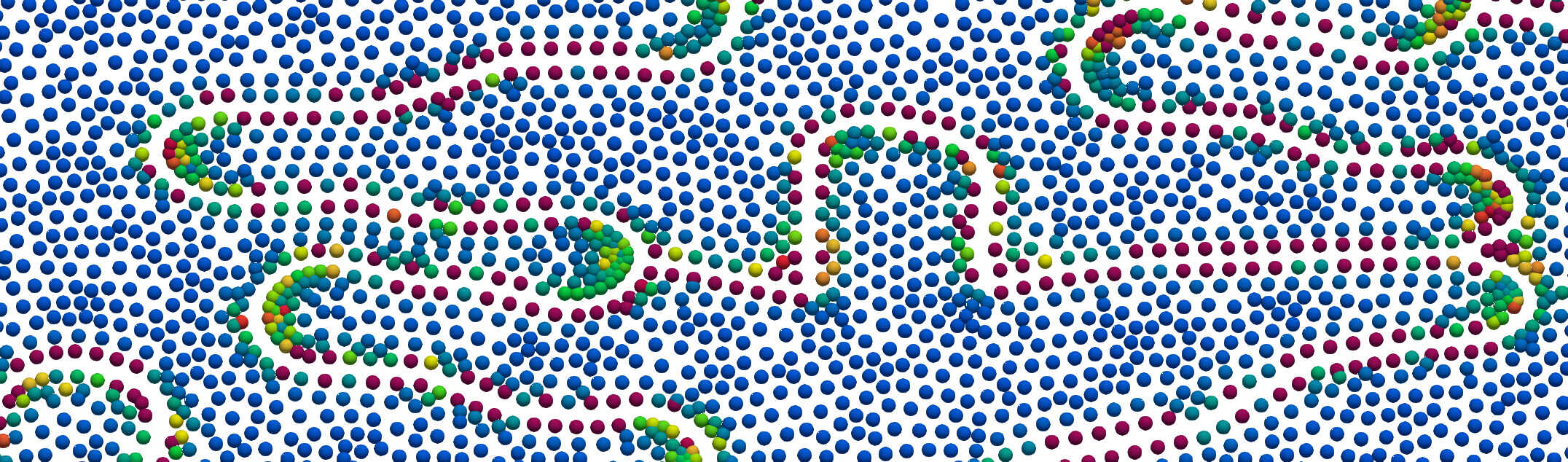}}}\label{fig:RRT10ZoomUni}
         }
	\caption{RRT for all mesh refinements and two indentations. Details of yellow dashed zoombox in Fig. \ref{fig:TAWSSRef25Zoom}.}
    \label{fig:RRT75-10CSZoom}
\end{figure}
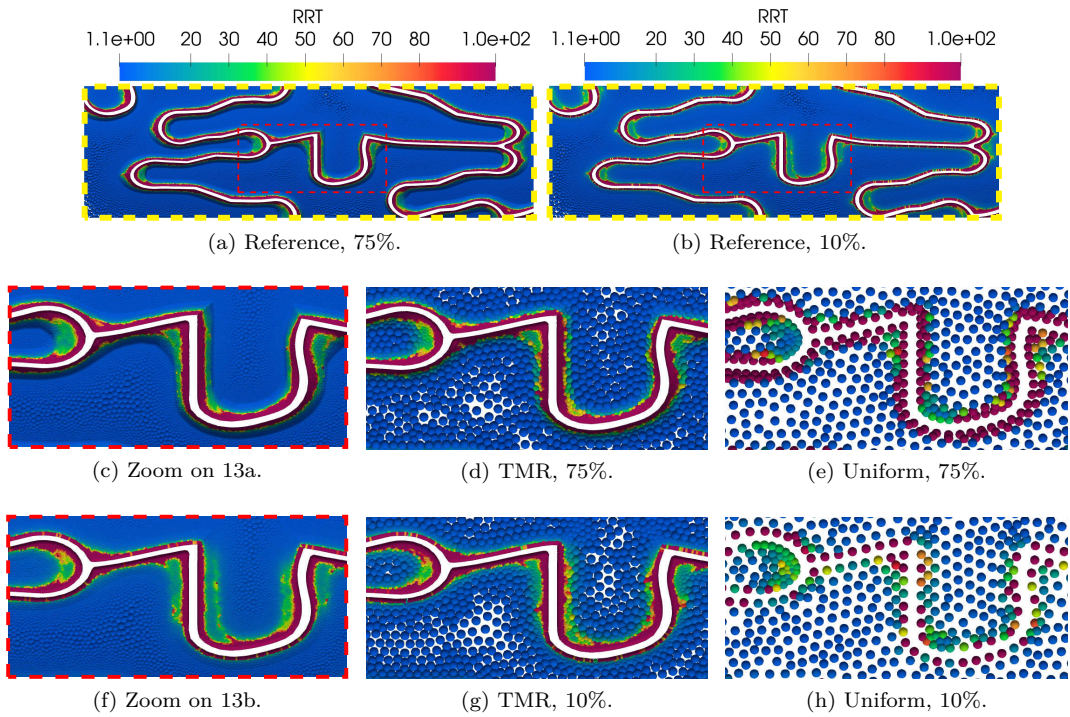

The physiological healthy RRT value is approximately 1.3 $\frac{1}{Pa}$ \cite{manjunatha2024silico}. Higher values entail both low WSS and recirculation. For the reference cases in Figures \ref{fig:RRT75} and \ref{fig:RRT75Zoom}, we observe that 75\% indentation have larger areas of critically high RRT values, while 10\% indentation shows a more spread-out distribution of RRT (Figures \ref{fig:RRT10} and \ref{fig:RRT10Zoom}). For both indentations, reference physiological values of RRT are observed far away from the stent. Figures \ref{fig:RRT75ZoomTMR} and \ref{fig:RRT10ZoomTMR} highlight how targeted mesh refinement is able to correctly capture the same RRT values very close to the stent strut and the complex microdynamics in the sorrounding area. In the case of uniform meshes (Figures \ref{fig:RRT75ZoomUni} and \ref{fig:RRT10ZoomUni}) RRT values are underestimated, especially in the stent proximity.

\begin{figure}
    \centering
    \resizebox*{14cm}{!}{\includegraphics[draft=\draftmode]{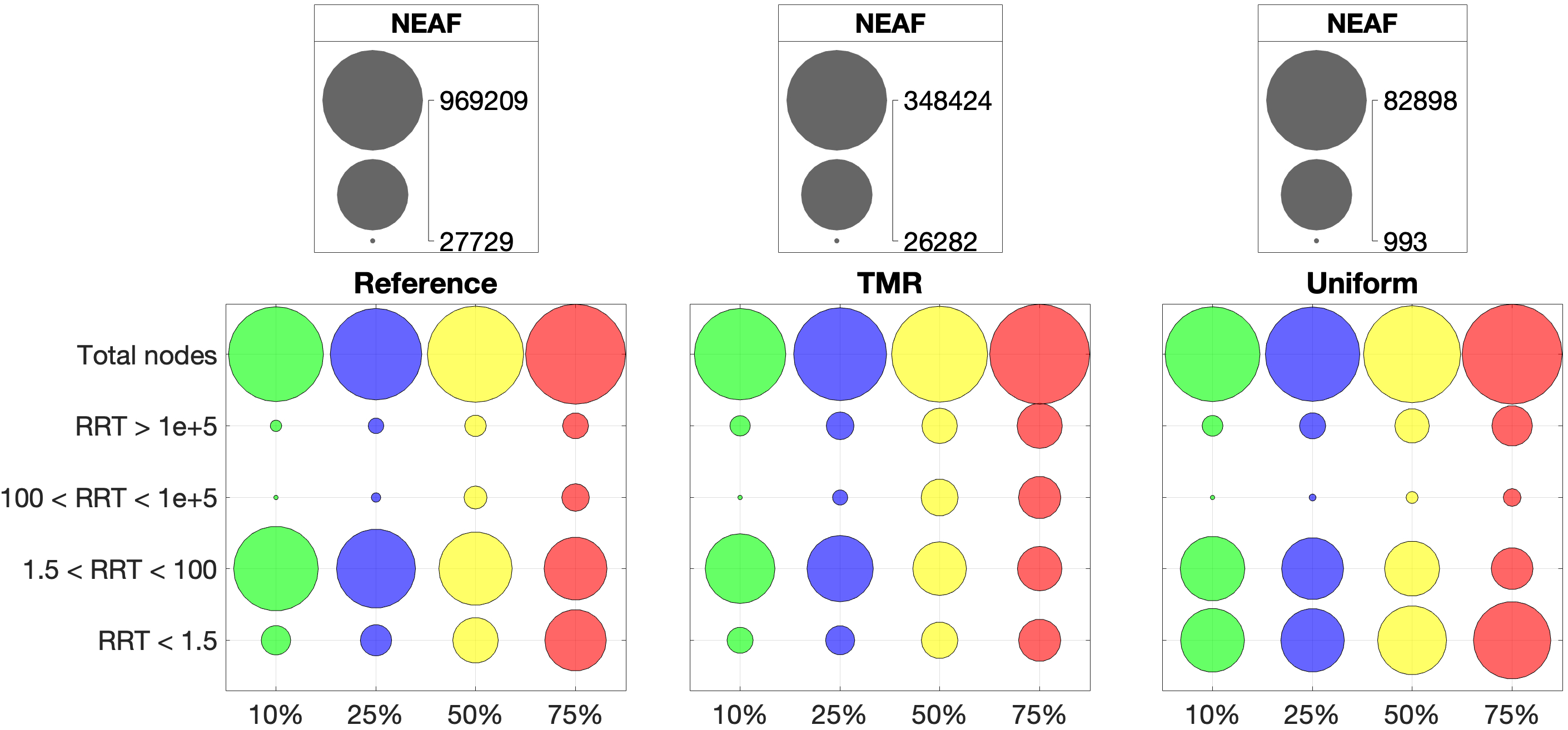}}
    \caption{NEAF bubble chart of RRT for all mesh refinements and indentations.}
    \label{fig:RRTBubble}
\end{figure}

Figure \ref{fig:RRTBubble} shows the NEAF for different RRT ranges. Most RRT values are classified between 0 and 100 $\frac{1}{Pa}$. Values of RRT between 1.5 and 100 $\frac{1}{Pa}$ represent the areas of low TAWSS between 0.1 and 0.4 Pa and mild recirculation. For each indentation, we notice that the reference NEAF of 1.5 $<$ RRT $<$ 100 decreases for increasing indentation percentage, and that NEAF for $\rrt >$ 100 increases consistently from 10\% to 75\% indentation. This pattern is also observed for all indentations with targeted mesh refinement, but the uniform meshes seem to miss most of the microdynamics in the range 100 $<$ RRT $<$ 1e+5 compared to the targeted mesh refinement and reference cases.\\

\begin{figure}
    \centering
    \subfloat[Reference, 75\%.]{
    \resizebox*{4.4cm}{!}{\includegraphics[draft=\draftmode]{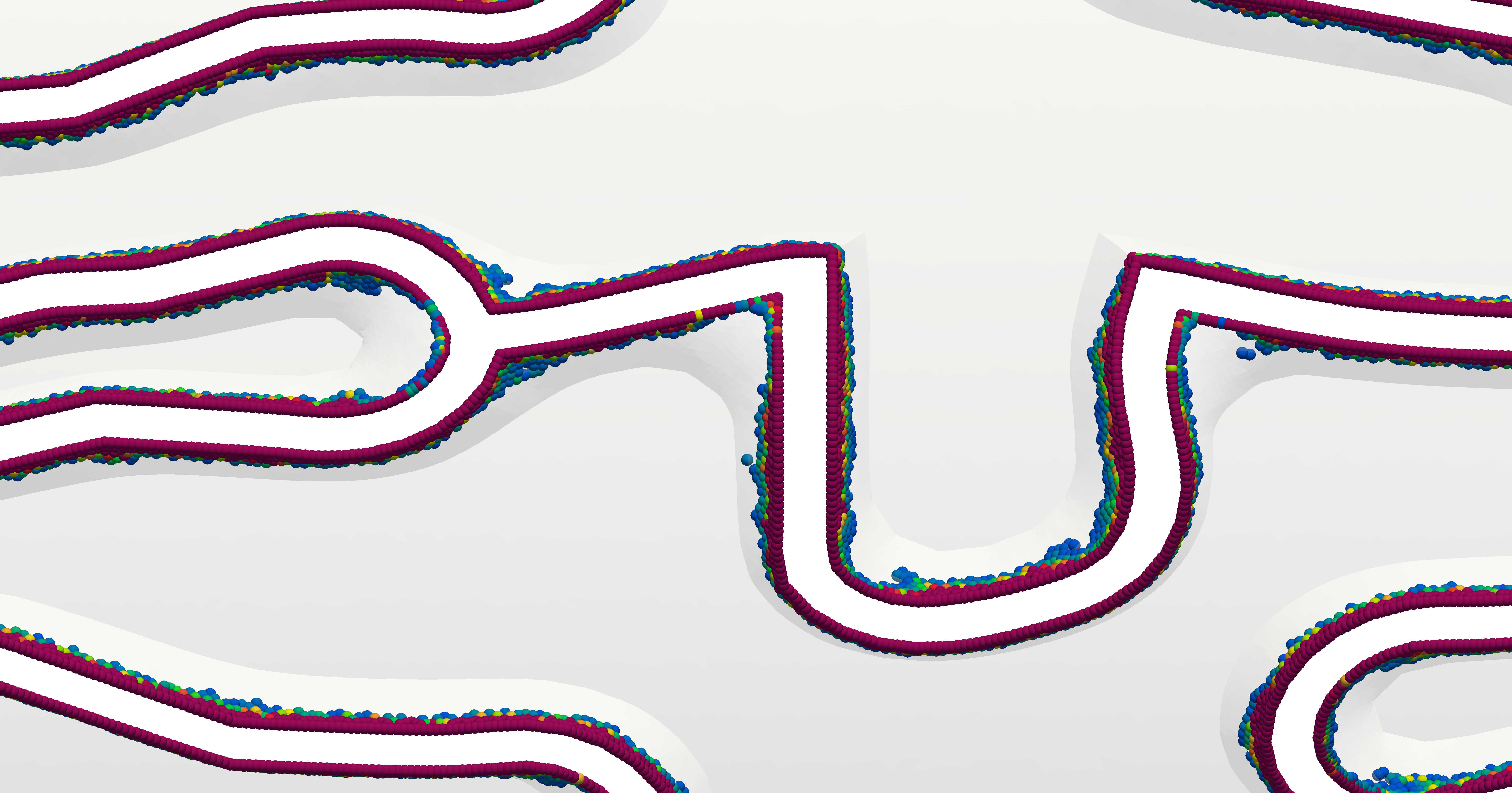}}
    \label{fig:LimitValues75}}
    \subfloat[TMR, 75\%.]{
    \resizebox*{4cm}{!}{\includegraphics[draft=\draftmode]{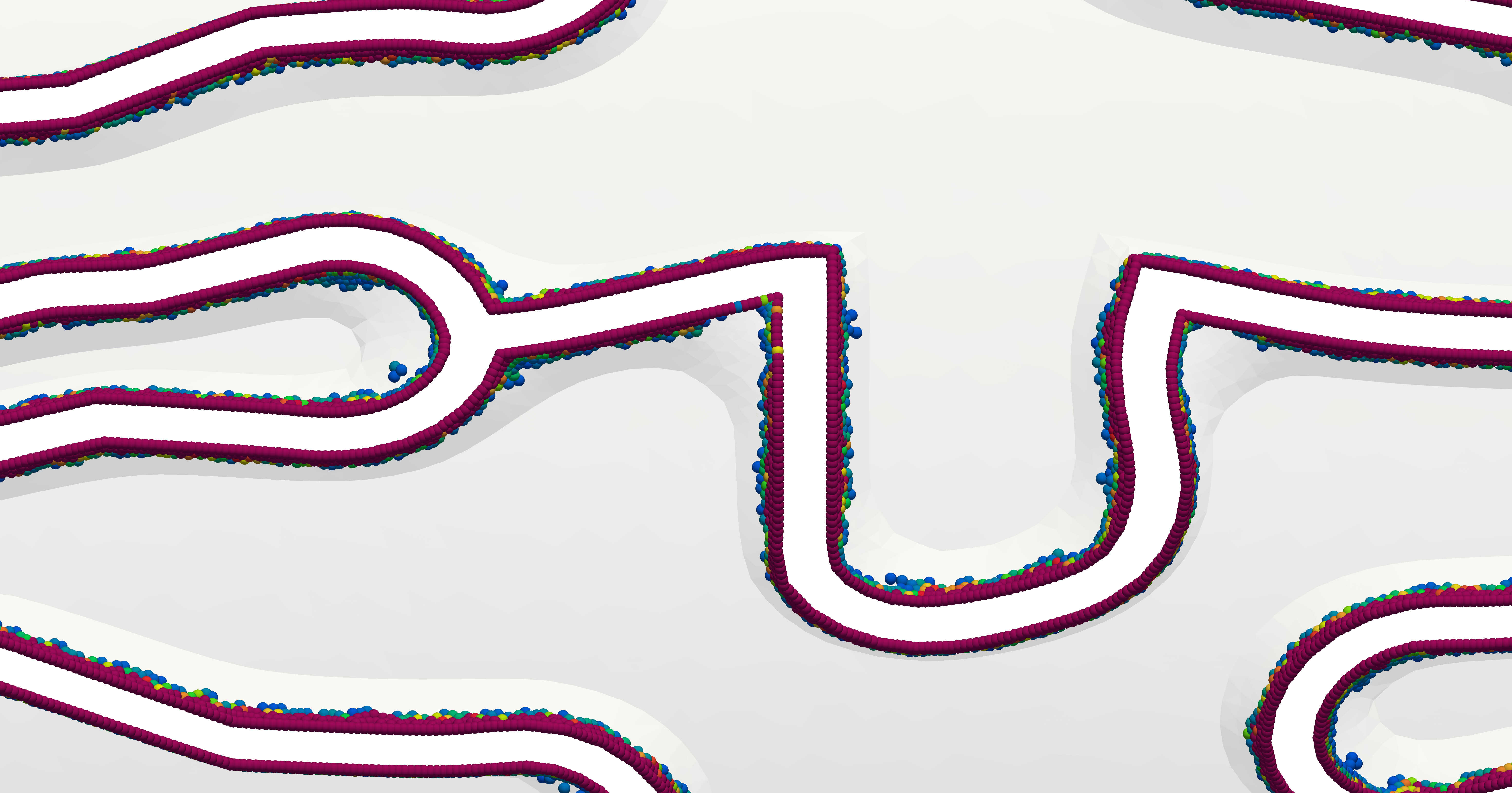}}
    }
    \subfloat[Uniform, 75\%.]{
    \resizebox*{4.4cm}{!}{\includegraphics[draft=\draftmode]{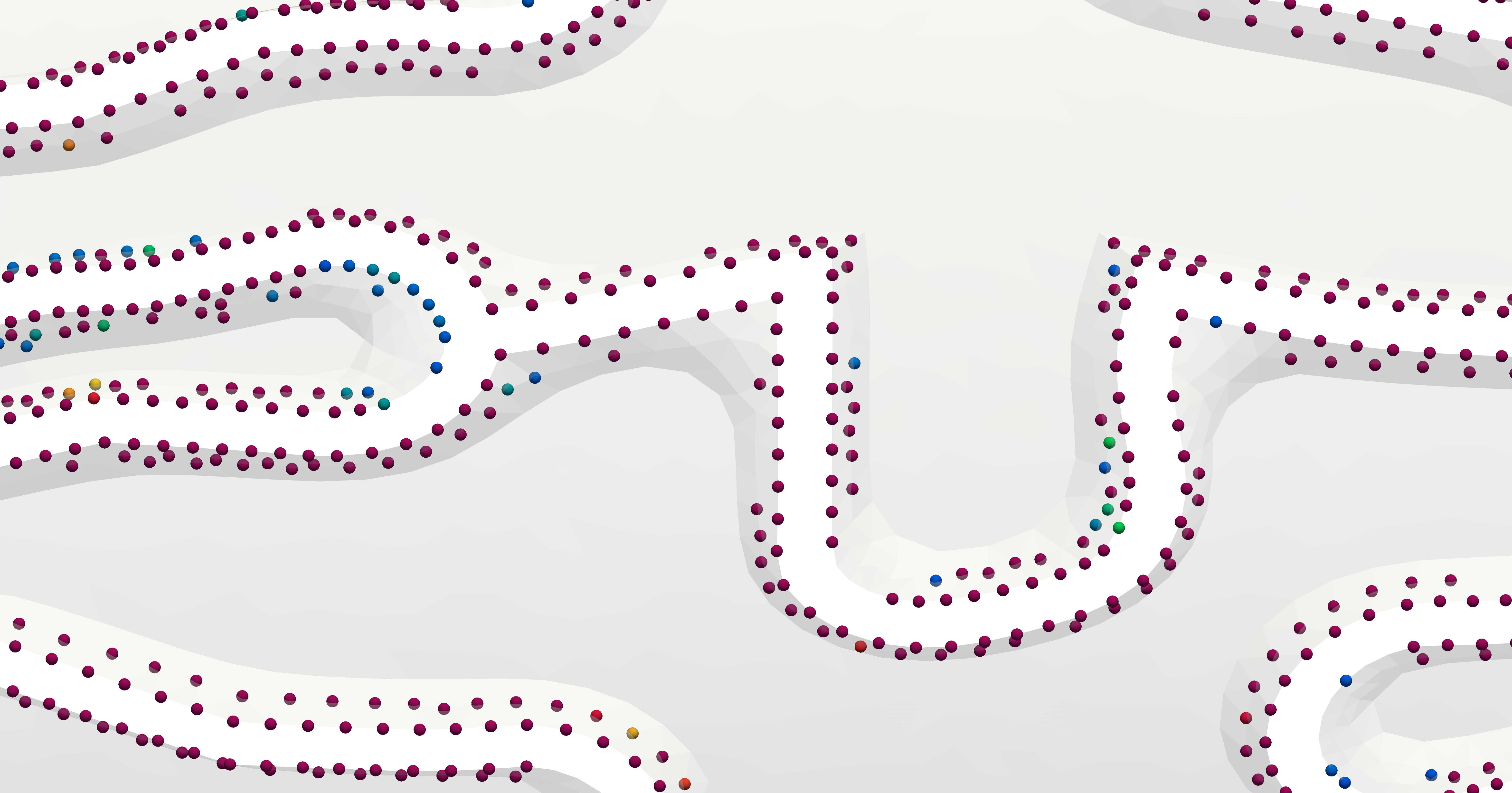}}
    }\\
    \subfloat[Reference, 10\%.]{
    \resizebox*{4.4cm}{!}{\includegraphics[draft=\draftmode]{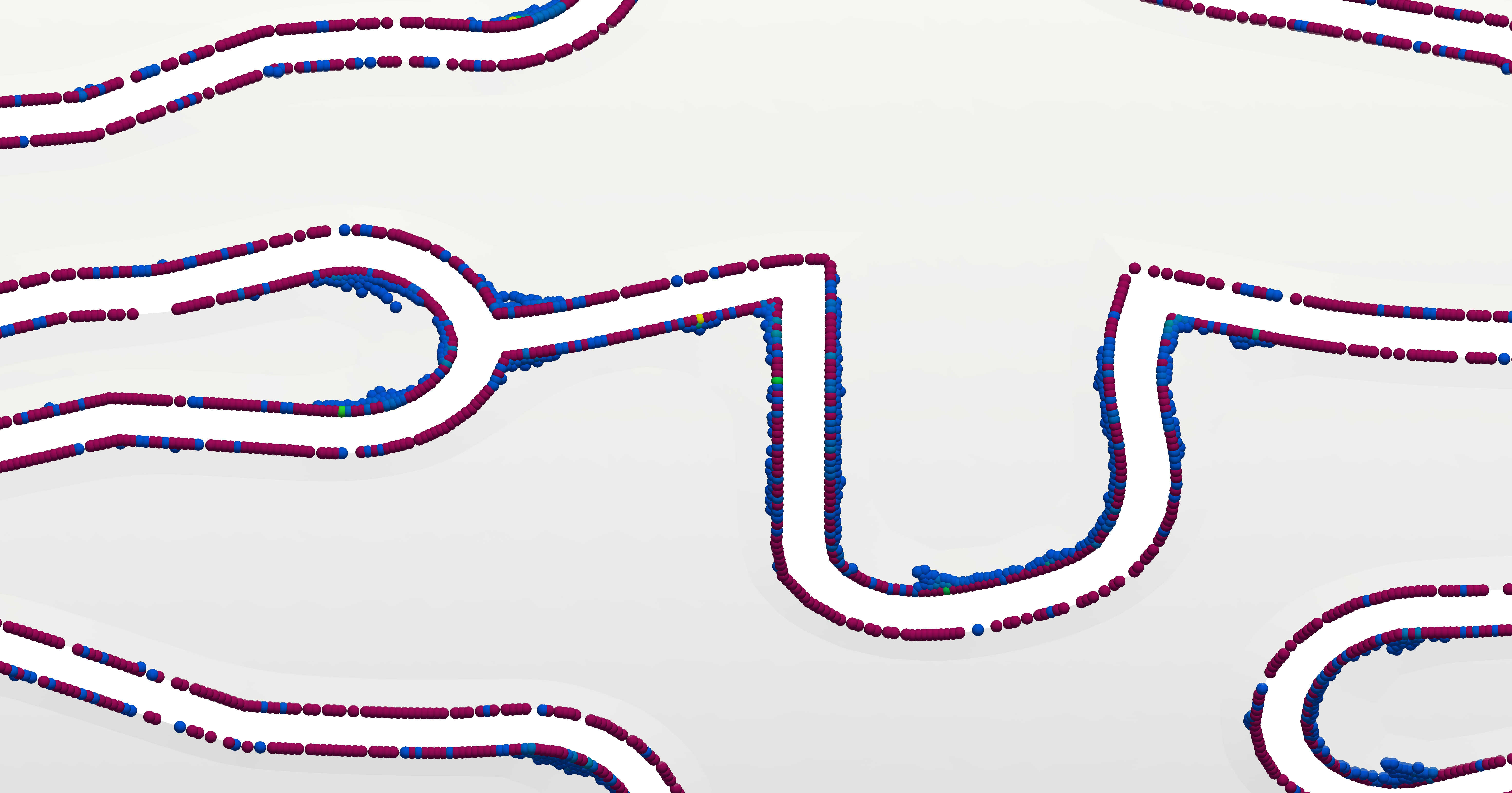}}
    \label{fig:LimitValues10}}
    \subfloat[TMR, 10\%.]{
    \resizebox*{4.4cm}{!}{\includegraphics[draft=\draftmode]{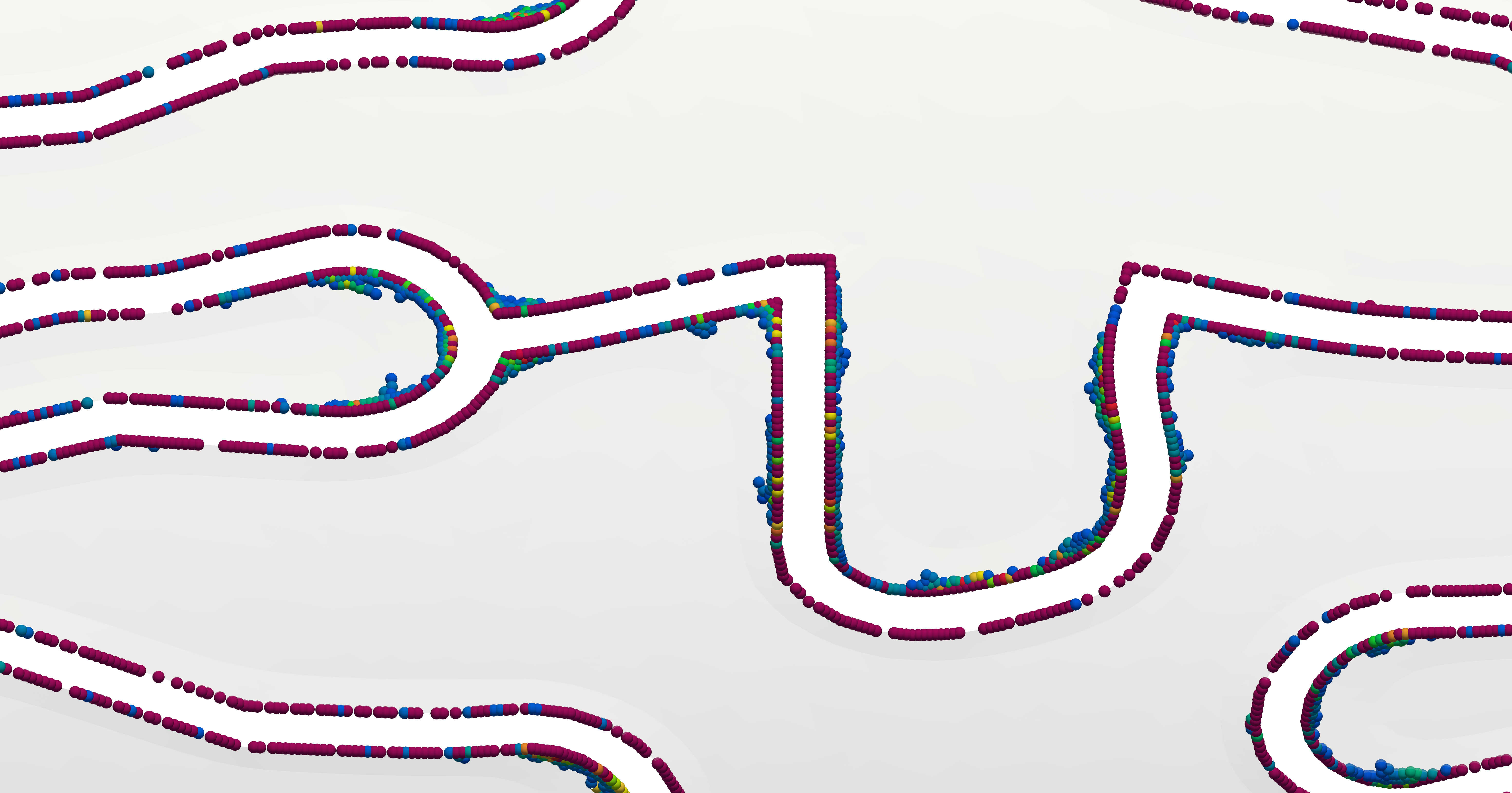}}
    }
    \subfloat[Uniform, 10\%.]{
    \resizebox*{4.4cm}{!}{\includegraphics[draft=\draftmode]{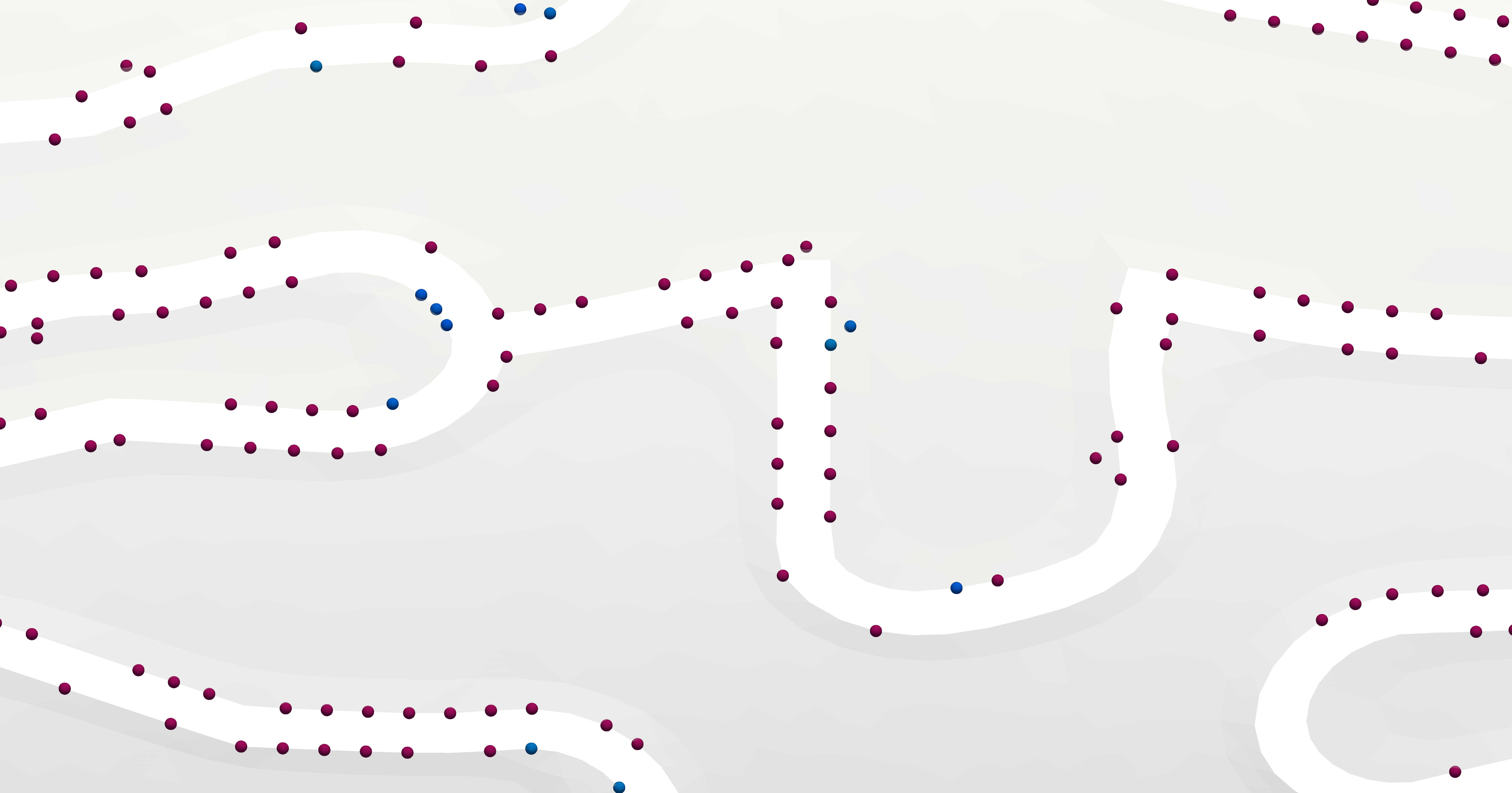}}
    }
    \caption{Areas of artery wall with limit values of hemodynamic indicators for all mesh refinements and two indentations. Zoom detail from Fig. \ref{fig:RRT75-10CSZoom}.}\label{fig:LimitPlots}
\end{figure}

Limit values of $\rrt > $ 1e+5 are represented by values of $\tawss = 0$ and $\osi = 0.5$. Figure \ref{fig:LimitPlots} highlight the areas with limit values for all mesh refinements with 75\% and 10\% indentation. These values surround the stent profile for both indentations. Looking at the reference cases, higher indentations show larger areas of limit values compared to 10\% indentation. The targeted mesh refinement reproduces very closely the limit values in the same areas of the artery wall. In the uniform case, large areas of limit values are missed, with underestimated RRT. The reference NEAF of limit values for RRT $>$ 1e+5 steadily increases for higher indentation percentages as shown in Fig. \ref{fig:LimitBubble} for all mesh refinements. However, NEAF values for reference and targeted mesh refinement are almost identical, while uniform meshes display much smaller NEAF for each indentation.

%% file: sections/4-discussion.tex
The analysis of blood velocity streamlines in \cite{manjunatha2024silico} show that small vortices are located near the stent struts. For this reason, we choose three mesh refinements with focus on the transition areas. Furthermore, hemodynamic indicators are more practical and can more precisely highlight the critical areas, compared to a qualitative analysis of streamlines.

From \cite{Cornelissen2023Development}, we expect to see WSS values closer to physiological ones in stented arteries with low indentation. In apparent contradiction, Fig. \ref{fig:WSS-low0.4} shows that for 10\% indentation we have larger areas of low WSS. This is due to the fact that low indentation means that the protruding stent shields the artery wall, which is therefore less exposed to the blood flow. However, Figure \ref{fig:WSS-lowZoom} shows that in the stent vicinity, the critical WSS magnitude increases for decreasing indentation with the following order (from lowest to highest values): 75\%, 50\%, 25\% and 10\%. Values of WSS can temporarily exceed the physiological range, but relevant side-effects are observed when low values of WSS are persistent. Thus, we make use of time-averaged indicators to highlight the critical areas over one cycle. High indentations have sharp variations of hemodynamic indicators from limit to physiological values, while lower indentations gradually span the whole values range. Most non-physiological values are concentrated in the transition area, especially for higher indentations. Thus, targeted mesh refinement is provides good accuracy thanks to the refinement in the stent vicinity.  

For a more quantitative overview, we investigate the empirical frequency of hemodynamic indicators, with a focus on critical values. Plotting frequency as histograms in Figure \ref{fig:TAWSS04hist}, we notice that higher indentations have larger areas of sub-threshold TAWSS, i.e., higher NEAF for $\tawss = 0$, and that the decaying order observed for WSS is maintained. Targeted mesh refinement shows very similar frequency trends for all indentations, compared to the reference case. The uniform mesh fails to maintain the frequency decaying order for TAWSS values close to 0.4 Pa. Frequency of critical OSI and RRT values also increase with indentation percentage. From Figures \ref{fig:LimitPlots} and \ref{fig:LimitBubble}, the targeted mesh refinement is able to reproduce very closely the reference frequency for limit values and to detect the same reference areas on the artery wall. Uniform meshes seem to be too coarse to accurately track areas at critical risk of ISR and these results are confirmed when evaluating the corresponding NEAF for all indentations.

Some limitations of this work are that the geometries do not take into account the stiffness variation of the artery due to stenosis and calcification and that we consider the artery wall to be fully rigid. Within the scope of this project we plan to fully couple the hemodynamics simulations to the moving wall in an FSI fashion to model both ISR and artery pulsation. Additionally, the authors are investigating model reduction techniques based on \cite{zunino2016integrated} and \cite{brandes2020hierarchically} to further reduce the computational effort.

%% file: sections/5-conclusion.tex
In this paper we investigate the effects of stent indentation on hemodynamic indicators in coronary arteries. We aim at providing a generally applicable in-silico study of risk factors of ISR, and thrombosis, based on stent implantation. To the best of our knowledge, there have been no numerical studies of hemodynamics in stented coronary arteries focusing on indentation effects. We investigate four indentation percentages with particular focus on sub-threshold TAWSS and limit values of OSI and RRT.

Assuming an idealized coronary artery makes this study independent of patient-specific parameters and allows to draw conclusions for a generic stent implantation procedure. This choice is also advantageous for calculating indentation percentages based on the initial diameter and for control on the extent of the transition area.

The results show that non-physiological values of hemodynamic indicators are detected close to the stent struts, mostly in the transition areas. Higher indentation percentages have higher NEAF of sub-threshold TAWSS, as well as critical OSI and RRT. The numerical results are in line with the experimental study \cite{Cornelissen2023Development} where higher indentation correlates with higher risk of induced inflammation, and ISR. Thus, hemodynamic indicators can be used as in-silico quantitative risk factors of ISR, in principle, for any stent design and artery geometry.

The study on three mesh refinements showed that targeted mesh refinement can moderately reduce computational costs while preserving enough accuracy. In particular, reference and targeted mesh refinement cases show nearly identical trends of NEAF for low TAWSS and limit values of OSI and RRT, while uniform meshes fail to detect such pattern. Hence, a certain level of mesh refinement is needed for accurate representation of hemodynamics in the stent vicinity.

%% file: main.bbl
\begin{thebibliography}{10}
\expandafter\ifx\csname url\endcsname\relax
  \def\url#1{\texttt{#1}}\fi
\expandafter\ifx\csname urlprefix\endcsname\relax\def\urlprefix{URL }\fi
\expandafter\ifx\csname href\endcsname\relax
  \def\href#1#2{#2} \def\path#1{#1}\fi

\bibitem{townsend2016cardiovascular}
N.~Townsend, L.~Wilson, P.~Bhatnagar, K.~Wickramasinghe, M.~Rayner, M.~Nichols, Cardiovascular disease in europe: epidemiological update 2016, European Heart Journal 37~(42) (2016) 3232--3245.

\bibitem{roth2020global}
G.~A. Roth, G.~A. Mensah, C.~O. Johnson, G.~Addolorato, E.~Ammirati, L.~M. Baddour, N.~C. Barengo, A.~Z. Beaton, E.~J. Benjamin, C.~P. Benziger, et~al., Global burden of cardiovascular diseases and risk factors, 1990--2019: update from the gbd 2019 study, Journal of the American College of Cardiology 76~(25) (2020) 2982--3021.

\bibitem{farooq2011restenosis}
V.~Farooq, B.~D. Gogas, P.~W. Serruys, Restenosis: delineating the numerous causes of drug-eluting stent restenosis, Circulation: Cardiovascular Interventions 4~(2) (2011) 195--205.

\bibitem{holmes2010stent}
D.~R. Holmes, D.~J. Kereiakes, S.~Garg, P.~W. Serruys, G.~J. Dehmer, S.~G. Ellis, D.~O. Williams, T.~Kimura, D.~J. Moliterno, Stent thrombosis, Journal of the American College of Cardiology 56~(17) (2010) 1357--1365.

\bibitem{khan2012drug}
W.~Khan, S.~Farah, A.~J. Domb, Drug eluting stents: developments and current status, Journal of controlled release 161~(2) (2012) 703--712.

\bibitem{douglas2012drug}
J.~S. Douglas, Drug-eluting stent restenosis: a need for new technology?, JACC: Cardiovascular Interventions 5~(7) (2012) 738--740.

\bibitem{buccheri2016understanding}
D.~Buccheri, D.~Piraino, G.~Andolina, B.~Cortese, Understanding and managing in-stent restenosis: a review of clinical data, from pathogenesis to treatment, Journal of thoracic disease 8~(10) (2016) E1150.

\bibitem{MCQUEEN2021do}
A.~McQueen, J.~Escuer, A.~Aggarwal, S.~Kennedy, C.~McCormick, K.~Oldroyd, S.~McGinty, Do we really understand how drug eluted from stents modulates arterial healing?, International Journal of Pharmaceutics 601 (2021) 120575.
\newblock \href {https://doi.org/https://doi.org/10.1016/j.ijpharm.2021.120575} {\path{doi:https://doi.org/10.1016/j.ijpharm.2021.120575}}.

\bibitem{takayama2011stent}
T.~Takayama, T.~Hiro, A.~Hirayama, Stent thrombosis and drug-eluting stents, Journal of cardiology 58~(2) (2011) 92--98.

\bibitem{zunino2009numerical}
P.~Zunino, C.~D’Angelo, L.~Petrini, C.~Vergara, C.~Capelli, F.~Migliavacca, Numerical simulation of drug eluting coronary stents: mechanics, fluid dynamics and drug release, Computer Methods in Applied Mechanics and Engineering 198~(45-46) (2009) 3633--3644.

\bibitem{antoniadis2015biomechanical}
A.~P. Antoniadis, P.~Mortier, G.~Kassab, G.~Dubini, N.~Foin, Y.~Murasato, A.~A. Giannopoulos, S.~Tu, K.~Iwasaki, Y.~Hikichi, et~al., Biomechanical modeling to improve coronary artery bifurcation stenting: expert review document on techniques and clinical implementation, Cardiovascular Interventions 8~(10) (2015) 1281--1296.

\bibitem{wang2015three}
H.~Wang, J.~Liu, X.~Zheng, X.~Rong, X.~Zheng, H.~Peng, Z.~Silber-Li, M.~Li, L.~Liu, Three-dimensional virtual surgery models for percutaneous coronary intervention (pci) optimization strategies, Scientific Reports 5~(1) (2015) 10945.

\bibitem{yang2017ale}
Y.~Yang, T.~Richter, W.~J{\"a}ger, M.~Neuss-Radu, An ale approach to mechano-chemical processes in fluid--structure interactions, International Journal for Numerical Methods in Fluids 84~(4) (2017) 199--220.

\bibitem{zun2017comparison}
P.~S. Zun, T.~Anikina, A.~Svitenkov, A.~G. Hoekstra, A comparison of fully-coupled 3d in-stent restenosis simulations to in-vivo data, Frontiers in Physiology 8 (2017) 284.

\bibitem{manjunatha2022multiphysics}
K.~Manjunatha, M.~Behr, F.~Vogt, S.~Reese, A multiphysics modeling approach for in-stent restenosis: Theoretical aspects and finite element implementation, Computers in Biology and Medicine 150 (2022) 106166.

\bibitem{romarowski2019novel}
R.~Romarowski, E.~Faggiano, M.~Conti, A.~Reali, S.~Morganti, F.~Auricchio, A novel computational framework to predict patient-specific hemodynamics after tevar: Integration of structural and fluid-dynamics analysis by image elaboration, Computers \& Fluids 179 (2019) 806--819.

\bibitem{silva2020modeling}
T.~Silva, W.~J{\"a}ger, M.~Neuss-Radu, A.~Sequeira, Modeling of the early stage of atherosclerosis with emphasis on the regulation of the endothelial permeability, Journal of Theoretical Biology 496 (2020) 110229.

\bibitem{forti2017monolithic}
D.~Forti, M.~Bukac, A.~Quaini, S.~Canic, S.~Deparis, A monolithic approach to fluid--composite structure interaction, Journal of Scientific Computing 72 (2017) 396--421.

\bibitem{ladisa2003three}
J.~F. LaDisa, I.~Guler, L.~E. Olson, D.~A. Hettrick, J.~R. Kersten, D.~C. Warltier, P.~S. Pagel, Three-dimensional computational fluid dynamics modeling of alterations in coronary wall shear stress produced by stent implantation, Annals of biomedical engineering 31 (2003) 972--980.

\bibitem{hachem2023reinforcement}
E.~Hachem, P.~Meliga, A.~Goetz, P.~J. Rico, J.~Viquerat, A.~Larcher, R.~Valette, A.~Sanches, V.~Lannelongue, H.~Ghraieb, et~al., Reinforcement learning for patient-specific optimal stenting of intracranial aneurysms, Scientific Reports 13~(1) (2023) 7147.

\bibitem{de2013computational}
G.~De~Santis, B.~Trachet, M.~Conti, M.~De~Beule, U.~Morbiducci, P.~Mortier, P.~Segers, P.~Verdonck, B.~Verhegghe, A computational study of the hemodynamic impact of open-versus closed-cell stent design in carotid artery stenting, Artificial Organs 37~(7) (2013) E96--E106.

\bibitem{chiastra2013computational}
C.~Chiastra, S.~Morlacchi, D.~Gallo, U.~Morbiducci, R.~C{\'a}rdenes, I.~Larrabide, F.~Migliavacca, Computational fluid dynamic simulations of image-based stented coronary bifurcation models, Journal of The Royal Society Interface 10~(84) (2013) 20130193.

\bibitem{calo2008multiphysics}
V.~Calo, N.~Brasher, Y.~Bazilevs, T.~Hughes, Multiphysics model for blood flow and drug transport with application to patient-specific coronary artery flow, Computational Mechanics 43 (2008) 161--177.

\bibitem{kolachalama2009luminal}
V.~B. Kolachalama, A.~R. Tzafriri, D.~Y. Arifin, E.~R. Edelman, Luminal flow patterns dictate arterial drug deposition in stent-based delivery, Journal of Controlled Release 133~(1) (2009) 24--30.

\bibitem{vergara2008multiscale}
C.~Vergara, P.~Zunino, Multiscale boundary conditions for drug release from cardiovascular stents, Multiscale Modeling \& Simulation 7~(2) (2008) 565--588.

\bibitem{gijsen1999influence}
F.~J. Gijsen, F.~N. van~de Vosse, J.~Janssen, The influence of the non-{N}ewtonian properties of blood on the flow in large arteries: steady flow in a carotid bifurcation model, Journal of Biomechanics 32~(6) (1999) 601--608.

\bibitem{leuprecht2001computer}
A.~Leuprecht, K.~Perktold, Computer simulation of non-newtonian effects on blood flow in large arteries, Computer Methods in Biomechanics and Biomedical Engineering 4~(2) (2001) 149--163.

\bibitem{behr2006models}
M.~Behr, D.~Arora, O.~Coronado, M.~Pasquali, Models and finite element techniques for blood flow simulation, International Journal of Computational Fluid Dynamics 20~(3-4) (2006) 175--181.

\bibitem{behbahani2009review}
M.~Behbahani, M.~Behr, M.~Hormes, U.~Steinseifer, D.~Arora, O.~Coronado, M.~Pasquali, A review of computational fluid dynamics analysis of blood pumps, European Journal of Applied Mathematics 20~(4) (2009) 363--397.

\bibitem{hassler2019variational}
S.~Ha{\ss}ler, L.~Pauli, M.~Behr, The variational multiscale formulation for the fully-implicit log-morphology equation as a tensor-based blood damage model, International Journal for Numerical Methods in Biomedical Engineering 35~(12) (2019) e3262.

\bibitem{marsden2014recent}
A.~L. Marsden, Y.~Bazilevs, C.~C. Long, M.~Behr, Recent advances in computational methodology for simulation of mechanical circulatory assist devices, Wiley Interdisciplinary Eeviews: Systems Biology and Medicine 6~(2) (2014) 169--188.

\bibitem{Sasse}
J.~Sasse, \href{https://publications.rwth-aachen.de/record/889071}{Shearthinning constitutive model in stented arteries}, Master's thesis, RWTH Aachen University (2020).
\newline\urlprefix\url{https://publications.rwth-aachen.de/record/889071}

\bibitem{hsiao2012hemodynamic}
H.-M. Hsiao, K.-H. Lee, Y.-C. Liao, Y.-C. Cheng, Hemodynamic simulation of intra-stent blood flow, Procedia Engineering 36 (2012) 128--136.

\bibitem{morbiducci2020wall}
U.~Morbiducci, V.~Mazzi, M.~Domanin, G.~De~Nisco, C.~Vergara, D.~A. Steinman, D.~Gallo, Wall shear stress topological skeleton independently predicts long-term restenosis after carotid bifurcation endarterectomy, Annals of biomedical engineering 48 (2020) 2936--2949.

\bibitem{park2016vivo}
H.~Park, J.~H. Park, S.~J. Lee, In vivo measurement of hemodynamic information in stenosed rat blood vessels using x-ray piv, Scientific reports 6~(1) (2016) 37985.

\bibitem{zingaro2021hemodynamics}
A.~Zingaro, F.~Menghini, A.~Quarteroni, et~al., Hemodynamics of the heart’s left atrium based on a variational multiscale-les numerical method, European Journal of Mechanics-B/Fluids 89 (2021) 380--400.

\bibitem{kronborg2023triple}
J.~Kronborg, J.~Hoffman, The triple decomposition of the velocity gradient tensor as a standardized real schur form, Physics of Fluids 35~(3) (2023).

\bibitem{rayz2010flow}
V.~Rayz, L.~Boussel, L.~Ge, J.~Leach, A.~Martin, M.~Lawton, C.~McCulloch, D.~Saloner, Flow residence time and regions of intraluminal thrombus deposition in intracranial aneurysms, Annals of Biomedical Engineering 38 (2010) 3058--3069.

\bibitem{nakazawa2008delayed}
G.~Nakazawa, A.~V. Finn, M.~Joner, E.~Ladich, R.~Kutys, E.~K. Mont, H.~K. Gold, A.~P. Burke, F.~D. Kolodgie, R.~Virmani, Delayed arterial healing and increased late stent thrombosis at culprit sites after drug-eluting stent placement for acute myocardial infarction patients: an autopsy study, Circulation 118~(11) (2008) 1138--1145.

\bibitem{cecchi2011role}
E.~Cecchi, C.~Giglioli, S.~Valente, C.~Lazzeri, G.~F. Gensini, R.~Abbate, L.~Mannini, Role of hemodynamic shear stress in cardiovascular disease, Atherosclerosis 214~(2) (2011) 249--256.

\bibitem{koskinas2012role}
K.~C. Koskinas, Y.~S. Chatzizisis, A.~P. Antoniadis, G.~D. Giannoglou, Role of endothelial shear stress in stent restenosis and thrombosis: pathophysiologic mechanisms and implications for clinical translation, Journal of the American College of Cardiology 59~(15) (2012) 1337--1349.

\bibitem{brindise2017hemodynamics}
M.~C. Brindise, C.~Chiastra, F.~Burzotta, F.~Migliavacca, P.~P. Vlachos, Hemodynamics of stent implantation procedures in coronary bifurcations: an in vitro study, Annals of Biomedical Engineering 45 (2017) 542--553.

\bibitem{jenei2016wall}
C.~Jenei, E.~Balogh, G.~T. Szab{\'o}, C.~A. D{\'e}zsi, Z.~K{\H{o}}szegi, Wall shear stress in the development of in-stent restenosis revisited. a critical review of clinical data on shear stress after intracoronary stent implantation, Cardiology Journal 23~(4) (2016) 365--373.

\bibitem{wentzel2001relationship}
J.~J. Wentzel, R.~Krams, J.~C. Schuurbiers, J.~A. Oomen, J.~Kloet, W.~J. van~der Giessen, P.~W. Serruys, C.~J. Slager, Relationship between neointimal thickness and shear stress after {W}allstent implantation in human coronary arteries, Circulation 103~(13) (2001) 1740--1745.

\bibitem{stone2003effect}
P.~H. Stone, A.~U. Coskun, S.~Kinlay, M.~E. Clark, M.~Sonka, A.~Wahle, O.~J. Ilegbusi, Y.~Yeghiazarians, J.~J. Popma, J.~Orav, et~al., Effect of endothelial shear stress on the progression of coronary artery disease, vascular remodeling, and in-stent restenosis in humans: in vivo 6-month follow-up study, Circulation 108~(4) (2003) 438--444.

\bibitem{he2020mechanistic}
R.~He, L.~Zhao, V.~V. Silberschmidt, Y.~Liu, Mechanistic evaluation of long-term in-stent restenosis based on models of tissue damage and growth, Biomechanics and Modeling in Mechanobiology 19 (2020) 1425--1446.

\bibitem{soulis2011relative}
J.~V. Soulis, O.~P. Lampri, D.~K. Fytanidis, G.~D. Giannoglou, Relative residence time and oscillatory shear index of non-newtonian flow models in aorta, in: 2011 10th international workshop on biomedical engineering, IEEE, 2011, pp. 1--4.

\bibitem{john2017influence}
L.~John, P.~Pust{\v{e}}jovsk{\'a}, O.~Steinbach, On the influence of the wall shear stress vector form on hemodynamic indicators, Computing and Visualization in Science 18 (2017) 113--122.

\bibitem{auricchio2013patient}
F.~Auricchio, M.~Conti, A.~Ferrara, S.~Morganti, A.~Reali, Patient-specific finite element analysis of carotid artery stenting: a focus on vessel modeling, International Journal for Numerical Methods in Biomedical Engineering 29~(6) (2013) 645--664.

\bibitem{zunino2016integrated}
P.~Zunino, J.~Tamba{\v{c}}a, E.~Cutr{\`\i}, S.~{\v{C}}ani{\'c}, L.~Formaggia, F.~Migliavacca, Integrated stent models based on dimension reduction: review and future perspectives, Annals of Biomedical Engineering 44 (2016) 604--617.

\bibitem{kim2010patient}
H.~J. Kim, I.~Vignon-Clementel, J.~Coogan, C.~Figueroa, K.~Jansen, C.~Taylor, Patient-specific modeling of blood flow and pressure in human coronary arteries, Annals of Biomedical Engineering 38 (2010) 3195--3209.

\bibitem{takizawa2012patient}
K.~Takizawa, K.~Schjodt, A.~Puntel, N.~Kostov, T.~E. Tezduyar, Patient-specific computer modeling of blood flow in cerebral arteries with aneurysm and stent, Computational Mechanics 50 (2012) 675--686.

\bibitem{colombo2020computing}
M.~Colombo, M.~Bologna, M.~Garbey, S.~Berceli, Y.~He, J.~F.~R. Matas, F.~Migliavacca, C.~Chiastra, Computing patient-specific hemodynamics in stented femoral artery models obtained from computed tomography using a validated 3{D} reconstruction method, Medical Engineering \& PPhysics 75 (2020) 23--35.

\bibitem{conti2016carotid}
M.~Conti, C.~Long, M.~Marconi, R.~Berchiolli, Y.~Bazilevs, A.~Reali, Carotid artery hemodynamics before and after stenting: A patient specific {CFD} study, Computers \& Fluids 141 (2016) 62--74.

\bibitem{Cornelissen2023Development}
A.~Cornelissen, R.~A. Florescu, S.~Reese, M.~Behr, A.~Ranno, K.~Manjunatha, N.~Schaaps, C.~B{\"o}hm, E.~A. Liehn, L.~Zhao, et~al., In-vivo assessment of vascular injury for the prediction of in-stent restenosis, International Journal of Cardiology 388 (2023) 131151.

\bibitem{schwartz1992restenosis}
R.~S. Schwartz, K.~C. Huber, J.~G. Murphy, W.~D. Edwards, A.~R. Camrud, R.~E. Vlietstra, D.~R. Holmes, Restenosis and the proportional neointimal response to coronary artery injury: results in a porcine model, Journal of the American College of Cardiology 19~(2) (1992) 267--274.

\bibitem{ding2009xience}
N.~Ding, S.~D. Pacetti, F.-W. TANG, M.~Gada, W.~Roorda, Xience {V}™ stent design and rationale, Journal of Interventional Cardiology 22 (2009) S18--S27.

\bibitem{Poncin2004}
P.~Poncin, J.~Proft, Stent tubing: Understanding the desired attributes, in: Materials \& Processes for Medical Devices Conference, 8-10 September, 2003, pp. 253--259.

\bibitem{dodge1992lumen}
J.~T. Dodge~Jr, B.~G. Brown, E.~L. Bolson, H.~T. Dodge, Lumen diameter of normal human coronary arteries. influence of age, sex, anatomic variation, and left ventricular hypertrophy or dilation., Circulation 86~(1) (1992) 232--246.

\bibitem{manjunatha2024silico}
K.~Manjunatha, A.~Ranno, J.~Shi, N.~Schaaps, P.~Nilcham, A.~Cornelissen, F.~Vogt, M.~Behr, S.~Reese, In silico reproduction of the pathophysiology of in-stent restenosis, arXiv:2401.03961 (2024).
\newblock \href {http://arxiv.org/abs/2401.03961} {\path{arXiv:2401.03961}}.

\bibitem{moore2002fluid}
J.~E. Moore, J.~L. Berry, Fluid and solid mechanical implications of vascular stenting, Annals of biomedical engineering 30 (2002) 498--508.

\bibitem{bertolotti2001numerical}
C.~Bertolotti, V.~Deplano, J.~Fuseri, P.~Dupouy, Numerical and experimental models of post-operative realistic flows in stenosed coronary bypasses, Journal of Biomechanics 34~(8) (2001) 1049--1064.

\bibitem{pauli2017stabilized}
L.~H. Pauli, M.~Behr, On stabilized space-time fem for anisotropic meshes: incompressible navier--stokes equations and applications to blood flow in medical devices, International Journal for Numerical Methods in Fluids 85~(3) (2017) 189--209.

\bibitem{donea2003finite}
J.~Donea, A.~Huerta, Finite element methods for flow problems, John Wiley \& Sons, 2003.

\bibitem{forti2015semi}
D.~Forti, L.~Ded{\`e}, Semi-implicit bdf time discretization of the navier--stokes equations with vms-les modeling in a high performance computing framework, Computers \& Fluids 117 (2015) 168--182.

\bibitem{saad1986gmres}
Y.~Saad, M.~H. Schultz, Gmres: A generalized minimal residual algorithm for solving nonsymmetric linear systems, SIAM Journal on Scientific and Statistical Computing 7~(3) (1986) 856--869.

\bibitem{saad2003iterative}
Y.~Saad, Iterative methods for sparse linear systems, SIAM, 2003.

\bibitem{krause2018jureca}
D.~Krause, P.~Th{\"o}rnig, Jureca: modular supercomputer at j{\"u}lich supercomputing centre, Journal of large-scale research facilities JLSRF 4 (2018) A132--A132.

\bibitem{benard2006computational}
N.~Benard, R.~Perrault, D.~Coisne, Computational approach to estimating the effects of blood properties on changes in intra-stent flow, Annals of Biomedical Engineering 34 (2006) 1259--1271.

\bibitem{eilers2010splines}
P.~H. Eilers, B.~D. Marx, Splines, knots, and penalties, Wiley Interdisciplinary Reviews: Computational Statistics 2~(6) (2010) 637--653.

\bibitem{matlabcftool}
I.~The~MathWorks, \href{https://mathworks.com/help/curvefit}{Curve Fitting Toolbox}, Natick, Massachusetts, United State (2021).
\newline\urlprefix\url{https://mathworks.com/help/curvefit}

\bibitem{brandes2020hierarchically}
Y.~A. Brandes Costa~Barbosa, S.~Perotto, Hierarchically reduced models for the stokes problem in patient-specific artery segments, International Journal of Computational Fluid Dynamics 34~(2) (2020) 160--171.

\end{thebibliography}
